\documentclass[aps,amsmath,twocolumn,amssymb,floatfix,showpacs,superscriptaddress,nofootinbib,longbibliography]{revtex4-1}
\usepackage{mathtools}
\usepackage{braket}
\usepackage[dvipsnames]{xcolor}
\usepackage{float}
\usepackage{subfigure}
\usepackage{dsfont}
\usepackage{tocbasic}
\usepackage{braket}
\usepackage[dvipsnames]{xcolor}
\usepackage{float}
\usepackage{subfigure}
\usepackage{tikz}
\usepackage[colorlinks=true,linktoc=page,linkcolor=magenta,citecolor=blue,urlcolor=purple]{hyperref}
\usepackage{multirow}

\mathchardef\mhyphen="2D 

\newcommand{\ie}{{i.e.,\,\,}}
\newcommand{\eg}{{e.g.,~}}
\newcommand{\ua}{{\uparrow }}
\newcommand{\da}{{\downarrow }}

\newcommand\bea{\begin{eqnarray}}
	\newcommand\eea{\end{eqnarray}}
\newcommand\beq{\begin{equation}}  
	\newcommand\eeq{\end{equation}}

\newcommand{\non}{\nonumber}

\newcommand{\mc}{\mathcal}

\newcommand{\tcr}{\textcolor{red}} 
\usepackage[normalem]{ulem}
\definecolor{lime}{HTML}{A6CE39}
\usepackage{sidecap,tikz}
\DeclareRobustCommand{\orcidicon}{\hspace{-1.0mm}
	\begin{tikzpicture}
		\draw[lime, fill=lime] (0.0,0.0) 
		circle [radius=0.15] 
		node[white] {{\fontfamily{qag}\selectfont \tiny \,ID}};
		\draw[white, fill=white] (-0.0525,0.095) 
		circle [radius=0.007];
	\end{tikzpicture}
	\hspace{-3.0mm}
}
\foreach \x in {A, ..., Z}{\expandafter\xdef\csname orcid\x\endcsname{\noexpand\href{https://orcid.org/\csname orcidauthor\x\endcsname}{\noexpand\orcidicon}}
}

\AtBeginDocument{%
	\newwrite\bibnotes
	\def\bibnotesext{Notes.bib}
	\immediate\openout\bibnotes=\jobname\bibnotesext
	\immediate\write\bibnotes{@CONTROL{REVTEX41Control}}
	\immediate\write\bibnotes{@CONTROL{%
			apsrev41Control,author="08",editor="1",pages="1",title="1",year="1"}}
	\if@filesw
	\immediate\write\@auxout{\string\citation{apsrev41Control}}%
	\fi
}%
\begin{document}

	
	\title{Topological superconductivity and  superconducting diode effect mediated via  unconventional magnet and Ising spin-orbit coupling}  
	
	\author{Amartya Pal\orcidA}
	\affiliation{Institute of Physics, Sachivalaya Marg, Bhubaneswar-751005, India}
	\affiliation{Homi Bhabha National Institute, Training School Complex, Anushakti Nagar, Mumbai 400094, India}

	\author{Debashish Mondal\orcidD{}}
	\affiliation{Institute of Physics, Sachivalaya Marg, Bhubaneswar-751005, India}
	\affiliation{Homi Bhabha National Institute, Training School Complex, Anushakti Nagar, Mumbai 400094, India}

	\author{Tanay Nag\orcidB{}}
	\email{tanay.nag@hyderabad.bits-pilani.ac.in}
	\affiliation{Department of Physics, BITS Pilani-Hyderabad Campus, Telangana 500078, India}
	
	\author{Arijit Saha\orcidC{}}
	\email{arijit@iopb.res.in}
	\affiliation{Institute of Physics, Sachivalaya Marg, Bhubaneswar-751005, India}
	\affiliation{Homi Bhabha National Institute, Training School Complex, Anushakti Nagar, Mumbai 400094, India}

\begin{abstract}
	We propose a theoretical framework in which a one-dimensional (1D) tight-binding model incorporating unconventional magnetic order together with Rashba and Ising spin–orbit couplings are considered to realize two key phenomena in condensed matter systems: topological superconductivity and the superconducting diode effect (SDE). We first elucidate the underlying band topology of the normal-state Hamiltonian and subsequently introduce an on-site attractive Hubbard interaction. Performing a a self-consistent mean-field analysis, we establish superconducting order parameters in both the conventional Bardeen–Cooper–Schrieffer (BCS) and finite-momentum Fulde–Ferrell–Larkin–Ovchinnikov (FFLO) pairing channels. Intriguingly, both pairing states can support topological superconductivity, characterized by a nontrivial winding number, and lead to the emergence of four zero-energy Majorana modes localized at the ends of the 1D chain. The FFLO state further gives rise to an intrinsic field-free SDE, manifested as a nonreciprocal supercurrent and quantified by the diode efficiency $\eta$. Notably, our model yields a large diode efficiency $\eta \sim 65\%$, highlighting its potential for realising topological superconductivity and highly efficient superconducting devices.
\end{abstract}

	%
	\maketitle
	%
\section{Introduction} 
The interplay of magnetism and superconductivity gives birth to a diverse range of intriguing phenomena in condensed matter systems. Topological superconductivity~\cite{Kitaev_2001,Alicea_2012,Alicea2011_NatPhys,Leijnse_2012,Pientka2013_Shiba,Nadj2013_Shiba}, one of the prominent outcomes of such interplay, has inspired decades-long research owing to its connection to Majorana zero modes (MZMs). It is well known that MZMs are charge-neutral zero energy quasi-particle excitations in topological superconductors (TSCs) obeying non-abelian braiding statistics. This property enables them to be the potential candidate for realising fault-tolerant topological quantum computation~\cite{Beenakker2013search,Kitaev2003_annals_of_phys,NayakRMP2008,Yazdani_hunt_majorana}. Numerous realistic platforms have been proposed in one dimension for engineering TSCs both from theoretical and experimental perspectives \eg Rashba spin-orbit coupled nanowire with external Zeeman field in close proximity to a regular $s$-wave superconductor (SC)~\cite{Alicea_2012,Alicea2011_NatPhys,LutchynPRL2010,Leijnse_2012,TewariPRL2012,Rokhinson2012,LawPRL2009,Mourik2012Science,Das2012_NatPhys,Albrecht2016,ChenSciAdv2017,Mondal2024,Mondal2023_NW,Arouca2024}, magnetic impurities deposited on the surface of various SCs~\cite{Pientka2013_Shiba,Nadj2013_Shiba,Klinovaja2013_shiba,Pientka2014,Christensen2016,Sharma2016_shiba,Chatterjee2023,Chatterjee2024_PRBLa,Chatterjee2024_PRBLb,Subhadarshini2024,Subhadarshini2025,Mondal_2023_Shiba,Yazdani_2015,Soldini2023,Wang2021PRL} etc. 

On the other hand, nonreciprocal charge transport builds the basis of modern electronic devices such as diodes, transistors, and rectifiers~\cite{Scaff1947,Shockley1949,Nagaosa2024_diode}. However, the finite resistivity in these systems inevitably leads to Joule heating and power loss. As a possible remedy, superconducting diode effect (SDE)~\cite{Nadeem2023,Jiang2022,Daido2022_SDE_PRL} has recently emerged as a novel phenomena for realising energy-efficient nonreciprocal superconducting rectification devices.  In superconducting diodes, based on the principle of SDE, current flow becomes dissipationless only in one direction while resistive in the opposite direction~\cite{Nadeem2023,Jiang2022,Daido2022_SDE_PRL,LiangFu2022_PNAS,He_2022}. The magnetochiral anisotropy manifests itself in SDE when the critical current density, 
above which superconducting order is completely destroyed, is direction dependent \ie $J_c^+ \ne J_c^-$ where, $J_c^+ (J_c^-)$ denotes the critical Cooper pair (CP) current flowing along the forward (reverse) direction. More specifically, SDE is realised when the magnitude to probe superconducting current lies between $J_c^+$ and $J_c^-$~\cite{Jiang2022}. Typically, SDE arises due to the combined breaking of inversion and time-reversal symmetries (TRS). In one of the microscopic mechanism, the phenomena of SDE is intimately connected with finite-momentum superconductivity namely Fulde–Ferrell–Larkin–Ovchinnikov (FFLO) pairing state~\cite{Fulde1964,Larkin_1964,LiangFu2022_PNAS} harboring the CPs with a finite center-of-mass momentum.
Experimental realization of SDE has been reported in Nb/Ta/V superlattices~\cite{Ando2020_SDE_Expt}, van der Waals heterostructure~\cite{Wu2022}, small-twist-angle trilayer graphene~\cite{Lin2022}, topological semimetals~\cite{Pal2022} etc. This triggers further extensive theoretical~\cite{Bergeret,Sigrist2022,Banerjee2024_PRL,LiangFu2022_JDE,LiangFu2021_PNAS,Hasan2024,Legg2022_SDE_MCA,Daido_2022_PRB,Debnath2024_diode,Chatterjee_2024_thermal,Sayan_Mondal_2025,Bhowmik2025a,Bhowmik2025b,Debnath2026a,Pal2026_pWM} and experimental activity~\cite{Bauriedl2022,Ghosh2024} in this field.

Unconventional magnetism has recently become a major frontier in condensed matter research.
One of such intriguing magnetic order appears to be the altermagnets (AMs) representing a class of collinear compensated antiferromagnets with broken TRS~\cite{Smejkal_PRX_1,Smejkal_PRX_2,BhowalPRX2024,Bai_PRL_2023,Lee2024MnTe,Lin2025}. In AMs, opposite spin-sublattices are connected through a rotation rather than translation or reflection, leading to spin-split energy bands caused by nonrelativistic spin-orbit coupling (SOC) while maintaining zero magnetization. 
Discovery of AMs has sparked extentsive theoretical research interests \eg TSCs~\cite{Ghorashi2024PRL,Mondal2025PRBL,Pal2025Flq_Josephson,Li2024,Li_PRBL_2024,Zhu2023,Yin2025PRB,Alam2025}, unconventional superconducting orders~\cite{Maeda2025,Fukaya_2025,Zhang2024,Lu2025PRL,Fukaya2025PRB_JJ,Sun2025_PRB,Ouassou2023PRL}, SDEs~\cite{Banerjee2024_PRB,Chakraborty2025_PRL,Samanta2025,Debnath_2025}, spintronics~\cite{FuPeiHo2025} etc. AMs exhibit significant advantages in realising TSCs due to  
vanishing magnetization and enhanced bulk topological superconducting gap compared to the external Zeeman field setups~\cite{Ghorashi2024PRL,Mondal2025PRBL,Pal2025Flq_Josephson}. 
However, it suffers from accidental zero modes, initially claimed to be MZMs~\cite{Ghorashi2024PRL} while later clarified to be nontopological modes lacking topological protection~\cite{Mondal2025PRBL,Pal2025Flq_Josephson}. Moreover, recent experiments on Ising superconductivity suggest unconventional superconducting phases with Ising SOC~\cite{Wickramaratne2023,Zhang2025_Ising,Lu2015}. Motivated by these facts, in this paper, we ask the following intriguing questions, (i) Can we formulate a unified model that captures the interplay between relativistic spin-orbit coupling (SOC), such as Rashba and Ising types, and non-relativistic SOC arising from unconventional magnetism, 
in a way that supports topological phases in both the normal and superconducting states? (ii) Implementing the above model, how to generate field-free SDE in one-dimension without any external magnetic field? 

\begin{figure}[h]
	\includegraphics[width=\columnwidth]{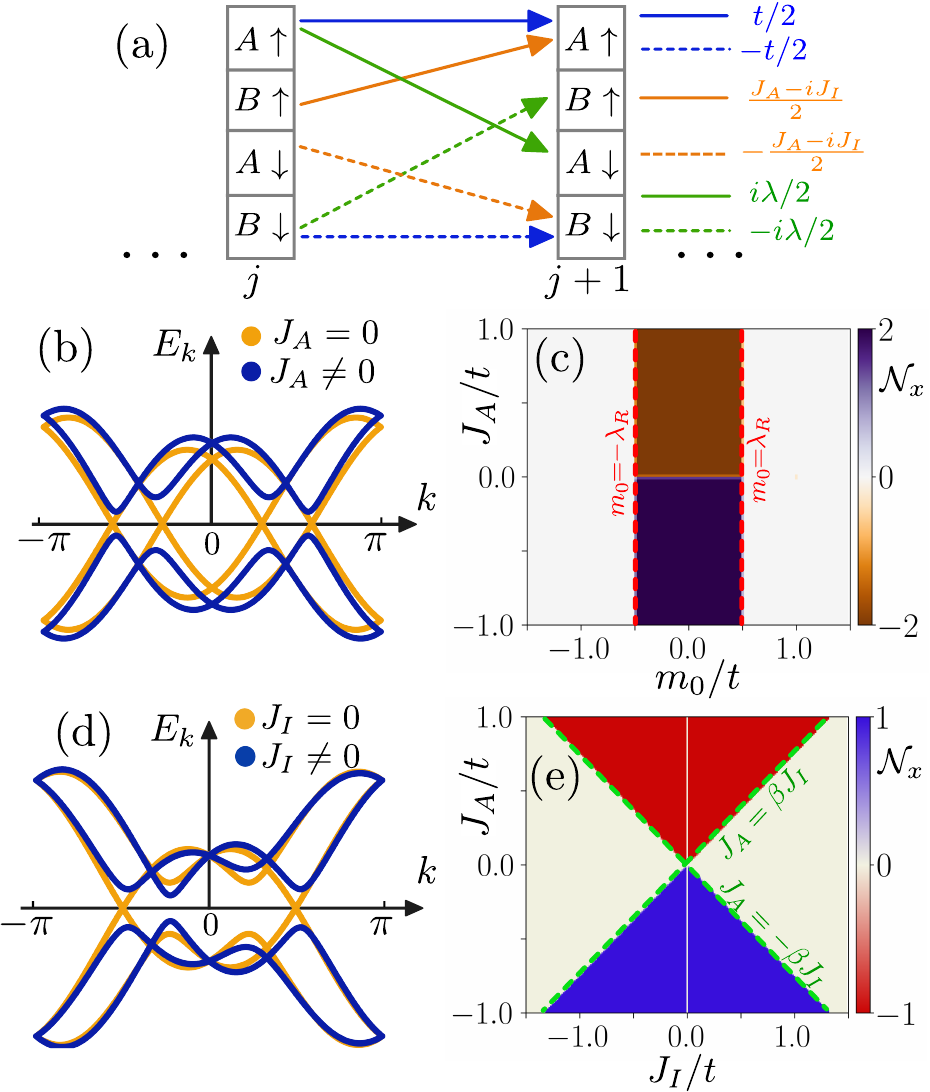}
	\caption{{\bf Band topology of the normal state Hamiltonian:} (a) Schematic illustration of a set of hopping terms belonging to$\mc{H}(k)$ between the sites $j$ and $j+1$ are depicted. (b)  Bulk spectrum $E_k$ as a function of $k$ is depicted with $(m_0,\lambda_R,J_I)=(0.2t,0.5t,0)$, $J_A=0$ (orange) and $J_A=0.6t$ (blue). (c) Winding number $\mc{N}_x$ is shown in the  ($m_0-J_A$) plane with $(\lambda_R,J_I)=(0.5t,0)$. (d) Bulk spectrum $E_k$ with respect to $k$  is shown with $(m_0,\lambda_R,J_A)=(0.5t,0.5t,0.4t)$, $J_I=0$ (orange) and $J_I=0.25t$ (blue). (e) Winding number $\mc{N}_x$ is displayed in the  $(J_I-J_A)$ plane choosing $(m_0,\lambda_R)=(0.5t,0.5t)$.}
	\label{Fig.1}
\end{figure}

In this paper, we intend to answer the following questions by proposing  a 1D model Hamiltonian that gives rise to field-free generation of topological superconductivity and SDE without nontopological zero modes. The key ingredients include an unconventional magnetic order resembling $d$-wave AMs, Rashba and Ising SOC in a semiconducting 1D nanowire. First, we analyse the band topology of the normal state Hamiltonian and establish the presence of a topological insulating phase characterized by $Z$ invariant (see Fig.\,\ref{Fig.1}). Then, assuming the presence of onsite attractive Hubbard interaction, we systematically perform a mean-field decomposition in the BCS and FFLO pairing channels and self-consistently compute the ground state superconducting orders. Our self-consistent analysis reveals the onset of topological superconductivity in the BCS channel in the absence of Ising SOC (see Fig.\,\ref{Fig.2}) and in the FFLO channel in presence of Ising SOC (see Fig.\,\ref{Fig.3}). We also topologically characterize the TSC phase in both BCS and the FFLO channel. As a manifestantion of the FFLO pairing state, 
we investigate the emergence of SDE in our system. Importantly, by suitably tuning the parameters of the model, a significant superconducting diode efficiency $\eta\sim 65\%$ 
can be achieved as illustrated in Fig.\,\ref{Fig.4}. This offers a possible platform for realizing field-free energy-efficient superconducting electronic device.
 
\section{Band topology of the normal state Hamiltonian}
We begin by introducing a 1D model Hamiltonian that realizes a topological insulating phase driven by unconventional magnetism and SOC (schematically shown in Fig.~\ref{Fig.1}(a)). The corresponding tight-binding Hamiltonian in the momentum space (assuming lattice spacing, $a=1$) can be written as~\cite{Ghorashi2024PRL,Mondal2025PRBL}, 
$	H = \sum_{k=-\pi}^{\pi} \psi_k^\dagger\,\, \mc{H}(k) \,\,\psi_k \non$
where,
\begin{eqnarray}
	\mc{H}(k) &=& (t \cos k -m_0) \sigma_0 \tau_z + J_A \cos k \sigma_z \tau_x \non \\  &&+ \, \lambda_R \sin k  \sigma_y \tau_0 + J_I \sin k \sigma_z \tau_x\ , 	\label{Eq.normal_state_Ham}
\end{eqnarray}
with $\psi_k = \{ c_{kA\ua}, c_{kB\ua}, c_{kA\da}, c_{kB\da}\}^T$ and $c_{\alpha,s}$ ($c^\dagger_{k\alpha,s}$) representing the annhilation (creation) operator for an electron with momentum $k$ and spin $s=\{\ua,\da\}$ in the orbital $\alpha=\{A,B\}$. The Pauli matrices $\sigma$ and $\tau$ act on the spin and orbital degrees of freedom, respectively. The model parameters $t$, $m_0$, $J_A$ denote the strength of nearest neighbour hopping amplitude, staggered on-site mass term, and the unconventional magnetic order (as it creates momentum dependent spin-splitting along with inter-orbital exchange) while $\lambda_R$ and $J_I$ indicate the magnitude of Rashba and Ising SOC with in-plane and out-of-plane spin polarization, respectively. In principle, Ising SOC can be induced via coupling to a transition metal dichalcogenides hosting strong out-of-plane spin polarization~\cite{Skliannyi2025,Xie2023,Zhao2023_Ising}. Both SOC terms individually break inversion symmetry. In supplementary material (SM)~\cite{supp}, we have explicitly presented the vanishing magnetization of the normal state. Importantly, the magnetic order ($J_A\ne0$) breaks the time reversal symmetry (TRS) $\mc{T}=i\sigma_y \mc{K}$ with $\mc{K}$ being the complex conjugation operator: $\mc{T}\mc{H}(k)\mc{T}^{-1}\ne \mc{H}(-k)$. However, the Hamiltonian $\mc{H}(k)$ preserves the chiral symmetry $\mc{S}=\sigma_x \tau_x$: $\mc{S}\mc{H}(k)\mc{S}^{-1}=-\mc{H}(k)$. 

We first examine the effect of Rashba SOC where Ising SOC is always absent. We depict the corresponding band dispersions in Fig.\,\ref{Fig.1}(b) and analytically find that the spectrum becomes gapless for $J_A=0$ and $|m_0|\le \sqrt{\lambda_R^2 + t^2}$ at four isolated momenta in the Brillouin zone (BZ). A finite $J_A$ gaps out these points as shown in Fig.\,\ref{Fig.1}(b), indicating the onset of a topological insulating phase. Following the ten-fold classification of the gapped topological phases, the Hamiltonian $\mc{H}(k)$ belongs to the AIII topological class~\cite{ChiuRMP2016} with a $Z$ invariant as $\mc{H}(k)$ preserves $\mc{S}$ and breaks $\mc{T}$. The corresponding $Z$ topological index can be written in terms of the winding number defined as~ \cite{RyuNJP2010,ChiuRMP2016,Mondal2025PRBL,Benalcazar2022_PRL,Pal2025_WSM}, 
\begin{eqnarray}
	\mc{N}_x=\frac{i}{2\pi} \int_{\rm{BZ}} dk \operatorname{Tr}\left[q^{-1}(k) \partial_{k} q(k)\right] \label{Eq:winding_num}\ .
\end{eqnarray}
where, $q(k)$ can be obtained by recasting $\mc{H}(k)$ into antidiagonal form by utilizing the presence of Chiral symmetry in the system (see  (SM)~\cite{supp} for the details). We compute the winding number $\mc{N}_x$ and depict in the $J_A- m_0$ plane with $\lambda_R=0.5t$ in Fig.\,\ref{Fig.1}(c). Notably, we find $\mc{N}_x=\pm 2$ for $J_A\ne0$ and $|m_0|<\lambda_R$ establishing the emergence of topological insulating phase hosting four end localized zero energy modes. 
The number of such zero modes become four due to the two orbital degrees of freedom 
(see SM\,\cite{supp} for detailed analysis). The topological phase transition occurs at $m_0=\lambda_R$ where the bulk is gapless in the $m_0-J_A$ plane. This can be controlled by tuning the strength of $\lambda_R$ (see SM\,\cite{supp} for the derivation).

Then, we examine the interplay of both Rashba and Ising SOC where evolution of band structure as  $J_I$ increases from zero is studied. We always maintain $J_A$ finite to analyze topological aspects extensively.  We choose $m_0=\lambda_R, J_A\ne 0$, resulting in two gapless points at $k=\pm \pi/2$ as shown in Fig.\,\ref{Fig.1}(d). Now, inclusion of Ising SOC ($J_I\ne 0$) gaps out these two gapless points in the BZ. However, it is crucial to note that the bulk band structure in Fig.\,\ref{Fig.1}(d) becomes asymmetric about $k=0$, in sharp contrast to the bulk band structure depicted in Fig.\,\ref{Fig.1}(b), which is symmetric around $k=0$. This carries a striking consequence while demonstrating SDE, which we  discuss later. For topological characterization, we again compute $\mc{N}_x$ in the $J_I-J_A$ plane maintaining $m_0=\lambda_R$ and depict the corresponding variation 
in Fig.\,\ref{Fig.1}(e). We observe $\mc{N}_x=\pm 1$ depending on the sign of $J_A$  with phase boundary located  at $J_A=\pm \beta J_I$ in the $J_I-J_A$ plane with $\beta=\frac{t^2-m_0^2}{2m_0t}$ 
(see SM~\cite{supp} for the derivation). These results establish that the interplay between unconventional magnetism and Ising SOC can drive topological phase transition with tunable number of zero modes.

\section{Realizing TSC utilizing self-consistent mean-field analysis}
We next tune the system to the phase boundary between the trivial and nontrivial topological insulating phases, where the normal state hosts a finite density of states near the Fermi level. We now seek possible superconducting instabilities in the normal state, assuming the presence of onsite intra-orbital attractive electron-electron ($e\mhyphen e$) Hubbard interaction of the form, 
\begin{equation}
	H_U = -U\!\!\!\!\!\!\! \sum_{i,\alpha=\{A,B\}}\!\!\!\!\!\! n_{i\alpha\ua}n_{i\alpha \da} \ ,
	\label{Eq. H_U_real_space}
\end{equation}
where, $n_{i\alpha s}=c^\dagger_{i\alpha s} c_{i\alpha s}$ with $c^\dagger_{i\alpha s}$ being the electron creation operator at site $i$ with the orbital `$\alpha$' with spin `$s$'. Here, `$U(>0)$' is the strength of onsite attractive Hubbard interaction. Such ($e\mhyphen e$) correlations can naturally emerge through a three dimensional bulk $s$-wave SC placed in close proximity to the 1D nanowire described by the Hamiltonian $\mc{H}(k)$~\cite{Bhowmik2025a,Bhowmik2025b}. Thus, CPs from the SC can leak into the nanowire via the tunneling process, inducing weak pairing correlations~\cite{Beenakker1997}. Employing translational invariance, we rewrite the interaction term (Eq.\,\eqref{Eq. H_U_real_space}) in the momentum space as,
\begin{equation}
	H_U = -\frac{U}{N} \!\!\sum_{k_1,k_2,k_3,\alpha}\!\!\!\! c^\dagger_{k_1+k_3 \alpha \ua} c^\dagger_{k_2-k_3 \alpha \da} c_{k_2 \alpha \da} c_{k_1 \alpha \ua}\ , \label{Eq. H_U_mtm_space}
\end{equation}
where, $N$ is the number of $k$-points in the BZ. 

For the purpose of our study, we define the FFLO channel as $k_2=-k_1+ 2q$ with `$2q$' being the center-of-mass momentum of the CPs. We introduce the FFLO order parameter as,
	$ \Delta_{q}^\alpha = -\frac{U}{N} \sum_k \braket{c_{-k+q \alpha \da} c_{k+q \alpha\ua}}$.
Then we systematically perform the mean-field decomposition of the interaction term, $H_U$ (see Eq.\,\eqref{Eq. H_U_mtm_space}) considering both conventional $q=0$  $s$-wave BCS and $q\ne0$ finite momentum FFLO pairing channel. For simplicity, we consider equal magnitude of intra-orbital pairing \ie  $\Delta_{q}^A=\Delta_{q}^B=\Delta$ and assume the absence of inter-orbital pairing amplitude. The resulting mean-field Bogoliubov-de Gennes (BdG) Hamiltonian is written as, 
\begin{equation}
	H_{\rm BdG}\,\,=\,\,\sum_k \Psi^\dagger_{kq} \mc{H}_{\rm BdG} (k,q) \Psi_
	{kq} + \frac{2N}{U}| \Delta|^2 \,+\,\, {\rm constant}\ ,
\end{equation}
where, $\Psi_k = (c_{k+qA\ua},c_{k+q B\ua}, c_{k+qA\da}, c_{k+q B\da},  c^\dagger_{-k+q A\ua},\\ c^\dagger_{-k+qB\ua},  c^\dagger_{-k+qA\da},c^\dagger_{-k+q B\da})^T$ denotes the Nambu spinor, and

\begin{equation}
	\mc{H}_{\rm BdG}(k,q) =  \frac{1}{2} \begin{bmatrix}  
		\mc{H}(k+q) -\mu & -i\sigma_y \Delta \\  
		i\sigma_y\Delta & -\mc{H}^*(-k+q) + \mu
	\end{bmatrix}	\ .	\label{Eq.H_BdG}
\end{equation}

We introduce the chemical potential $\mu$ to tune the Fermi energy of the normal state of the system. Here, $\mc{H}(k+q)$ is governed by the normal state Hamiltonian as mentioned in Eq.\,\eqref{Eq.normal_state_Ham}. The condensation energy density for the superconducting state is defined as, $\Omega (q,\Delta) = F(q,\Delta) - F(q,0)$ with $F(q,\Delta)$ is the free energy density of the SC at zero temperature obtained using the relation~\cite{Coleman_2015}, 
$\displaystyle{F(q,\Delta) = \frac{1}{N} \sum_{n,k, E_{nk}<0}\!\!\!\! E_{nk} + \frac{2\Delta^2}{U}}$
with $E_{nk}$ denoting the energy eigenvalues of the Hamiltonain $\mc{H}_{{\rm BdG}} (k,q)$. 

By minimizing the condensation energy $\Omega(q,\Delta)$ with respect to both the pairing momentum $q$ and the pairing amplitude $\Delta$, we determine the true superconducting ground state $(\Delta_{0},q_{0})$. This procedure captures the energetically most favorable pairing state amenable to the system, ensuring a fully self-consistent treatment of superconductivity. We set $U=1.5t$ to be within the weak coupling BCS regime (see SM~\cite{supp} for a detailed discussion on the validity), which leads to $\Delta_{0}\sim 0.4t,q_0=0$ with $J_A=J_I=\mu=0$ and $m_0=\lambda_R=0.5t,t=1$. 

\begin{figure}
	\includegraphics[scale=0.47]{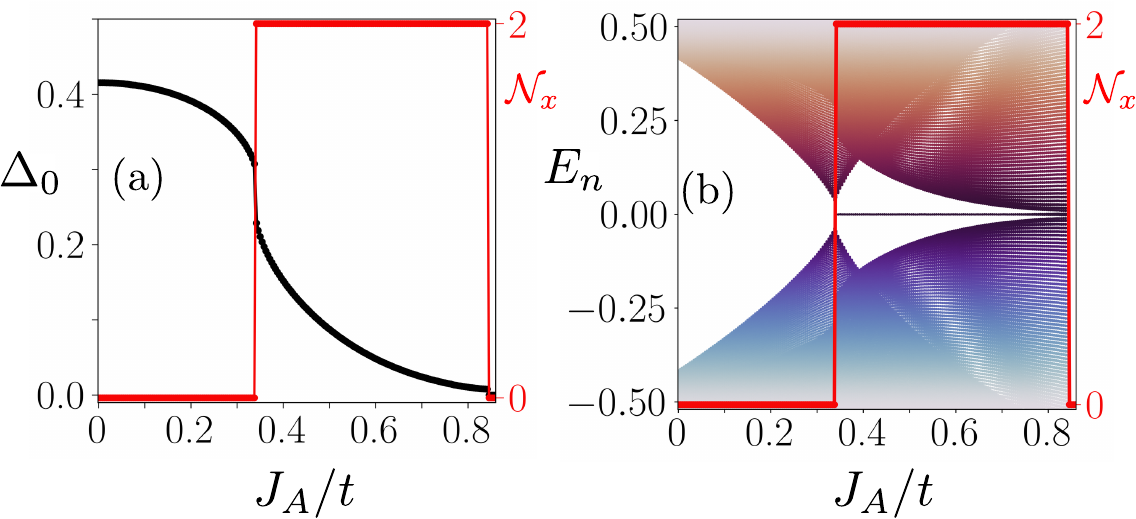}
	\caption{{\bf {Topological superconductivity in the $s$-wave channel:}} (a) Self-consistently obtained superconducting pairing amplitude, $\Delta_{0}$ (left axis) and winding number $\mc{N}_x$ (right axis) are depicted as a function of $J_A/t$. (b) Real space energy eigenvalues $E_{n}$ (left axis) of $\mc{H}_{\rm BdG}$ (Eq.\,\eqref{Eq.H_BdG}) for a 1D finite system of 250 lattice sites, is shown with respect to $J_A/t$ with the self-consistently obtained $\Delta_{0}$ values presented in panel (a). The same winding number $\mc{N}_x$ (right axis) is shown to support the topological regime indicated in $E_{n}$. Other model parameters are chosen as $(m_0,\lambda_R,J_I,\mu)=(0.5t,0.5t,0,0)$.}
	\label{Fig.2}
\end{figure}

We first explore the effect due to only $J_A$ (but $J_{I}=0$) on the superconducting state. We present the variation of $\Delta_{0}$ as a function of $J_A$ in Fig.\,\ref{Fig.2}(a) (left axis). We observe that the superconducting pairing amplitude $\Delta_{0}$ is continuously suppressed with increasing $J_A$ and eventually vanishes beyond a critical value of $J_A$ where the system becomes metallic. This is because the exchange field associated with $J_A$ splits the spin up and down bands, thereby reducing the available phase space for the formation of spin-singlet pairing in the BCS channel. Note that, $q_0=0$ because in absence of $J_I$ the normal state dispersion (see Fig.\,\ref{Fig.1}(b)) remains symmetric with respect to $k=0$ and energy states with momentum $k$ and $-k$ are equally accessible.

We further investigate the topological aspects of the superconducting state in  the BCS channel. Since, chiral symmetry is preserved even in the superconducting state, we again compute $\mc{N}_x$ (defined in Eq.\,\eqref{Eq:winding_num}) to identify the topological superconducting regime. Using the self-consistent values of $\Delta_{0}$ we compute $\mc{N}_x$ as a function of $J_A$ as shown in Fig.\,\ref{Fig.2}(a) (right axis). Remarkably, we find the system undergoes a topological phase transition, becoming a TSC harboring MZMs with $\mc{N}_x=2$. Note that, Ising SOC is still absent and the topological superconductivity emerges out of gapless normal state. Thus, Ising SOC is not necessary to realise TSC in the BCS channel.  Notably, $\Delta_{0}$ changes discontinuously at the topological phase transition point, as evident in Fig.~\ref{Fig.2}(a) (left axis). Such behaviour may arise from the bulk gap closing-and-reopening of $\mc{H}_{\rm BdG}(k)$ at the topological phase transition point. We also highlight that the topological regime is significantly enhanced while using the self-consistent solutions compared to the non-self-consistency scenario (see SM~\cite{supp}). For completeness, we compute the energy eigenvalues of $\mc{H}_{{\rm BdG}} (k,q)$ in real space under open boundary condition and depict the corresponding eigenvalue spectrum as a function of $J_A$ in Fig.\,\ref{Fig.2}(b) (left axis). We refer to SM~\cite{supp} for details on lattice regularized version of $H_{\rm BdG}$. We find the appearance of four MZMs in the TSC phase, consistent with winding number $\mc{N}_x=2$ (see Fig.\,\ref{Fig.2}(b) (right axis)). The four MZMs with two being localized at the same end of the 1D system, do not couple to each other as they appear due to simultaneous gap closing at different bands~\cite{Paul2018} (see SM~\cite{supp} 
for further details).
These findings demonstrate that self-consistent treatment of superconductivity is instrumental to capture the precise topological phase boundaries and the 
robust emergence of MZMs.

\section{Realizing SDE and TSC phase in the FFLO channel: mean field approach}
\begin{figure}
	\includegraphics[scale=0.52]{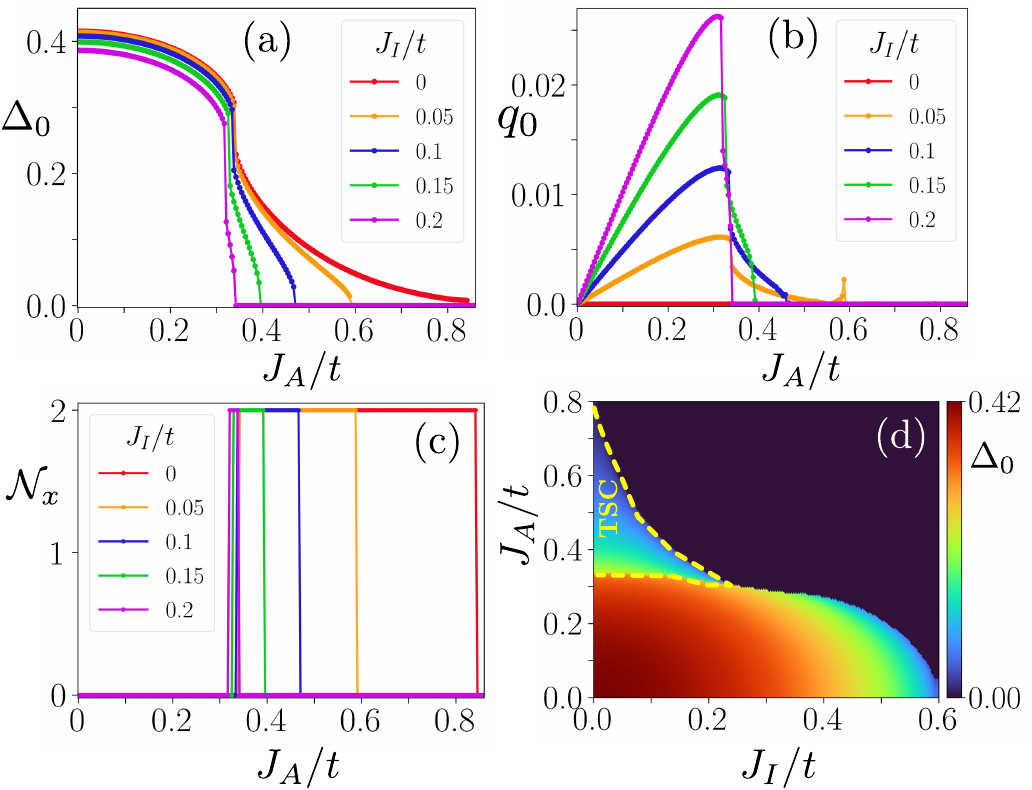}
	\caption{{\bf{Topological superconductivity in the FFLO channel}:}  In panels (a) and (b), we showcase self-consistently obtained true FFLO order parameters $\Delta_{0}$ and $q_0$, respectively, as a function of $J_A/t$ for various values of $J_I/t$. (c) Winding number $\mc{N}_x$ is depicted with respect to $J_A/t$ for the same set of $J_I/t$ values mentioned in panel~(a). (d) Variation of $\Delta_{0}$ in the FFLO channel is shown in the $J_A/t-J_I/t$ plane, and the region covered by the yellow dashed line highlight the TSC phase in $J_A/t-J_I/t$ plane. Other model paramaters are chosen as $(m_0,\lambda_R,\mu)=(0.5t,0.5t,0)$.}
	\label{Fig.3}
\end{figure}

We now explore the combined effect of Ising SOC and unconventional magnetic order ($J_{A},J_{I}\neq 0$), which gives rise to a finite-momentum CP ground state.
We demonstrate the variation of absolute values of $\Delta_{0}$ and $q_0$, obtained self-consistently, as a function of $J_A$ for various strengths of $J_I$ in Fig.\,\ref{Fig.3}(a) and Fig.\,\ref{Fig.3}(b) respectively. Emergence of the FFLO order parameter is clearly favoured only when both $J_A$ and $J_I$ are present 
beacuse the band dispersion (Fig.\,\ref{Fig.1}(d)) is assymetric with respect to $k=0$ when both $J_A,J_I\ne0$. This fabricates conventional BCS ground state energetically less favorable due to the absence of electrons with momenta `$k$' and `$-k$' at the same energy. Thus, the FFLO order is energetically more favorable compared to the BCS state resulting in a superconducting ground state with finite momentum CP.

Before analysing SDE, we topologically characterize the superconducting phase in the FFLO channel by computing the winding number $\mc{N}_x$ with the self-consistent values of $(q_0,\Delta_{0})$. 
As displayed in Fig.\,\ref{Fig.3}(c), a topological superconducting phase with $\mc{N}_x=2$ appears for intermediate values of $J_A$ even when $J_I\ne0$. Moreover, comparing the the panels (b) and (c) of Fig.\,\ref{Fig.3}, the coexistence of topological order and finite momentum FFLO pairing for intermediate values of $J_A$ with $J_I\ne0$ is clearly visible, thus establishing the 
TSC phase in the FFLO channel. Similar to the BCS channel, the appearance of topological order in the FFLO channel is associated with the discontinuity in both $\Delta_{0}$ and $q_0$ (see Fig.\,\ref{Fig.3}(a)-(b)). We believe such nontrivial behaviour arises due to bulk bandgap inversion associated with the topological phase transition. However, a more formal and systematic study is required to understand the underlying reason for this phenomena.  
To this end, we showcase the variation of $\Delta_{0}$ over the entire $J_A$ and $J_I$ plane in Fig.\,\ref{Fig.3}(d) and highlight the region with emergent topological superconducting phase chacaterized by nonzero value of $\mc{N}_x$. Note that, the regime of TSC phase in the FFLO channel diminishes as one increases the value of $J_{I}$. 

One of the prominent manisfestation of the bulk FFLO superconducting order is the emergence of SDE, characterized by a nonreciprocal behaviour of the supercurrent density $J(q) $, defined as~\cite{Daido2022_SDE_PRL,LiangFu2022_PNAS}, 
\begin{equation}
	J(q) = 2e\frac{\partial \Omega(q,\Delta^q_{0})}{\partial q}\ ,
\end{equation}
with the convention that positive (negative) value of $J(q)$ indicates supercurrent flowing along the forward (reverse) direction. Here, $\Omega(q,\Delta_{0}^q)$ is obtained for a given value of $q$ with $\Delta_0^q$ being computed by minimizing $\Omega(q,\Delta)$ with respect to $\Delta$ \ie $\Omega(q,\Delta_{0}^q)\equiv {\rm min}_\Delta \Omega(q,\Delta)$. Hence, $\Delta_{0}^q$ is dependent on the value of $q$. For the true FFLO ground state \ie $q=q_0$, $\Delta_{0}^q=\Delta_{0}$ which implies $J(q_0)$  is identically zero by definition. Furthermore, we also inroduce the critical supercurrent $J_c^+~(J_c^-)$, defined as the maximum supercurrent flowing along the forward (reverse) direction above which the superconductivity is destroyed completely, \ie $J_c^+ \equiv {\rm max}_q J(q)$ and $J_c^- \equiv |{\rm min}_q J(q)|$. Then SDE can be observed only when $J_c^+ \ne J_c^-$ and such nonreciprocal behaviour is quantified by the diode efficiency factor, $\eta$ defined 
as~\cite{Bhowmik2025a,Bhowmik2025b},
\begin{equation}
	\eta =\frac{|J_c^+ - J_c^-|}{J_c^+ + J_c^-}\ .
\end{equation} 

\begin{figure}
	\includegraphics[scale=0.53]{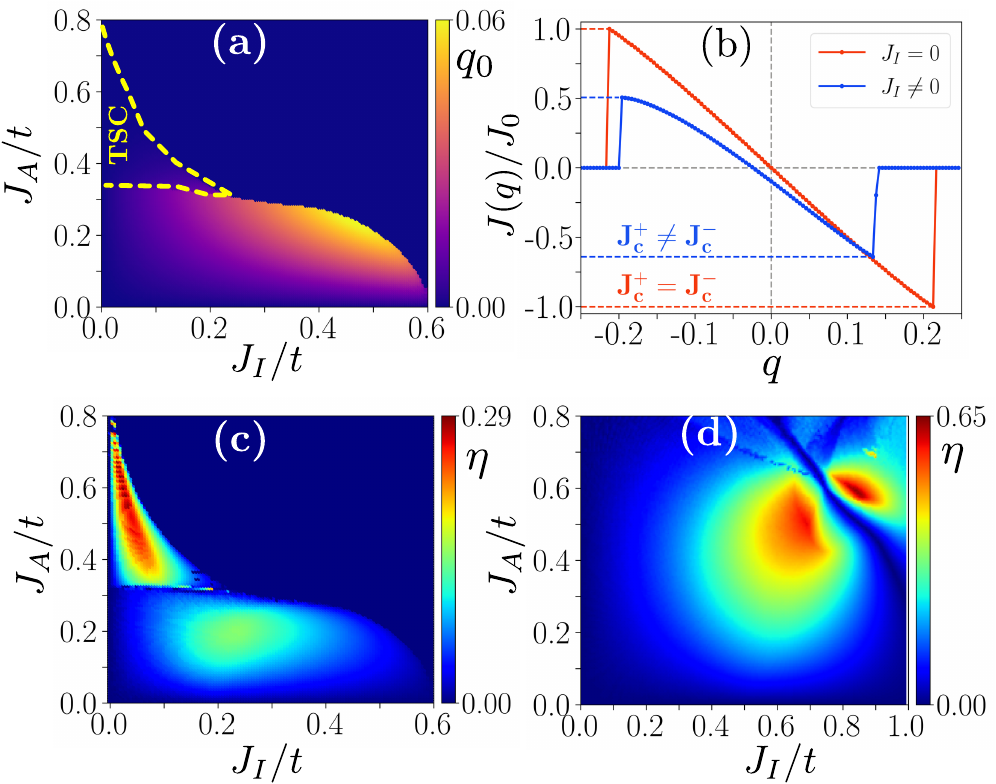}
	\caption{{\bf {Non-reciprocal supercurrent and SDE efficiency:}} Panel (a) depicts the variation of $q_0$ in the $J_I/t-J_A/t$ plane. The TSC phase in the FFLO pairing state is highlighted by the yellow dashed line. (b) The nonreciprocal nature of the supercurrent $J(q)$, normalized by $J_0\equiv J_c^+(J_I=0)$ is shown as a function of $q$ in the presence of Ising SOC, $J_I$. Panels (c) and (d) highlight the diode efficiency $\eta$ 
		in the $J_I/t-J_A/t$ plane with $(m_0,\lambda_R,\mu)=(0.5t,0.5t,0)$ in panel (c) and $(m_0,\lambda_R,\mu)=(0.15t,0.15t,0.7t)$ in panel (d) respectively.}
	\label{Fig.4}
\end{figure}

In Fig.\,\ref{Fig.4}(a), we first display the true finite momentum $q_0$ of the CPs corresponding to the FFLO pairing 
by varying both $J_A$ and $J_I$.  We then also highlight the parameter space exhibiting the topological superconductivity hosting MZMs. To emphasize the role of Ising SOC in realizing SDE, we present the variation of $J(q)$ as a function of $q$ in Fig.\,\ref{Fig.4}(b) choosing both $J_I=0$ and $J_I\ne0$. We explicitly observe that $J_c^+ \ne J_c^-$ when $J_I\ne 0$ which establishes the nonreciprocal behaviour of critical supercurrent density, implying the emergence of SDE. Although, nonzero values of $\lambda_R$ and $J_A$ breaks inversion symmetry and TRS of the system, it is not sufficient to realize SDE. A finite value of $J_I$ is required to observe the SDE which highlight the pivotal role of Ising SOC in the system. Interestingly, SDE can be observed without applying any external magnetic field. Such field-free superconducting diodes can have significant technological advantages in terms of device miniaturization and application point of view.

We further compute the diode efficiency $\eta$ in the plane of $J_I$ and $J_A$ and present the corresponding result in Fig.\,\ref{Fig.4}(c). Notably, we observe the emergence of maximum diode efficiency of $\eta \sim 29\%$ in the $J_I-J_A$ plane where the system exhibits topological superconductivity. Until this point, we focuss on only varying $J_I$ and $J_A$ maintaining other model paramters $(\lambda_R,m_0,\mu)=(0.5t,0.5t,0)$ to be fixed. Hence, we allow other model parameters to \tcr{vary} in order to obtain the maximum diode efficiency possible in the system. Particularly, we tune $\mu$ which modifies the normal state density of states at the Fermi energy. Suitably choosing the model parameters as $m_0=\lambda_R=0.15t, \mu=0.7t$, we compute the supercurrent density and thereby obtain the diode efficiency $\eta$. As depicted in Fig.\,\ref{Fig.4}(d), we observe significant large value of diode efficiency $\eta\sim 65\%$ in the same parameter space. This may indicate that tuning only $\mu$, $\eta$ can be enhanced. However, this is not true in general as $\eta$ also strongly depends 
on other model paramaters which cause the band structure to modify leading to a modification in the density of left and right movers. (see SM~\cite{supp} for details). These results highlight the central role of Ising SOC and unconventional magnetic order in generating highly efficient, field-free superconducting diodes with coexisting topological superconductivity. 
Importantly, the MZMs are protected by the chiral symmetry which is broken for $\mu\ne 0$ in the FFLO channel. As a result, the boundary modes are 
gapped out from zero energy. However, phenomena of SDE do not require such symmetry-protection, thus remains unaffected in the absence of chiral symmetry~ (see SM~\cite{supp} for detailed discussions).

\section{Summary and discussion}
To summarize, in this paper, we introduce a tight-binding model that combines unconventional magnetic order with Rashba and Ising SOC, providing a unified framework for realizing topological insulator, TSC, and superconducting diode in a field-free manner. 1D model Hamiltonians manifesting topological properties in both normal and superconducting states lack in literature, and our proposed model fills this gap. Next, introducing an attractive Hubbard interaction and performing a fully self-consistent mean-field analysis, we demonstrate the emergence of topological superconductivity in both the BCS and FFLO pairing channels. In particular, the combined effect of Ising SOC and unconventional magnetic order stabilizes the FFLO order that leads to the emergence of an intrinsic, field-free SDE. The resultant nonreciprocal supercurrent response is quantified by the diode efficiency $\eta$ which attains remarkably high value $\eta\sim 65\%$ under optimal parameter regimes. These findings establish our model as a versatile platform to engineer and control topological superconductivity and nonreciprocal superconducting transport without the need of external magnetic fields~\cite{Bhowmik2025b}. In Ref.~\cite{Legg2022_SDE_MCA}, appearance of SDE has been proposed by applying two Zeeman fields, whereas we observe SDE utilizing two types of SOC (Rashba and Ising) which strengthen the originality of our model. 
Our proposed phenomena may be possible to realize in engineered heterostructures combining the key ingredients. For instance, an InSb nanowire~\cite{Mourik2012Science}, with strong Rashba SOC, can be proximitized by an altermagnet such as MnTe~\cite{Lee2024MnTe} and to a layer with strong Ising SOC such as MoS$_2$~\cite{Lu2015}. Superconducting correlation can then be induced via the proximity effect using a conventional BCS SC such as Nb~\cite{Mourik2012Science}.

As mentioned earlier, in contrast to the well-known models such as the Su–Schrieffer–Heeger~\cite{SSH_1979} and 1D Bernevig–Hughes–Zhang models~\cite{Bernevig2006_BHZ}, which exhibit topological insulating phases but becomes trivial 
$s$-wave superconductor once such superconducting pairing is introduced,

it's topological character in both normal and superconducting phases. We refer to the SM~\cite{supp} for the detailed discussion.

To this end, we note that in this work the mean field decomposition has been performed assuming equal magnitude of intra-orbital pairing and neglecting the presence of inter-orbital pairing. In future, one can relax these assumptions and can investigate the role of inter-orbital pairing in realising TSC and SDE. 
It would also be intriguing to explore beyond mean-field techniques such as density matrix renormalization
group (DMRG) or Quantum Monte Carlo with Rashba SOC~\cite{Faundez2025} to more accurately procure the superconducting ground state properties due to limitations of mean-field solutions.

{\it Note added.--} Recently, we became aware of a work that also discusses SDE due to the interplay of altermagnets and Ising SOC~\cite{Ruthvik2025}.

\section*{Acknowledgments} 
A.P., D.M., and A.S. acknowledge SAMKHYA: HighPerformance Computing Facility provided by the Institute of Physics, Bhubaneswar and the two workstations provided by the Institute of Physics, Bhubaneswar from DAE APEX Project, for numerical computations. T.N. and A.S. thank the Advanced Research Grant (ARG) from Anusandhan National Research Foundation Grant No. ANRF/ARG/2025/003163/PS. T.N. acknowledges NFSG from Grant No. BITS Pilani NFSG/HYD/2023/H0911

\section*{Data Availability} The data that support the findings of this article are not publicly available. The data are available from the authors upon reasonable request.

\bibliographystyle{apsrev4-2mod}
\bibliography{bibfile.bib}

\begin{thebibliography}{109}%
\makeatletter
\providecommand \@ifxundefined [1]{%
 \@ifx{#1\undefined}
}%
\providecommand \@ifnum [1]{%
 \ifnum #1\expandafter \@firstoftwo
 \else \expandafter \@secondoftwo
 \fi
}%
\providecommand \@ifx [1]{%
 \ifx #1\expandafter \@firstoftwo
 \else \expandafter \@secondoftwo
 \fi
}%
\providecommand \natexlab [1]{#1}%
\providecommand \enquote  [1]{``#1''}%
\providecommand \bibnamefont  [1]{#1}%
\providecommand \bibfnamefont [1]{#1}%
\providecommand \citenamefont [1]{#1}%
\providecommand \href@noop [0]{\@secondoftwo}%
\providecommand \href [0]{\begingroup \@sanitize@url \@href}%
\providecommand \@href[1]{\@@startlink{#1}\@@href}%
\providecommand \@@href[1]{\endgroup#1\@@endlink}%
\providecommand \@sanitize@url [0]{\catcode `\\12\catcode `\$12\catcode
  `\&12\catcode `\#12\catcode `\^12\catcode `\_12\catcode `\%12\relax}%
\providecommand \@@startlink[1]{}%
\providecommand \@@endlink[0]{}%
\providecommand \url  [0]{\begingroup\@sanitize@url \@url }%
\providecommand \@url [1]{\endgroup\@href {#1}{\urlprefix }}%
\providecommand \urlprefix  [0]{URL }%
\providecommand \Eprint [0]{\href }%
\providecommand \doibase [0]{http://dx.doi.org/}%
\providecommand \selectlanguage [0]{\@gobble}%
\providecommand \bibinfo  [0]{\@secondoftwo}%
\providecommand \bibfield  [0]{\@secondoftwo}%
\providecommand \translation [1]{[#1]}%
\providecommand \BibitemOpen [0]{}%
\providecommand \bibitemStop [0]{}%
\providecommand \bibitemNoStop [0]{.\EOS\space}%
\providecommand \EOS [0]{\spacefactor3000\relax}%
\providecommand \BibitemShut  [1]{\csname bibitem#1\endcsname}%
\let\auto@bib@innerbib\@empty
\bibitem [{\citenamefont {Kitaev}(2001)}]{Kitaev_2001}%
  \BibitemOpen
  \bibfield  {author} {\bibinfo {author} {\bibfnamefont {A.~Y.}\ \bibnamefont
  {Kitaev}},\ }\bibfield  {title} {\emph {\enquote {\bibinfo {title} {Unpaired
  Majorana fermions in quantum wires},}\ }}\href {\doibase
  10.1070/1063-7869/44/10s/s29} {\bibfield  {journal} {\bibinfo  {journal}
  {Physics-Uspekhi}\ }\textbf {\bibinfo {volume} {44}},\ \bibinfo {pages} {131}
  (\bibinfo {year} {2001})}\BibitemShut {NoStop}%
\bibitem [{\citenamefont {Alicea}(2012)}]{Alicea_2012}%
  \BibitemOpen
  \bibfield  {author} {\bibinfo {author} {\bibfnamefont {J.}~\bibnamefont
  {Alicea}},\ }\bibfield  {title} {\emph {\enquote {\bibinfo {title} {New
  directions in the pursuit of Majorana fermions in solid state systems},}\
  }}\href {\doibase 10.1088/0034-4885/75/7/076501} {\bibfield  {journal}
  {\bibinfo  {journal} {Reports on Progress in Physics}\ }\textbf {\bibinfo
  {volume} {75}},\ \bibinfo {pages} {076501} (\bibinfo {year}
  {2012})}\BibitemShut {NoStop}%
\bibitem [{\citenamefont {Alicea}\ \emph {et~al.}(2011)\citenamefont {Alicea},
  \citenamefont {Oreg}, \citenamefont {Refael}, \citenamefont {von Oppen},\
  and\ \citenamefont {Fisher}}]{Alicea2011_NatPhys}%
  \BibitemOpen
  \bibfield  {author} {\bibinfo {author} {\bibfnamefont {J.}~\bibnamefont
  {Alicea}}, \bibinfo {author} {\bibfnamefont {Y.}~\bibnamefont {Oreg}},
  \bibinfo {author} {\bibfnamefont {G.}~\bibnamefont {Refael}}, \bibinfo
  {author} {\bibfnamefont {F.}~\bibnamefont {von Oppen}}, \ and\ \bibinfo
  {author} {\bibfnamefont {M.~P.~A.}\ \bibnamefont {Fisher}},\ }\bibfield
  {title} {\emph {\enquote {\bibinfo {title} {Non-Abelian statistics and
  topological quantum information processing in 1D wire networks},}\ }}\href
  {\doibase 10.1038/nphys1915} {\bibfield  {journal} {\bibinfo  {journal}
  {Nature Physics}\ }\textbf {\bibinfo {volume} {7}},\ \bibinfo {pages} {412}
  (\bibinfo {year} {2011})}\BibitemShut {NoStop}%
\bibitem [{\citenamefont {Leijnse}\ and\ \citenamefont
  {Flensberg}(2012)}]{Leijnse_2012}%
  \BibitemOpen
  \bibfield  {author} {\bibinfo {author} {\bibfnamefont {M.}~\bibnamefont
  {Leijnse}}\ and\ \bibinfo {author} {\bibfnamefont {K.}~\bibnamefont
  {Flensberg}},\ }\bibfield  {title} {\emph {\enquote {\bibinfo {title}
  {Introduction to topological superconductivity and Majorana fermions},}\
  }}\href {\doibase 10.1088/0268-1242/27/12/124003} {\bibfield  {journal}
  {\bibinfo  {journal} {Semiconductor Science and Technology}\ }\textbf
  {\bibinfo {volume} {27}},\ \bibinfo {pages} {124003} (\bibinfo {year}
  {2012})}\BibitemShut {NoStop}%
\bibitem [{\citenamefont {Pientka}\ \emph {et~al.}(2013)\citenamefont
  {Pientka}, \citenamefont {Glazman},\ and\ \citenamefont {von
  Oppen}}]{Pientka2013_Shiba}%
  \BibitemOpen
  \bibfield  {author} {\bibinfo {author} {\bibfnamefont {F.}~\bibnamefont
  {Pientka}}, \bibinfo {author} {\bibfnamefont {L.~I.}\ \bibnamefont
  {Glazman}}, \ and\ \bibinfo {author} {\bibfnamefont {F.}~\bibnamefont {von
  Oppen}},\ }\bibfield  {title} {\emph {\enquote {\bibinfo {title} {Topological
  superconducting phase in helical Shiba chains},}\ }}\href {\doibase
  10.1103/PhysRevB.88.155420} {\bibfield  {journal} {\bibinfo  {journal} {Phys.
  Rev. B}\ }\textbf {\bibinfo {volume} {88}},\ \bibinfo {pages} {155420}
  (\bibinfo {year} {2013})}\BibitemShut {NoStop}%
\bibitem [{\citenamefont {Nadj-Perge}\ \emph {et~al.}(2013)\citenamefont
  {Nadj-Perge}, \citenamefont {Drozdov}, \citenamefont {Bernevig},\ and\
  \citenamefont {Yazdani}}]{Nadj2013_Shiba}%
  \BibitemOpen
  \bibfield  {author} {\bibinfo {author} {\bibfnamefont {S.}~\bibnamefont
  {Nadj-Perge}}, \bibinfo {author} {\bibfnamefont {I.~K.}\ \bibnamefont
  {Drozdov}}, \bibinfo {author} {\bibfnamefont {B.~A.}\ \bibnamefont
  {Bernevig}}, \ and\ \bibinfo {author} {\bibfnamefont {A.}~\bibnamefont
  {Yazdani}},\ }\bibfield  {title} {\emph {\enquote {\bibinfo {title} {Proposal
  for realizing Majorana fermions in chains of magnetic atoms on a
  superconductor},}\ }}\href {\doibase 10.1103/PhysRevB.88.020407} {\bibfield
  {journal} {\bibinfo  {journal} {Phys. Rev. B}\ }\textbf {\bibinfo {volume}
  {88}},\ \bibinfo {pages} {020407} (\bibinfo {year} {2013})}\BibitemShut
  {NoStop}%
\bibitem [{\citenamefont {Beenakker}(2013)}]{Beenakker2013search}%
  \BibitemOpen
  \bibfield  {author} {\bibinfo {author} {\bibfnamefont {C.}~\bibnamefont
  {Beenakker}},\ }\bibfield  {title} {\emph {\enquote {\bibinfo {title} {Search
  for Majorana fermions in superconductors},}\ }}\href {\doibase
  10.1146/annurev-conmatphys-030212-184337} {\bibfield  {journal} {\bibinfo
  {journal} {Annu. Rev. Condens. Matter Phys.}\ }\textbf {\bibinfo {volume}
  {4}},\ \bibinfo {pages} {113} (\bibinfo {year} {2013})}\BibitemShut {NoStop}%
\bibitem [{\citenamefont {Kitaev}(2003)}]{Kitaev2003_annals_of_phys}%
  \BibitemOpen
  \bibfield  {author} {\bibinfo {author} {\bibfnamefont {A.}~\bibnamefont
  {Kitaev}},\ }\bibfield  {title} {\emph {\enquote {\bibinfo {title}
  {Fault-tolerant quantum computation by anyons},}\ }}\href {\doibase
  https://doi.org/10.1016/S0003-4916(02)00018-0} {\bibfield  {journal}
  {\bibinfo  {journal} {Annals of Physics}\ }\textbf {\bibinfo {volume}
  {303}},\ \bibinfo {pages} {2} (\bibinfo {year} {2003})}\BibitemShut {NoStop}%
\bibitem [{\citenamefont {Nayak}\ \emph {et~al.}(2008)\citenamefont {Nayak},
  \citenamefont {Simon}, \citenamefont {Stern}, \citenamefont {Freedman},\ and\
  \citenamefont {Das~Sarma}}]{NayakRMP2008}%
  \BibitemOpen
  \bibfield  {author} {\bibinfo {author} {\bibfnamefont {C.}~\bibnamefont
  {Nayak}}, \bibinfo {author} {\bibfnamefont {S.~H.}\ \bibnamefont {Simon}},
  \bibinfo {author} {\bibfnamefont {A.}~\bibnamefont {Stern}}, \bibinfo
  {author} {\bibfnamefont {M.}~\bibnamefont {Freedman}}, \ and\ \bibinfo
  {author} {\bibfnamefont {S.}~\bibnamefont {Das~Sarma}},\ }\bibfield  {title}
  {\emph {\enquote {\bibinfo {title} {Non-Abelian anyons and topological
  quantum computation},}\ }}\href {\doibase 10.1103/RevModPhys.80.1083}
  {\bibfield  {journal} {\bibinfo  {journal} {Rev. Mod. Phys.}\ }\textbf
  {\bibinfo {volume} {80}},\ \bibinfo {pages} {1083} (\bibinfo {year}
  {2008})}\BibitemShut {NoStop}%
\bibitem [{\citenamefont {Yazdani}\ \emph {et~al.}(2023)\citenamefont
  {Yazdani}, \citenamefont {von Oppen}, \citenamefont {Halperin},\ and\
  \citenamefont {Yacoby}}]{Yazdani_hunt_majorana}%
  \BibitemOpen
  \bibfield  {author} {\bibinfo {author} {\bibfnamefont {A.}~\bibnamefont
  {Yazdani}}, \bibinfo {author} {\bibfnamefont {F.}~\bibnamefont {von Oppen}},
  \bibinfo {author} {\bibfnamefont {B.~I.}\ \bibnamefont {Halperin}}, \ and\
  \bibinfo {author} {\bibfnamefont {A.}~\bibnamefont {Yacoby}},\ }\bibfield
  {title} {\emph {\enquote {\bibinfo {title} {Hunting for Majoranas},}\ }}\href
  {\doibase 10.1126/science.ade0850} {\bibfield  {journal} {\bibinfo  {journal}
  {Science}\ }\textbf {\bibinfo {volume} {380}},\ \bibinfo {pages} {eade0850}
  (\bibinfo {year} {2023})}\BibitemShut {NoStop}%
\bibitem [{\citenamefont {Lutchyn}\ \emph {et~al.}(2010)\citenamefont
  {Lutchyn}, \citenamefont {Sau},\ and\ \citenamefont
  {Das~Sarma}}]{LutchynPRL2010}%
  \BibitemOpen
  \bibfield  {author} {\bibinfo {author} {\bibfnamefont {R.~M.}\ \bibnamefont
  {Lutchyn}}, \bibinfo {author} {\bibfnamefont {J.~D.}\ \bibnamefont {Sau}}, \
  and\ \bibinfo {author} {\bibfnamefont {S.}~\bibnamefont {Das~Sarma}},\
  }\bibfield  {title} {\emph {\enquote {\bibinfo {title} {Majorana Fermions and
  a Topological Phase Transition in Semiconductor-Superconductor
  Heterostructures},}\ }}\href {\doibase 10.1103/PhysRevLett.105.077001}
  {\bibfield  {journal} {\bibinfo  {journal} {Phys. Rev. Lett.}\ }\textbf
  {\bibinfo {volume} {105}},\ \bibinfo {pages} {077001} (\bibinfo {year}
  {2010})}\BibitemShut {NoStop}%
\bibitem [{\citenamefont {Tewari}\ and\ \citenamefont
  {Sau}(2012)}]{TewariPRL2012}%
  \BibitemOpen
  \bibfield  {author} {\bibinfo {author} {\bibfnamefont {S.}~\bibnamefont
  {Tewari}}\ and\ \bibinfo {author} {\bibfnamefont {J.~D.}\ \bibnamefont
  {Sau}},\ }\bibfield  {title} {\emph {\enquote {\bibinfo {title} {Topological
  Invariants for Spin-Orbit Coupled Superconductor Nanowires},}\ }}\href
  {\doibase 10.1103/PhysRevLett.109.150408} {\bibfield  {journal} {\bibinfo
  {journal} {Phys. Rev. Lett.}\ }\textbf {\bibinfo {volume} {109}},\ \bibinfo
  {pages} {150408} (\bibinfo {year} {2012})}\BibitemShut {NoStop}%
\bibitem [{\citenamefont {Rokhinson}\ \emph {et~al.}(2012)\citenamefont
  {Rokhinson}, \citenamefont {Liu},\ and\ \citenamefont
  {Furdyna}}]{Rokhinson2012}%
  \BibitemOpen
  \bibfield  {author} {\bibinfo {author} {\bibfnamefont {L.~P.}\ \bibnamefont
  {Rokhinson}}, \bibinfo {author} {\bibfnamefont {X.}~\bibnamefont {Liu}}, \
  and\ \bibinfo {author} {\bibfnamefont {J.~K.}\ \bibnamefont {Furdyna}},\
  }\bibfield  {title} {\emph {\enquote {\bibinfo {title} {The fractional a.c.
  Josephson effect in a semiconductor--superconductor nanowire as a signature
  of Majorana particles},}\ }}\href {\doibase 10.1038/nphys2429} {\bibfield
  {journal} {\bibinfo  {journal} {Nature Physics}\ }\textbf {\bibinfo {volume}
  {8}},\ \bibinfo {pages} {795} (\bibinfo {year} {2012})}\BibitemShut {NoStop}%
\bibitem [{\citenamefont {Law}\ \emph {et~al.}(2009)\citenamefont {Law},
  \citenamefont {Lee},\ and\ \citenamefont {Ng}}]{LawPRL2009}%
  \BibitemOpen
  \bibfield  {author} {\bibinfo {author} {\bibfnamefont {K.~T.}\ \bibnamefont
  {Law}}, \bibinfo {author} {\bibfnamefont {P.~A.}\ \bibnamefont {Lee}}, \ and\
  \bibinfo {author} {\bibfnamefont {T.~K.}\ \bibnamefont {Ng}},\ }\bibfield
  {title} {\emph {\enquote {\bibinfo {title} {Majorana Fermion Induced Resonant
  Andreev Reflection},}\ }}\href {\doibase 10.1103/PhysRevLett.103.237001}
  {\bibfield  {journal} {\bibinfo  {journal} {Phys. Rev. Lett.}\ }\textbf
  {\bibinfo {volume} {103}},\ \bibinfo {pages} {237001} (\bibinfo {year}
  {2009})}\BibitemShut {NoStop}%
\bibitem [{\citenamefont {Mourik}\ \emph {et~al.}(2012)\citenamefont {Mourik},
  \citenamefont {Zuo}, \citenamefont {Frolov}, \citenamefont {Plissard},
  \citenamefont {Bakkers},\ and\ \citenamefont
  {Kouwenhoven}}]{Mourik2012Science}%
  \BibitemOpen
  \bibfield  {author} {\bibinfo {author} {\bibfnamefont {V.}~\bibnamefont
  {Mourik}}, \bibinfo {author} {\bibfnamefont {K.}~\bibnamefont {Zuo}},
  \bibinfo {author} {\bibfnamefont {S.~M.}\ \bibnamefont {Frolov}}, \bibinfo
  {author} {\bibfnamefont {S.~R.}\ \bibnamefont {Plissard}}, \bibinfo {author}
  {\bibfnamefont {E.~P. A.~M.}\ \bibnamefont {Bakkers}}, \ and\ \bibinfo
  {author} {\bibfnamefont {L.~P.}\ \bibnamefont {Kouwenhoven}},\ }\bibfield
  {title} {\emph {\enquote {\bibinfo {title} {Signatures of Majorana Fermions
  in Hybrid Superconductor-Semiconductor Nanowire Devices},}\ }}\href {\doibase
  10.1126/science.1222360} {\bibfield  {journal} {\bibinfo  {journal}
  {Science}\ }\textbf {\bibinfo {volume} {336}},\ \bibinfo {pages} {1003}
  (\bibinfo {year} {2012})}\BibitemShut {NoStop}%
\bibitem [{\citenamefont {Das}\ \emph {et~al.}(2012)\citenamefont {Das},
  \citenamefont {Ronen}, \citenamefont {Most}, \citenamefont {Oreg},
  \citenamefont {Heiblum},\ and\ \citenamefont {Shtrikman}}]{Das2012_NatPhys}%
  \BibitemOpen
  \bibfield  {author} {\bibinfo {author} {\bibfnamefont {A.}~\bibnamefont
  {Das}}, \bibinfo {author} {\bibfnamefont {Y.}~\bibnamefont {Ronen}}, \bibinfo
  {author} {\bibfnamefont {Y.}~\bibnamefont {Most}}, \bibinfo {author}
  {\bibfnamefont {Y.}~\bibnamefont {Oreg}}, \bibinfo {author} {\bibfnamefont
  {M.}~\bibnamefont {Heiblum}}, \ and\ \bibinfo {author} {\bibfnamefont
  {H.}~\bibnamefont {Shtrikman}},\ }\bibfield  {title} {\emph {\enquote
  {\bibinfo {title} {Zero-bias peaks and splitting in an Al--InAs nanowire
  topological superconductor as a signature of Majorana fermions},}\ }}\href
  {\doibase https://doi.org/10.1038/nphys2479} {\bibfield  {journal} {\bibinfo
  {journal} {Nature Physics}\ }\textbf {\bibinfo {volume} {8}},\ \bibinfo
  {pages} {887} (\bibinfo {year} {2012})}\BibitemShut {NoStop}%
\bibitem [{\citenamefont {Albrecht}\ \emph {et~al.}(2016)\citenamefont
  {Albrecht}, \citenamefont {Higginbotham}, \citenamefont {Madsen},
  \citenamefont {Kuemmeth}, \citenamefont {Jespersen}, \citenamefont
  {Nyg{\aa}rd}, \citenamefont {Krogstrup},\ and\ \citenamefont
  {Marcus}}]{Albrecht2016}%
  \BibitemOpen
  \bibfield  {author} {\bibinfo {author} {\bibfnamefont {S.~M.}\ \bibnamefont
  {Albrecht}}, \bibinfo {author} {\bibfnamefont {A.~P.}\ \bibnamefont
  {Higginbotham}}, \bibinfo {author} {\bibfnamefont {M.}~\bibnamefont
  {Madsen}}, \bibinfo {author} {\bibfnamefont {F.}~\bibnamefont {Kuemmeth}},
  \bibinfo {author} {\bibfnamefont {T.~S.}\ \bibnamefont {Jespersen}}, \bibinfo
  {author} {\bibfnamefont {J.}~\bibnamefont {Nyg{\aa}rd}}, \bibinfo {author}
  {\bibfnamefont {P.}~\bibnamefont {Krogstrup}}, \ and\ \bibinfo {author}
  {\bibfnamefont {C.~M.}\ \bibnamefont {Marcus}},\ }\bibfield  {title} {\emph
  {\enquote {\bibinfo {title} {Exponential protection of zero modes in Majorana
  islands},}\ }}\href {\doibase 10.1038/nature17162} {\bibfield  {journal}
  {\bibinfo  {journal} {Nature}\ }\textbf {\bibinfo {volume} {531}},\ \bibinfo
  {pages} {206} (\bibinfo {year} {2016})}\BibitemShut {NoStop}%
\bibitem [{\citenamefont {Chen}\ \emph {et~al.}(2017)\citenamefont {Chen},
  \citenamefont {Yu}, \citenamefont {Stenger}, \citenamefont {Hocevar},
  \citenamefont {Car}, \citenamefont {Plissard}, \citenamefont {Bakkers},
  \citenamefont {Stanescu},\ and\ \citenamefont {Frolov}}]{ChenSciAdv2017}%
  \BibitemOpen
  \bibfield  {author} {\bibinfo {author} {\bibfnamefont {J.}~\bibnamefont
  {Chen}}, \bibinfo {author} {\bibfnamefont {P.}~\bibnamefont {Yu}}, \bibinfo
  {author} {\bibfnamefont {J.}~\bibnamefont {Stenger}}, \bibinfo {author}
  {\bibfnamefont {M.}~\bibnamefont {Hocevar}}, \bibinfo {author} {\bibfnamefont
  {D.}~\bibnamefont {Car}}, \bibinfo {author} {\bibfnamefont {S.~R.}\
  \bibnamefont {Plissard}}, \bibinfo {author} {\bibfnamefont {E.~P. A.~M.}\
  \bibnamefont {Bakkers}}, \bibinfo {author} {\bibfnamefont {T.~D.}\
  \bibnamefont {Stanescu}}, \ and\ \bibinfo {author} {\bibfnamefont {S.~M.}\
  \bibnamefont {Frolov}},\ }\bibfield  {title} {\emph {\enquote {\bibinfo
  {title} {Experimental phase diagram of zero-bias conductance peaks in
  superconductor/semiconductor nanowire devices},}\ }}\href {\doibase
  10.1126/sciadv.1701476} {\bibfield  {journal} {\bibinfo  {journal} {Science
  Advances}\ }\textbf {\bibinfo {volume} {3}},\ \bibinfo {pages} {e1701476}
  (\bibinfo {year} {2017})}\BibitemShut {NoStop}%
\bibitem [{\citenamefont {Mondal}\ \emph
  {et~al.}(2025{\natexlab{a}})\citenamefont {Mondal}, \citenamefont {Kumari},
  \citenamefont {Nag},\ and\ \citenamefont {Saha}}]{Mondal2024}%
  \BibitemOpen
  \bibfield  {author} {\bibinfo {author} {\bibfnamefont {D.}~\bibnamefont
  {Mondal}}, \bibinfo {author} {\bibfnamefont {R.}~\bibnamefont {Kumari}},
  \bibinfo {author} {\bibfnamefont {T.}~\bibnamefont {Nag}}, \ and\ \bibinfo
  {author} {\bibfnamefont {A.}~\bibnamefont {Saha}},\ }\bibfield  {title}
  {\emph {\enquote {\bibinfo {title} {Transport signatures of single and
  multiple Floquet Majorana modes in a one-dimensional Rashba nanowire and
  Shiba chain},}\ }}\href {\doibase 10.1103/6wnf-b5g8} {\bibfield  {journal}
  {\bibinfo  {journal} {Phys. Rev. B}\ }\textbf {\bibinfo {volume} {111}},\
  \bibinfo {pages} {235441} (\bibinfo {year} {2025}{\natexlab{a}})}\BibitemShut
  {NoStop}%
\bibitem [{\citenamefont {Mondal}\ \emph
  {et~al.}(2023{\natexlab{a}})\citenamefont {Mondal}, \citenamefont {Ghosh},
  \citenamefont {Nag},\ and\ \citenamefont {Saha}}]{Mondal2023_NW}%
  \BibitemOpen
  \bibfield  {author} {\bibinfo {author} {\bibfnamefont {D.}~\bibnamefont
  {Mondal}}, \bibinfo {author} {\bibfnamefont {A.~K.}\ \bibnamefont {Ghosh}},
  \bibinfo {author} {\bibfnamefont {T.}~\bibnamefont {Nag}}, \ and\ \bibinfo
  {author} {\bibfnamefont {A.}~\bibnamefont {Saha}},\ }\bibfield  {title}
  {\emph {\enquote {\bibinfo {title} {Topological characterization and
  stability of Floquet Majorana modes in Rashba nanowires},}\ }}\href {\doibase
  10.1103/PhysRevB.107.035427} {\bibfield  {journal} {\bibinfo  {journal}
  {Phys. Rev. B}\ }\textbf {\bibinfo {volume} {107}},\ \bibinfo {pages}
  {035427} (\bibinfo {year} {2023}{\natexlab{a}})}\BibitemShut {NoStop}%
\bibitem [{\citenamefont {Arouca}\ \emph {et~al.}(2024)\citenamefont {Arouca},
  \citenamefont {Nag},\ and\ \citenamefont {Black-Schaffer}}]{Arouca2024}%
  \BibitemOpen
  \bibfield  {author} {\bibinfo {author} {\bibfnamefont {R.}~\bibnamefont
  {Arouca}}, \bibinfo {author} {\bibfnamefont {T.}~\bibnamefont {Nag}}, \ and\
  \bibinfo {author} {\bibfnamefont {A.~M.}\ \bibnamefont {Black-Schaffer}},\
  }\bibfield  {title} {\emph {\enquote {\bibinfo {title} {Mixed higher-order
  topology, and nodal and nodeless flat band topological phases in a
  superconducting multiorbital model},}\ }}\href {\doibase
  10.1103/PhysRevB.110.064520} {\bibfield  {journal} {\bibinfo  {journal}
  {Phys. Rev. B}\ }\textbf {\bibinfo {volume} {110}},\ \bibinfo {pages}
  {064520} (\bibinfo {year} {2024})}\BibitemShut {NoStop}%
\bibitem [{\citenamefont {Klinovaja}\ \emph {et~al.}(2013)\citenamefont
  {Klinovaja}, \citenamefont {Stano}, \citenamefont {Yazdani},\ and\
  \citenamefont {Loss}}]{Klinovaja2013_shiba}%
  \BibitemOpen
  \bibfield  {author} {\bibinfo {author} {\bibfnamefont {J.}~\bibnamefont
  {Klinovaja}}, \bibinfo {author} {\bibfnamefont {P.}~\bibnamefont {Stano}},
  \bibinfo {author} {\bibfnamefont {A.}~\bibnamefont {Yazdani}}, \ and\
  \bibinfo {author} {\bibfnamefont {D.}~\bibnamefont {Loss}},\ }\bibfield
  {title} {\emph {\enquote {\bibinfo {title} {Topological Superconductivity and
  Majorana Fermions in RKKY Systems},}\ }}\href {\doibase
  10.1103/PhysRevLett.111.186805} {\bibfield  {journal} {\bibinfo  {journal}
  {Phys. Rev. Lett.}\ }\textbf {\bibinfo {volume} {111}},\ \bibinfo {pages}
  {186805} (\bibinfo {year} {2013})}\BibitemShut {NoStop}%
\bibitem [{\citenamefont {Pientka}\ \emph {et~al.}(2014)\citenamefont
  {Pientka}, \citenamefont {Glazman},\ and\ \citenamefont {von
  Oppen}}]{Pientka2014}%
  \BibitemOpen
  \bibfield  {author} {\bibinfo {author} {\bibfnamefont {F.}~\bibnamefont
  {Pientka}}, \bibinfo {author} {\bibfnamefont {L.~I.}\ \bibnamefont
  {Glazman}}, \ and\ \bibinfo {author} {\bibfnamefont {F.}~\bibnamefont {von
  Oppen}},\ }\bibfield  {title} {\emph {\enquote {\bibinfo {title}
  {Unconventional topological phase transitions in helical Shiba chains},}\
  }}\href {\doibase 10.1103/PhysRevB.89.180505} {\bibfield  {journal} {\bibinfo
   {journal} {Phys. Rev. B}\ }\textbf {\bibinfo {volume} {89}},\ \bibinfo
  {pages} {180505} (\bibinfo {year} {2014})}\BibitemShut {NoStop}%
\bibitem [{\citenamefont {Christensen}\ \emph {et~al.}(2016)\citenamefont
  {Christensen}, \citenamefont {Schecter}, \citenamefont {Flensberg},
  \citenamefont {Andersen},\ and\ \citenamefont {Paaske}}]{Christensen2016}%
  \BibitemOpen
  \bibfield  {author} {\bibinfo {author} {\bibfnamefont {M.~H.}\ \bibnamefont
  {Christensen}}, \bibinfo {author} {\bibfnamefont {M.}~\bibnamefont
  {Schecter}}, \bibinfo {author} {\bibfnamefont {K.}~\bibnamefont {Flensberg}},
  \bibinfo {author} {\bibfnamefont {B.~M.}\ \bibnamefont {Andersen}}, \ and\
  \bibinfo {author} {\bibfnamefont {J.}~\bibnamefont {Paaske}},\ }\bibfield
  {title} {\emph {\enquote {\bibinfo {title} {Spiral magnetic order and
  topological superconductivity in a chain of magnetic adatoms on a
  two-dimensional superconductor},}\ }}\href {\doibase
  10.1103/PhysRevB.94.144509} {\bibfield  {journal} {\bibinfo  {journal} {Phys.
  Rev. B}\ }\textbf {\bibinfo {volume} {94}},\ \bibinfo {pages} {144509}
  (\bibinfo {year} {2016})}\BibitemShut {NoStop}%
\bibitem [{\citenamefont {Sharma}\ and\ \citenamefont
  {Tewari}(2016)}]{Sharma2016_shiba}%
  \BibitemOpen
  \bibfield  {author} {\bibinfo {author} {\bibfnamefont {G.}~\bibnamefont
  {Sharma}}\ and\ \bibinfo {author} {\bibfnamefont {S.}~\bibnamefont
  {Tewari}},\ }\bibfield  {title} {\emph {\enquote {\bibinfo {title}
  {Yu-Shiba-Rusinov states and topological superconductivity in Ising paired
  superconductors},}\ }}\href {\doibase 10.1103/PhysRevB.94.094515} {\bibfield
  {journal} {\bibinfo  {journal} {Phys. Rev. B}\ }\textbf {\bibinfo {volume}
  {94}},\ \bibinfo {pages} {094515} (\bibinfo {year} {2016})}\BibitemShut
  {NoStop}%
\bibitem [{\citenamefont {Chatterjee}\ \emph {et~al.}(2023)\citenamefont
  {Chatterjee}, \citenamefont {Pradhan}, \citenamefont {Nandy},\ and\
  \citenamefont {Saha}}]{Chatterjee2023}%
  \BibitemOpen
  \bibfield  {author} {\bibinfo {author} {\bibfnamefont {P.}~\bibnamefont
  {Chatterjee}}, \bibinfo {author} {\bibfnamefont {S.}~\bibnamefont {Pradhan}},
  \bibinfo {author} {\bibfnamefont {A.~K.}\ \bibnamefont {Nandy}}, \ and\
  \bibinfo {author} {\bibfnamefont {A.}~\bibnamefont {Saha}},\ }\bibfield
  {title} {\emph {\enquote {\bibinfo {title} {Tailoring the phase transition
  from topological superconductor to trivial superconductor induced by magnetic
  textures of a spin chain on a $p$-wave superconductor},}\ }}\href {\doibase
  10.1103/PhysRevB.107.085423} {\bibfield  {journal} {\bibinfo  {journal}
  {Phys. Rev. B}\ }\textbf {\bibinfo {volume} {107}},\ \bibinfo {pages}
  {085423} (\bibinfo {year} {2023})}\BibitemShut {NoStop}%
\bibitem [{\citenamefont {Chatterjee}\ \emph
  {et~al.}(2024{\natexlab{a}})\citenamefont {Chatterjee}, \citenamefont
  {Ghosh}, \citenamefont {Nandy},\ and\ \citenamefont
  {Saha}}]{Chatterjee2024_PRBLa}%
  \BibitemOpen
  \bibfield  {author} {\bibinfo {author} {\bibfnamefont {P.}~\bibnamefont
  {Chatterjee}}, \bibinfo {author} {\bibfnamefont {A.~K.}\ \bibnamefont
  {Ghosh}}, \bibinfo {author} {\bibfnamefont {A.~K.}\ \bibnamefont {Nandy}}, \
  and\ \bibinfo {author} {\bibfnamefont {A.}~\bibnamefont {Saha}},\ }\bibfield
  {title} {\emph {\enquote {\bibinfo {title} {Second-order topological
  superconductor via noncollinear magnetic texture},}\ }}\href {\doibase
  10.1103/PhysRevB.109.L041409} {\bibfield  {journal} {\bibinfo  {journal}
  {Phys. Rev. B}\ }\textbf {\bibinfo {volume} {109}},\ \bibinfo {pages}
  {L041409} (\bibinfo {year} {2024}{\natexlab{a}})}\BibitemShut {NoStop}%
\bibitem [{\citenamefont {Chatterjee}\ \emph
  {et~al.}(2024{\natexlab{b}})\citenamefont {Chatterjee}, \citenamefont
  {Banik}, \citenamefont {Bera}, \citenamefont {Ghosh}, \citenamefont
  {Pradhan}, \citenamefont {Saha},\ and\ \citenamefont
  {Nandy}}]{Chatterjee2024_PRBLb}%
  \BibitemOpen
  \bibfield  {author} {\bibinfo {author} {\bibfnamefont {P.}~\bibnamefont
  {Chatterjee}}, \bibinfo {author} {\bibfnamefont {S.}~\bibnamefont {Banik}},
  \bibinfo {author} {\bibfnamefont {S.}~\bibnamefont {Bera}}, \bibinfo {author}
  {\bibfnamefont {A.~K.}\ \bibnamefont {Ghosh}}, \bibinfo {author}
  {\bibfnamefont {S.}~\bibnamefont {Pradhan}}, \bibinfo {author} {\bibfnamefont
  {A.}~\bibnamefont {Saha}}, \ and\ \bibinfo {author} {\bibfnamefont {A.~K.}\
  \bibnamefont {Nandy}},\ }\bibfield  {title} {\emph {\enquote {\bibinfo
  {title} {Topological superconductivity by engineering noncollinear magnetism
  in magnet/superconductor heterostructures: A realistic prescription for the
  two-dimensional Kitaev model},}\ }}\href {\doibase
  10.1103/PhysRevB.109.L121301} {\bibfield  {journal} {\bibinfo  {journal}
  {Phys. Rev. B}\ }\textbf {\bibinfo {volume} {109}},\ \bibinfo {pages}
  {L121301} (\bibinfo {year} {2024}{\natexlab{b}})}\BibitemShut {NoStop}%
\bibitem [{\citenamefont {Subhadarshini}\ \emph {et~al.}(2024)\citenamefont
  {Subhadarshini}, \citenamefont {Pal}, \citenamefont {Chatterjee},\ and\
  \citenamefont {Saha}}]{Subhadarshini2024}%
  \BibitemOpen
  \bibfield  {author} {\bibinfo {author} {\bibfnamefont {M.}~\bibnamefont
  {Subhadarshini}}, \bibinfo {author} {\bibfnamefont {A.}~\bibnamefont {Pal}},
  \bibinfo {author} {\bibfnamefont {P.}~\bibnamefont {Chatterjee}}, \ and\
  \bibinfo {author} {\bibfnamefont {A.}~\bibnamefont {Saha}},\ }\bibfield
  {title} {\emph {\enquote {\bibinfo {title} {Multiple topological phase
  transitions unveiling gapless topological superconductivity in
  magnet/unconventional superconductor hybrid platform},}\ }}\href {\doibase
  10.1063/5.0199275} {\bibfield  {journal} {\bibinfo  {journal} {Applied
  Physics Letters}\ }\textbf {\bibinfo {volume} {124}},\ \bibinfo {pages}
  {183102} (\bibinfo {year} {2024})}\BibitemShut {NoStop}%
\bibitem [{\citenamefont {Subhadarshini}\ \emph {et~al.}(2025)\citenamefont
  {Subhadarshini}, \citenamefont {Pal}, \citenamefont {Chatterjee},\ and\
  \citenamefont {Saha}}]{Subhadarshini2025}%
  \BibitemOpen
  \bibfield  {author} {\bibinfo {author} {\bibfnamefont {M.}~\bibnamefont
  {Subhadarshini}}, \bibinfo {author} {\bibfnamefont {A.}~\bibnamefont {Pal}},
  \bibinfo {author} {\bibfnamefont {P.}~\bibnamefont {Chatterjee}}, \ and\
  \bibinfo {author} {\bibfnamefont {A.}~\bibnamefont {Saha}},\ }\bibfield
  {title} {\emph {\enquote {\bibinfo {title} {Identifying Majorana edge and end
  modes in a Josephson junction of a $p$-wave superconductor with a magnetic
  barrier},}\ }}\href {\doibase 10.1103/6knk-m4vx} {\bibfield  {journal}
  {\bibinfo  {journal} {Phys. Rev. B}\ }\textbf {\bibinfo {volume} {112}},\
  \bibinfo {pages} {115439} (\bibinfo {year} {2025})}\BibitemShut {NoStop}%
\bibitem [{\citenamefont {Mondal}\ \emph
  {et~al.}(2023{\natexlab{b}})\citenamefont {Mondal}, \citenamefont {Ghosh},
  \citenamefont {Nag},\ and\ \citenamefont {Saha}}]{Mondal_2023_Shiba}%
  \BibitemOpen
  \bibfield  {author} {\bibinfo {author} {\bibfnamefont {D.}~\bibnamefont
  {Mondal}}, \bibinfo {author} {\bibfnamefont {A.~K.}\ \bibnamefont {Ghosh}},
  \bibinfo {author} {\bibfnamefont {T.}~\bibnamefont {Nag}}, \ and\ \bibinfo
  {author} {\bibfnamefont {A.}~\bibnamefont {Saha}},\ }\bibfield  {title}
  {\emph {\enquote {\bibinfo {title} {Engineering anomalous Floquet Majorana
  modes and their time evolution in a helical Shiba chain},}\ }}\href {\doibase
  10.1103/PhysRevB.108.L081403} {\bibfield  {journal} {\bibinfo  {journal}
  {Phys. Rev. B}\ }\textbf {\bibinfo {volume} {108}},\ \bibinfo {pages}
  {L081403} (\bibinfo {year} {2023}{\natexlab{b}})}\BibitemShut {NoStop}%
\bibitem [{\citenamefont {Yazdani}(2015)}]{Yazdani_2015}%
  \BibitemOpen
  \bibfield  {author} {\bibinfo {author} {\bibfnamefont {A.}~\bibnamefont
  {Yazdani}},\ }\bibfield  {title} {\emph {\enquote {\bibinfo {title}
  {Visualizing Majorana fermions in a chain of magnetic atoms on a
  superconductor},}\ }}\href {\doibase 10.1088/0031-8949/2015/T164/014012}
  {\bibfield  {journal} {\bibinfo  {journal} {Physica Scripta}\ }\textbf
  {\bibinfo {volume} {2015}},\ \bibinfo {pages} {014012} (\bibinfo {year}
  {2015})}\BibitemShut {NoStop}%
\bibitem [{\citenamefont {Soldini}\ \emph {et~al.}(2023)\citenamefont
  {Soldini}, \citenamefont {K{\"u}ster}, \citenamefont {Wagner}, \citenamefont
  {Das}, \citenamefont {Aldarawsheh}, \citenamefont {Thomale}, \citenamefont
  {Lounis}, \citenamefont {Parkin}, \citenamefont {Sessi},\ and\ \citenamefont
  {Neupert}}]{Soldini2023}%
  \BibitemOpen
  \bibfield  {author} {\bibinfo {author} {\bibfnamefont {M.~O.}\ \bibnamefont
  {Soldini}}, \bibinfo {author} {\bibfnamefont {F.}~\bibnamefont {K{\"u}ster}},
  \bibinfo {author} {\bibfnamefont {G.}~\bibnamefont {Wagner}}, \bibinfo
  {author} {\bibfnamefont {S.}~\bibnamefont {Das}}, \bibinfo {author}
  {\bibfnamefont {A.}~\bibnamefont {Aldarawsheh}}, \bibinfo {author}
  {\bibfnamefont {R.}~\bibnamefont {Thomale}}, \bibinfo {author} {\bibfnamefont
  {S.}~\bibnamefont {Lounis}}, \bibinfo {author} {\bibfnamefont {S.~S.~P.}\
  \bibnamefont {Parkin}}, \bibinfo {author} {\bibfnamefont {P.}~\bibnamefont
  {Sessi}}, \ and\ \bibinfo {author} {\bibfnamefont {T.}~\bibnamefont
  {Neupert}},\ }\bibfield  {title} {\emph {\enquote {\bibinfo {title}
  {Two-dimensional Shiba lattices as a possible platform for crystalline
  topological superconductivity},}\ }}\href {\doibase
  10.1038/s41567-023-02104-5} {\bibfield  {journal} {\bibinfo  {journal}
  {Nature Physics}\ }\textbf {\bibinfo {volume} {19}},\ \bibinfo {pages} {1848}
  (\bibinfo {year} {2023})}\BibitemShut {NoStop}%
\bibitem [{\citenamefont {Wang}\ \emph {et~al.}(2021)\citenamefont {Wang},
  \citenamefont {Wiebe}, \citenamefont {Zhong}, \citenamefont {Gu},\ and\
  \citenamefont {Wiesendanger}}]{Wang2021PRL}%
  \BibitemOpen
  \bibfield  {author} {\bibinfo {author} {\bibfnamefont {D.}~\bibnamefont
  {Wang}}, \bibinfo {author} {\bibfnamefont {J.}~\bibnamefont {Wiebe}},
  \bibinfo {author} {\bibfnamefont {R.}~\bibnamefont {Zhong}}, \bibinfo
  {author} {\bibfnamefont {G.}~\bibnamefont {Gu}}, \ and\ \bibinfo {author}
  {\bibfnamefont {R.}~\bibnamefont {Wiesendanger}},\ }\bibfield  {title} {\emph
  {\enquote {\bibinfo {title} {Spin-Polarized Yu-Shiba-Rusinov States in an
  Iron-Based Superconductor},}\ }}\href {\doibase
  10.1103/PhysRevLett.126.076802} {\bibfield  {journal} {\bibinfo  {journal}
  {Phys. Rev. Lett.}\ }\textbf {\bibinfo {volume} {126}},\ \bibinfo {pages}
  {076802} (\bibinfo {year} {2021})}\BibitemShut {NoStop}%
\bibitem [{\citenamefont {Scaff}\ and\ \citenamefont {Ohl}(1947)}]{Scaff1947}%
  \BibitemOpen
  \bibfield  {author} {\bibinfo {author} {\bibfnamefont {J.~H.}\ \bibnamefont
  {Scaff}}\ and\ \bibinfo {author} {\bibfnamefont {R.~S.}\ \bibnamefont
  {Ohl}},\ }\bibfield  {title} {\emph {\enquote {\bibinfo {title} {Development
  of silicon crystal rectifiers for microwave radar receivers},}\ }}\href
  {\doibase 10.1002/j.1538-7305.1947.tb01310.x} {\bibfield  {journal} {\bibinfo
   {journal} {The Bell System Technical Journal}\ }\textbf {\bibinfo {volume}
  {26}},\ \bibinfo {pages} {1} (\bibinfo {year} {1947})}\BibitemShut {NoStop}%
\bibitem [{\citenamefont {Shockley}(1949)}]{Shockley1949}%
  \BibitemOpen
  \bibfield  {author} {\bibinfo {author} {\bibfnamefont {W.}~\bibnamefont
  {Shockley}},\ }\bibfield  {title} {\emph {\enquote {\bibinfo {title} {The
  theory of p-n junctions in semiconductors and p-n junction transistors},}\
  }}\href {\doibase 10.1002/j.1538-7305.1949.tb03645.x} {\bibfield  {journal}
  {\bibinfo  {journal} {The Bell System Technical Journal}\ }\textbf {\bibinfo
  {volume} {28}},\ \bibinfo {pages} {435} (\bibinfo {year} {1949})}\BibitemShut
  {NoStop}%
\bibitem [{\citenamefont {Nagaosa}\ and\ \citenamefont
  {Yanase}(2024)}]{Nagaosa2024_diode}%
  \BibitemOpen
  \bibfield  {author} {\bibinfo {author} {\bibfnamefont {N.}~\bibnamefont
  {Nagaosa}}\ and\ \bibinfo {author} {\bibfnamefont {Y.}~\bibnamefont
  {Yanase}},\ }\bibfield  {title} {\emph {\enquote {\bibinfo {title}
  {Nonreciprocal Transport and Optical Phenomena in Quantum Materials},}\
  }}\href {\doibase https://doi.org/10.1146/annurev-conmatphys-032822-033734}
  {\bibfield  {journal} {\bibinfo  {journal} {Annual Review of Condensed Matter
  Physics}\ }\textbf {\bibinfo {volume} {15}},\ \bibinfo {pages} {63} (\bibinfo
  {year} {2024})}\BibitemShut {NoStop}%
\bibitem [{\citenamefont {Nadeem}\ \emph {et~al.}(2023)\citenamefont {Nadeem},
  \citenamefont {Fuhrer},\ and\ \citenamefont {Wang}}]{Nadeem2023}%
  \BibitemOpen
  \bibfield  {author} {\bibinfo {author} {\bibfnamefont {M.}~\bibnamefont
  {Nadeem}}, \bibinfo {author} {\bibfnamefont {M.~S.}\ \bibnamefont {Fuhrer}},
  \ and\ \bibinfo {author} {\bibfnamefont {X.}~\bibnamefont {Wang}},\
  }\bibfield  {title} {\emph {\enquote {\bibinfo {title} {The superconducting
  diode effect},}\ }}\href {\doibase 10.1038/s42254-023-00632-w} {\bibfield
  {journal} {\bibinfo  {journal} {Nature Reviews Physics}\ }\textbf {\bibinfo
  {volume} {5}},\ \bibinfo {pages} {558} (\bibinfo {year} {2023})}\BibitemShut
  {NoStop}%
\bibitem [{\citenamefont {Jiang}\ and\ \citenamefont {Hu}(2022)}]{Jiang2022}%
  \BibitemOpen
  \bibfield  {author} {\bibinfo {author} {\bibfnamefont {K.}~\bibnamefont
  {Jiang}}\ and\ \bibinfo {author} {\bibfnamefont {J.}~\bibnamefont {Hu}},\
  }\bibfield  {title} {\emph {\enquote {\bibinfo {title} {Superconducting diode
  effects},}\ }}\href {\doibase 10.1038/s41567-022-01701-0} {\bibfield
  {journal} {\bibinfo  {journal} {Nature Physics}\ }\textbf {\bibinfo {volume}
  {18}},\ \bibinfo {pages} {1145} (\bibinfo {year} {2022})}\BibitemShut
  {NoStop}%
\bibitem [{\citenamefont {Daido}\ \emph {et~al.}(2022)\citenamefont {Daido},
  \citenamefont {Ikeda},\ and\ \citenamefont {Yanase}}]{Daido2022_SDE_PRL}%
  \BibitemOpen
  \bibfield  {author} {\bibinfo {author} {\bibfnamefont {A.}~\bibnamefont
  {Daido}}, \bibinfo {author} {\bibfnamefont {Y.}~\bibnamefont {Ikeda}}, \ and\
  \bibinfo {author} {\bibfnamefont {Y.}~\bibnamefont {Yanase}},\ }\bibfield
  {title} {\emph {\enquote {\bibinfo {title} {Intrinsic Superconducting Diode
  Effect},}\ }}\href {\doibase 10.1103/PhysRevLett.128.037001} {\bibfield
  {journal} {\bibinfo  {journal} {Phys. Rev. Lett.}\ }\textbf {\bibinfo
  {volume} {128}},\ \bibinfo {pages} {037001} (\bibinfo {year}
  {2022})}\BibitemShut {NoStop}%
\bibitem [{\citenamefont {Yuan}\ and\ \citenamefont
  {Fu}(2022)}]{LiangFu2022_PNAS}%
  \BibitemOpen
  \bibfield  {author} {\bibinfo {author} {\bibfnamefont {N.~F.~Q.}\
  \bibnamefont {Yuan}}\ and\ \bibinfo {author} {\bibfnamefont {L.}~\bibnamefont
  {Fu}},\ }\bibfield  {title} {\emph {\enquote {\bibinfo {title} {Supercurrent
  diode effect and finite-momentum superconductors},}\ }}\href {\doibase
  10.1073/pnas.2119548119} {\bibfield  {journal} {\bibinfo  {journal}
  {Proceedings of the National Academy of Sciences}\ }\textbf {\bibinfo
  {volume} {119}},\ \bibinfo {pages} {e2119548119} (\bibinfo {year}
  {2022})}\BibitemShut {NoStop}%
\bibitem [{\citenamefont {He}\ \emph {et~al.}(2022)\citenamefont {He},
  \citenamefont {Tanaka},\ and\ \citenamefont {Nagaosa}}]{He_2022}%
  \BibitemOpen
  \bibfield  {author} {\bibinfo {author} {\bibfnamefont {J.~J.}\ \bibnamefont
  {He}}, \bibinfo {author} {\bibfnamefont {Y.}~\bibnamefont {Tanaka}}, \ and\
  \bibinfo {author} {\bibfnamefont {N.}~\bibnamefont {Nagaosa}},\ }\bibfield
  {title} {\emph {\enquote {\bibinfo {title} {A phenomenological theory of
  superconductor diodes},}\ }}\href {\doibase 10.1088/1367-2630/ac6766}
  {\bibfield  {journal} {\bibinfo  {journal} {New Journal of Physics}\ }\textbf
  {\bibinfo {volume} {24}},\ \bibinfo {pages} {053014} (\bibinfo {year}
  {2022})}\BibitemShut {NoStop}%
\bibitem [{\citenamefont {Fulde}\ and\ \citenamefont
  {Ferrell}(1964)}]{Fulde1964}%
  \BibitemOpen
  \bibfield  {author} {\bibinfo {author} {\bibfnamefont {P.}~\bibnamefont
  {Fulde}}\ and\ \bibinfo {author} {\bibfnamefont {R.~A.}\ \bibnamefont
  {Ferrell}},\ }\bibfield  {title} {\emph {\enquote {\bibinfo {title}
  {Superconductivity in a Strong Spin-Exchange Field},}\ }}\href {\doibase
  10.1103/PhysRev.135.A550} {\bibfield  {journal} {\bibinfo  {journal} {Phys.
  Rev.}\ }\textbf {\bibinfo {volume} {135}},\ \bibinfo {pages} {A550} (\bibinfo
  {year} {1964})}\BibitemShut {NoStop}%
\bibitem [{\citenamefont {Larkin}\ and\ \citenamefont
  {Ovchinnikov}(1964)}]{Larkin_1964}%
  \BibitemOpen
  \bibfield  {author} {\bibinfo {author} {\bibfnamefont {A.~I.}\ \bibnamefont
  {Larkin}}\ and\ \bibinfo {author} {\bibfnamefont {Y.~N.}\ \bibnamefont
  {Ovchinnikov}},\ }\bibfield  {title} {\emph {\enquote {\bibinfo {title}
  {{Nonuniform state of superconductors}},}\ }}\href@noop {} {\bibfield
  {journal} {\bibinfo  {journal} {Zh. Eksp. Teor. Fiz.}\ }\textbf {\bibinfo
  {volume} {47}},\ \bibinfo {pages} {1136} (\bibinfo {year}
  {1964})}\BibitemShut {NoStop}%
\bibitem [{\citenamefont {Ando}\ \emph {et~al.}(2020)\citenamefont {Ando},
  \citenamefont {Miyasaka}, \citenamefont {Li}, \citenamefont {Ishizuka},
  \citenamefont {Arakawa}, \citenamefont {Shiota}, \citenamefont {Moriyama},
  \citenamefont {Yanase},\ and\ \citenamefont {Ono}}]{Ando2020_SDE_Expt}%
  \BibitemOpen
  \bibfield  {author} {\bibinfo {author} {\bibfnamefont {F.}~\bibnamefont
  {Ando}}, \bibinfo {author} {\bibfnamefont {Y.}~\bibnamefont {Miyasaka}},
  \bibinfo {author} {\bibfnamefont {T.}~\bibnamefont {Li}}, \bibinfo {author}
  {\bibfnamefont {J.}~\bibnamefont {Ishizuka}}, \bibinfo {author}
  {\bibfnamefont {T.}~\bibnamefont {Arakawa}}, \bibinfo {author} {\bibfnamefont
  {Y.}~\bibnamefont {Shiota}}, \bibinfo {author} {\bibfnamefont
  {T.}~\bibnamefont {Moriyama}}, \bibinfo {author} {\bibfnamefont
  {Y.}~\bibnamefont {Yanase}}, \ and\ \bibinfo {author} {\bibfnamefont
  {T.}~\bibnamefont {Ono}},\ }\bibfield  {title} {\emph {\enquote {\bibinfo
  {title} {Observation of superconducting diode effect},}\ }}\href {\doibase
  10.1038/s41586-020-2590-4} {\bibfield  {journal} {\bibinfo  {journal}
  {Nature}\ }\textbf {\bibinfo {volume} {584}},\ \bibinfo {pages} {373}
  (\bibinfo {year} {2020})}\BibitemShut {NoStop}%
\bibitem [{\citenamefont {Wu}\ \emph {et~al.}(2022)\citenamefont {Wu},
  \citenamefont {Wang}, \citenamefont {Xu}, \citenamefont {Sivakumar},
  \citenamefont {Pasco}, \citenamefont {Filippozzi}, \citenamefont {Parkin},
  \citenamefont {Zeng}, \citenamefont {McQueen},\ and\ \citenamefont
  {Ali}}]{Wu2022}%
  \BibitemOpen
  \bibfield  {author} {\bibinfo {author} {\bibfnamefont {H.}~\bibnamefont
  {Wu}}, \bibinfo {author} {\bibfnamefont {Y.}~\bibnamefont {Wang}}, \bibinfo
  {author} {\bibfnamefont {Y.}~\bibnamefont {Xu}}, \bibinfo {author}
  {\bibfnamefont {P.~K.}\ \bibnamefont {Sivakumar}}, \bibinfo {author}
  {\bibfnamefont {C.}~\bibnamefont {Pasco}}, \bibinfo {author} {\bibfnamefont
  {U.}~\bibnamefont {Filippozzi}}, \bibinfo {author} {\bibfnamefont {S.~S.~P.}\
  \bibnamefont {Parkin}}, \bibinfo {author} {\bibfnamefont {Y.-J.}\
  \bibnamefont {Zeng}}, \bibinfo {author} {\bibfnamefont {T.}~\bibnamefont
  {McQueen}}, \ and\ \bibinfo {author} {\bibfnamefont {M.~N.}\ \bibnamefont
  {Ali}},\ }\bibfield  {title} {\emph {\enquote {\bibinfo {title} {The
  field-free Josephson diode in a van der Waals heterostructure},}\ }}\href
  {\doibase 10.1038/s41586-022-04504-8} {\bibfield  {journal} {\bibinfo
  {journal} {Nature}\ }\textbf {\bibinfo {volume} {604}},\ \bibinfo {pages}
  {653} (\bibinfo {year} {2022})}\BibitemShut {NoStop}%
\bibitem [{\citenamefont {Lin}\ \emph {et~al.}(2022)\citenamefont {Lin},
  \citenamefont {Siriviboon}, \citenamefont {Scammell}, \citenamefont {Liu},
  \citenamefont {Rhodes}, \citenamefont {Watanabe}, \citenamefont {Taniguchi},
  \citenamefont {Hone}, \citenamefont {Scheurer},\ and\ \citenamefont
  {Li}}]{Lin2022}%
  \BibitemOpen
  \bibfield  {author} {\bibinfo {author} {\bibfnamefont {J.-X.}\ \bibnamefont
  {Lin}}, \bibinfo {author} {\bibfnamefont {P.}~\bibnamefont {Siriviboon}},
  \bibinfo {author} {\bibfnamefont {H.~D.}\ \bibnamefont {Scammell}}, \bibinfo
  {author} {\bibfnamefont {S.}~\bibnamefont {Liu}}, \bibinfo {author}
  {\bibfnamefont {D.}~\bibnamefont {Rhodes}}, \bibinfo {author} {\bibfnamefont
  {K.}~\bibnamefont {Watanabe}}, \bibinfo {author} {\bibfnamefont
  {T.}~\bibnamefont {Taniguchi}}, \bibinfo {author} {\bibfnamefont
  {J.}~\bibnamefont {Hone}}, \bibinfo {author} {\bibfnamefont {M.~S.}\
  \bibnamefont {Scheurer}}, \ and\ \bibinfo {author} {\bibfnamefont {J.~I.~A.}\
  \bibnamefont {Li}},\ }\bibfield  {title} {\emph {\enquote {\bibinfo {title}
  {Zero-field superconducting diode effect in small-twist-angle trilayer
  graphene},}\ }}\href {\doibase 10.1038/s41567-022-01700-1} {\bibfield
  {journal} {\bibinfo  {journal} {Nature Physics}\ }\textbf {\bibinfo {volume}
  {18}},\ \bibinfo {pages} {1221} (\bibinfo {year} {2022})}\BibitemShut
  {NoStop}%
\bibitem [{\citenamefont {Pal}\ \emph {et~al.}(2022)\citenamefont {Pal},
  \citenamefont {Chakraborty}, \citenamefont {Sivakumar}, \citenamefont
  {Davydova}, \citenamefont {Gopi}, \citenamefont {Pandeya}, \citenamefont
  {Krieger}, \citenamefont {Zhang}, \citenamefont {Date}, \citenamefont {Ju},
  \citenamefont {Yuan}, \citenamefont {Schr{\"o}ter}, \citenamefont {Fu},\ and\
  \citenamefont {Parkin}}]{Pal2022}%
  \BibitemOpen
  \bibfield  {author} {\bibinfo {author} {\bibfnamefont {B.}~\bibnamefont
  {Pal}}, \bibinfo {author} {\bibfnamefont {A.}~\bibnamefont {Chakraborty}},
  \bibinfo {author} {\bibfnamefont {P.~K.}\ \bibnamefont {Sivakumar}}, \bibinfo
  {author} {\bibfnamefont {M.}~\bibnamefont {Davydova}}, \bibinfo {author}
  {\bibfnamefont {A.~K.}\ \bibnamefont {Gopi}}, \bibinfo {author}
  {\bibfnamefont {A.~K.}\ \bibnamefont {Pandeya}}, \bibinfo {author}
  {\bibfnamefont {J.~A.}\ \bibnamefont {Krieger}}, \bibinfo {author}
  {\bibfnamefont {Y.}~\bibnamefont {Zhang}}, \bibinfo {author} {\bibfnamefont
  {M.}~\bibnamefont {Date}}, \bibinfo {author} {\bibfnamefont {S.}~\bibnamefont
  {Ju}}, \bibinfo {author} {\bibfnamefont {N.}~\bibnamefont {Yuan}}, \bibinfo
  {author} {\bibfnamefont {N.~B.~M.}\ \bibnamefont {Schr{\"o}ter}}, \bibinfo
  {author} {\bibfnamefont {L.}~\bibnamefont {Fu}}, \ and\ \bibinfo {author}
  {\bibfnamefont {S.~S.~P.}\ \bibnamefont {Parkin}},\ }\bibfield  {title}
  {\emph {\enquote {\bibinfo {title} {Josephson diode effect from Cooper pair
  momentum in a topological semimetal},}\ }}\href {\doibase
  10.1038/s41567-022-01699-5} {\bibfield  {journal} {\bibinfo  {journal}
  {Nature Physics}\ }\textbf {\bibinfo {volume} {18}},\ \bibinfo {pages} {1228}
  (\bibinfo {year} {2022})}\BibitemShut {NoStop}%
\bibitem [{\citenamefont {Ili\ifmmode~\acute{c}\else \'{c}\fi{}}\ and\
  \citenamefont {Bergeret}(2022)}]{Bergeret}%
  \BibitemOpen
  \bibfield  {author} {\bibinfo {author} {\bibfnamefont {S.}~\bibnamefont
  {Ili\ifmmode~\acute{c}\else \'{c}\fi{}}}\ and\ \bibinfo {author}
  {\bibfnamefont {F.~S.}\ \bibnamefont {Bergeret}},\ }\bibfield  {title} {\emph
  {\enquote {\bibinfo {title} {Theory of the Supercurrent Diode Effect in
  Rashba Superconductors with Arbitrary Disorder},}\ }}\href {\doibase
  10.1103/PhysRevLett.128.177001} {\bibfield  {journal} {\bibinfo  {journal}
  {Phys. Rev. Lett.}\ }\textbf {\bibinfo {volume} {128}},\ \bibinfo {pages}
  {177001} (\bibinfo {year} {2022})}\BibitemShut {NoStop}%
\bibitem [{\citenamefont {Zinkl}\ \emph {et~al.}(2022)\citenamefont {Zinkl},
  \citenamefont {Hamamoto},\ and\ \citenamefont {Sigrist}}]{Sigrist2022}%
  \BibitemOpen
  \bibfield  {author} {\bibinfo {author} {\bibfnamefont {B.}~\bibnamefont
  {Zinkl}}, \bibinfo {author} {\bibfnamefont {K.}~\bibnamefont {Hamamoto}}, \
  and\ \bibinfo {author} {\bibfnamefont {M.}~\bibnamefont {Sigrist}},\
  }\bibfield  {title} {\emph {\enquote {\bibinfo {title} {Symmetry conditions
  for the superconducting diode effect in chiral superconductors},}\ }}\href
  {\doibase 10.1103/PhysRevResearch.4.033167} {\bibfield  {journal} {\bibinfo
  {journal} {Phys. Rev. Res.}\ }\textbf {\bibinfo {volume} {4}},\ \bibinfo
  {pages} {033167} (\bibinfo {year} {2022})}\BibitemShut {NoStop}%
\bibitem [{\citenamefont {Banerjee}\ and\ \citenamefont
  {Scheurer}(2024{\natexlab{a}})}]{Banerjee2024_PRL}%
  \BibitemOpen
  \bibfield  {author} {\bibinfo {author} {\bibfnamefont {S.}~\bibnamefont
  {Banerjee}}\ and\ \bibinfo {author} {\bibfnamefont {M.~S.}\ \bibnamefont
  {Scheurer}},\ }\bibfield  {title} {\emph {\enquote {\bibinfo {title}
  {Enhanced Superconducting Diode Effect due to Coexisting Phases},}\ }}\href
  {\doibase 10.1103/PhysRevLett.132.046003} {\bibfield  {journal} {\bibinfo
  {journal} {Phys. Rev. Lett.}\ }\textbf {\bibinfo {volume} {132}},\ \bibinfo
  {pages} {046003} (\bibinfo {year} {2024}{\natexlab{a}})}\BibitemShut
  {NoStop}%
\bibitem [{\citenamefont {Davydova}\ \emph {et~al.}(2022)\citenamefont
  {Davydova}, \citenamefont {Prembabu},\ and\ \citenamefont
  {Fu}}]{LiangFu2022_JDE}%
  \BibitemOpen
  \bibfield  {author} {\bibinfo {author} {\bibfnamefont {M.}~\bibnamefont
  {Davydova}}, \bibinfo {author} {\bibfnamefont {S.}~\bibnamefont {Prembabu}},
  \ and\ \bibinfo {author} {\bibfnamefont {L.}~\bibnamefont {Fu}},\ }\bibfield
  {title} {\emph {\enquote {\bibinfo {title} {Universal Josephson diode
  effect},}\ }}\href {\doibase 10.1126/sciadv.abo0309} {\bibfield  {journal}
  {\bibinfo  {journal} {Science Advances}\ }\textbf {\bibinfo {volume} {8}},\
  \bibinfo {pages} {eabo0309} (\bibinfo {year} {2022})}\BibitemShut {NoStop}%
\bibitem [{\citenamefont {Yuan}\ and\ \citenamefont
  {Fu}(2021)}]{LiangFu2021_PNAS}%
  \BibitemOpen
  \bibfield  {author} {\bibinfo {author} {\bibfnamefont {N.~F.~Q.}\
  \bibnamefont {Yuan}}\ and\ \bibinfo {author} {\bibfnamefont {L.}~\bibnamefont
  {Fu}},\ }\bibfield  {title} {\emph {\enquote {\bibinfo {title} {Topological
  metals and finite-momentum superconductors},}\ }}\href {\doibase
  10.1073/pnas.2019063118} {\bibfield  {journal} {\bibinfo  {journal}
  {Proceedings of the National Academy of Sciences}\ }\textbf {\bibinfo
  {volume} {118}},\ \bibinfo {pages} {e2019063118} (\bibinfo {year}
  {2021})}\BibitemShut {NoStop}%
\bibitem [{\citenamefont {Hasan}\ \emph {et~al.}(2024)\citenamefont {Hasan},
  \citenamefont {Shaffer}, \citenamefont {Khodas},\ and\ \citenamefont
  {Levchenko}}]{Hasan2024}%
  \BibitemOpen
  \bibfield  {author} {\bibinfo {author} {\bibfnamefont {J.}~\bibnamefont
  {Hasan}}, \bibinfo {author} {\bibfnamefont {D.}~\bibnamefont {Shaffer}},
  \bibinfo {author} {\bibfnamefont {M.}~\bibnamefont {Khodas}}, \ and\ \bibinfo
  {author} {\bibfnamefont {A.}~\bibnamefont {Levchenko}},\ }\bibfield  {title}
  {\emph {\enquote {\bibinfo {title} {Supercurrent diode effect in helical
  superconductors},}\ }}\href {\doibase 10.1103/PhysRevB.110.024508} {\bibfield
   {journal} {\bibinfo  {journal} {Phys. Rev. B}\ }\textbf {\bibinfo {volume}
  {110}},\ \bibinfo {pages} {024508} (\bibinfo {year} {2024})}\BibitemShut
  {NoStop}%
\bibitem [{\citenamefont {Legg}\ \emph {et~al.}(2022)\citenamefont {Legg},
  \citenamefont {Loss},\ and\ \citenamefont {Klinovaja}}]{Legg2022_SDE_MCA}%
  \BibitemOpen
  \bibfield  {author} {\bibinfo {author} {\bibfnamefont {H.~F.}\ \bibnamefont
  {Legg}}, \bibinfo {author} {\bibfnamefont {D.}~\bibnamefont {Loss}}, \ and\
  \bibinfo {author} {\bibfnamefont {J.}~\bibnamefont {Klinovaja}},\ }\bibfield
  {title} {\emph {\enquote {\bibinfo {title} {Superconducting diode effect due
  to magnetochiral anisotropy in topological insulators and Rashba
  nanowires},}\ }}\href {\doibase 10.1103/PhysRevB.106.104501} {\bibfield
  {journal} {\bibinfo  {journal} {Phys. Rev. B}\ }\textbf {\bibinfo {volume}
  {106}},\ \bibinfo {pages} {104501} (\bibinfo {year} {2022})}\BibitemShut
  {NoStop}%
\bibitem [{\citenamefont {Daido}\ and\ \citenamefont
  {Yanase}(2022)}]{Daido_2022_PRB}%
  \BibitemOpen
  \bibfield  {author} {\bibinfo {author} {\bibfnamefont {A.}~\bibnamefont
  {Daido}}\ and\ \bibinfo {author} {\bibfnamefont {Y.}~\bibnamefont {Yanase}},\
  }\bibfield  {title} {\emph {\enquote {\bibinfo {title} {Superconducting diode
  effect and nonreciprocal transition lines},}\ }}\href {\doibase
  10.1103/PhysRevB.106.205206} {\bibfield  {journal} {\bibinfo  {journal}
  {Phys. Rev. B}\ }\textbf {\bibinfo {volume} {106}},\ \bibinfo {pages}
  {205206} (\bibinfo {year} {2022})}\BibitemShut {NoStop}%
\bibitem [{\citenamefont {Debnath}\ and\ \citenamefont
  {Dutta}(2024)}]{Debnath2024_diode}%
  \BibitemOpen
  \bibfield  {author} {\bibinfo {author} {\bibfnamefont {D.}~\bibnamefont
  {Debnath}}\ and\ \bibinfo {author} {\bibfnamefont {P.}~\bibnamefont
  {Dutta}},\ }\bibfield  {title} {\emph {\enquote {\bibinfo {title}
  {Gate-tunable Josephson diode effect in Rashba spin-orbit coupled quantum dot
  junctions},}\ }}\href {\doibase 10.1103/PhysRevB.109.174511} {\bibfield
  {journal} {\bibinfo  {journal} {Phys. Rev. B}\ }\textbf {\bibinfo {volume}
  {109}},\ \bibinfo {pages} {174511} (\bibinfo {year} {2024})}\BibitemShut
  {NoStop}%
\bibitem [{\citenamefont {Chatterjee}\ and\ \citenamefont
  {Dutta}(2024)}]{Chatterjee_2024_thermal}%
  \BibitemOpen
  \bibfield  {author} {\bibinfo {author} {\bibfnamefont {P.}~\bibnamefont
  {Chatterjee}}\ and\ \bibinfo {author} {\bibfnamefont {P.}~\bibnamefont
  {Dutta}},\ }\bibfield  {title} {\emph {\enquote {\bibinfo {title}
  {Quasiparticles-mediated thermal diode effect in Weyl Josephson junctions},}\
  }}\href {\doibase 10.1088/1367-2630/ad617a} {\bibfield  {journal} {\bibinfo
  {journal} {New Journal of Physics}\ }\textbf {\bibinfo {volume} {26}},\
  \bibinfo {pages} {073035} (\bibinfo {year} {2024})}\BibitemShut {NoStop}%
\bibitem [{\citenamefont {Mondal}\ \emph
  {et~al.}(2025{\natexlab{b}})\citenamefont {Mondal}, \citenamefont {Fu},\ and\
  \citenamefont {Cayao}}]{Sayan_Mondal_2025}%
  \BibitemOpen
  \bibfield  {author} {\bibinfo {author} {\bibfnamefont {S.}~\bibnamefont
  {Mondal}}, \bibinfo {author} {\bibfnamefont {P.-H.}\ \bibnamefont {Fu}}, \
  and\ \bibinfo {author} {\bibfnamefont {J.}~\bibnamefont {Cayao}},\ }\bibfield
   {title} {\emph {\enquote {\bibinfo {title} {Josephson diode effect with
  Andreev and Majorana bound states},}\ }}\href {\doibase 10.1103/79tj-c3y4}
  {\bibfield  {journal} {\bibinfo  {journal} {Phys. Rev. B}\ }\textbf {\bibinfo
  {volume} {112}},\ \bibinfo {pages} {144506} (\bibinfo {year}
  {2025}{\natexlab{b}})}\BibitemShut {NoStop}%
\bibitem [{\citenamefont {Bhowmik}\ \emph {et~al.}(2025)\citenamefont
  {Bhowmik}, \citenamefont {Samanta}, \citenamefont {Nandy}, \citenamefont
  {Saha},\ and\ \citenamefont {Ghosh}}]{Bhowmik2025a}%
  \BibitemOpen
  \bibfield  {author} {\bibinfo {author} {\bibfnamefont {S.}~\bibnamefont
  {Bhowmik}}, \bibinfo {author} {\bibfnamefont {D.}~\bibnamefont {Samanta}},
  \bibinfo {author} {\bibfnamefont {A.~K.}\ \bibnamefont {Nandy}}, \bibinfo
  {author} {\bibfnamefont {A.}~\bibnamefont {Saha}}, \ and\ \bibinfo {author}
  {\bibfnamefont {S.~K.}\ \bibnamefont {Ghosh}},\ }\bibfield  {title} {\emph
  {\enquote {\bibinfo {title} {Optimizing one dimensional superconducting
  diodes: interplay of Rashba spin-orbit coupling and magnetic fields},}\
  }}\href {\doibase 10.1038/s42005-025-02044-x} {\bibfield  {journal} {\bibinfo
   {journal} {Communications Physics}\ }\textbf {\bibinfo {volume} {8}},\
  \bibinfo {pages} {260} (\bibinfo {year} {2025})}\BibitemShut {NoStop}%
\bibitem [{\citenamefont {Bhowmik}\ and\ \citenamefont
  {Saha}(2025)}]{Bhowmik2025b}%
  \BibitemOpen
  \bibfield  {author} {\bibinfo {author} {\bibfnamefont {S.}~\bibnamefont
  {Bhowmik}}\ and\ \bibinfo {author} {\bibfnamefont {A.}~\bibnamefont {Saha}},\
  }\bibfield  {title} {\emph {\enquote {\bibinfo {title} {Topological Majorana
  zero modes and the superconducting diode effect driven by
  Fulde-Ferrell-Larkin-Ovchinnikov pairing in a helical Shiba chain},}\ }}\href
  {\doibase 10.1103/PhysRevB.111.L161402} {\bibfield  {journal} {\bibinfo
  {journal} {Phys. Rev. B}\ }\textbf {\bibinfo {volume} {111}},\ \bibinfo
  {pages} {L161402} (\bibinfo {year} {2025})}\BibitemShut {NoStop}%
\bibitem [{\citenamefont {Debnath}\ \emph {et~al.}(2026)\citenamefont
  {Debnath}, \citenamefont {Saha},\ and\ \citenamefont {Dutta}}]{Debnath2026a}%
  \BibitemOpen
  \bibfield  {author} {\bibinfo {author} {\bibfnamefont {D.}~\bibnamefont
  {Debnath}}, \bibinfo {author} {\bibfnamefont {A.}~\bibnamefont {Saha}}, \
  and\ \bibinfo {author} {\bibfnamefont {P.}~\bibnamefont {Dutta}},\ }\bibfield
   {title} {\emph {\enquote {\bibinfo {title} {Spin polarization and diode
  effect in thermoelectric current through altermagnet-based superconductor
  heterostructures},}\ }}\href {\doibase 10.1103/jd41-26xr} {\bibfield
  {journal} {\bibinfo  {journal} {Phys. Rev. B}\ }\textbf {\bibinfo {volume}
  {113}},\ \bibinfo {pages} {104508} (\bibinfo {year} {2026})}\BibitemShut
  {NoStop}%
\bibitem [{\citenamefont {Pal}\ \emph {et~al.}()\citenamefont {Pal},
  \citenamefont {Dutta},\ and\ \citenamefont {Saha}}]{Pal2026_pWM}%
  \BibitemOpen
  \bibfield  {author} {\bibinfo {author} {\bibfnamefont {A.}~\bibnamefont
  {Pal}}, \bibinfo {author} {\bibfnamefont {P.}~\bibnamefont {Dutta}}, \ and\
  \bibinfo {author} {\bibfnamefont {A.}~\bibnamefont {Saha}},\ }\bibfield
  {title} {\emph {\enquote {\bibinfo {title} {Emergent superconducting phases
  in unconventional $p$-wave magnets: Topological superconductivity, Bogoliubov
  Fermi surfaces and superconducting diode effect},}\ }}\href
  {https://arxiv.org/abs/2603.03221} {\ }\Eprint
  {http://arxiv.org/abs/2603.03221}{arXiv:2603.03221
  [cond-mat.supr-con]}\BibitemShut {NoStop}%
\bibitem [{\citenamefont {Bauriedl}\ \emph {et~al.}(2022)\citenamefont
  {Bauriedl}, \citenamefont {B{\"a}uml}, \citenamefont {Fuchs}, \citenamefont
  {Baumgartner}, \citenamefont {Paulik}, \citenamefont {Bauer}, \citenamefont
  {Lin}, \citenamefont {Lupton}, \citenamefont {Taniguchi}, \citenamefont
  {Watanabe}, \citenamefont {Strunk},\ and\ \citenamefont
  {Paradiso}}]{Bauriedl2022}%
  \BibitemOpen
  \bibfield  {author} {\bibinfo {author} {\bibfnamefont {L.}~\bibnamefont
  {Bauriedl}}, \bibinfo {author} {\bibfnamefont {C.}~\bibnamefont {B{\"a}uml}},
  \bibinfo {author} {\bibfnamefont {L.}~\bibnamefont {Fuchs}}, \bibinfo
  {author} {\bibfnamefont {C.}~\bibnamefont {Baumgartner}}, \bibinfo {author}
  {\bibfnamefont {N.}~\bibnamefont {Paulik}}, \bibinfo {author} {\bibfnamefont
  {J.~M.}\ \bibnamefont {Bauer}}, \bibinfo {author} {\bibfnamefont {K.-Q.}\
  \bibnamefont {Lin}}, \bibinfo {author} {\bibfnamefont {J.~M.}\ \bibnamefont
  {Lupton}}, \bibinfo {author} {\bibfnamefont {T.}~\bibnamefont {Taniguchi}},
  \bibinfo {author} {\bibfnamefont {K.}~\bibnamefont {Watanabe}}, \bibinfo
  {author} {\bibfnamefont {C.}~\bibnamefont {Strunk}}, \ and\ \bibinfo {author}
  {\bibfnamefont {N.}~\bibnamefont {Paradiso}},\ }\bibfield  {title} {\emph
  {\enquote {\bibinfo {title} {Supercurrent diode effect and magnetochiral
  anisotropy in few-layer NbSe2},}\ }}\href {\doibase
  10.1038/s41467-022-31954-5} {\bibfield  {journal} {\bibinfo  {journal}
  {Nature Communications}\ }\textbf {\bibinfo {volume} {13}},\ \bibinfo {pages}
  {4266} (\bibinfo {year} {2022})}\BibitemShut {NoStop}%
\bibitem [{\citenamefont {Ghosh}\ \emph {et~al.}(2024)\citenamefont {Ghosh},
  \citenamefont {Patil}, \citenamefont {Basu}, \citenamefont {{Kuldeep}},
  \citenamefont {Dutta}, \citenamefont {Jangade}, \citenamefont {Kulkarni},
  \citenamefont {Thamizhavel}, \citenamefont {Steiner}, \citenamefont {von
  Oppen},\ and\ \citenamefont {Deshmukh}}]{Ghosh2024}%
  \BibitemOpen
  \bibfield  {author} {\bibinfo {author} {\bibfnamefont {S.}~\bibnamefont
  {Ghosh}}, \bibinfo {author} {\bibfnamefont {V.}~\bibnamefont {Patil}},
  \bibinfo {author} {\bibfnamefont {A.}~\bibnamefont {Basu}}, \bibinfo {author}
  {\bibnamefont {{Kuldeep}}}, \bibinfo {author} {\bibfnamefont
  {A.}~\bibnamefont {Dutta}}, \bibinfo {author} {\bibfnamefont {D.~A.}\
  \bibnamefont {Jangade}}, \bibinfo {author} {\bibfnamefont {R.}~\bibnamefont
  {Kulkarni}}, \bibinfo {author} {\bibfnamefont {A.}~\bibnamefont
  {Thamizhavel}}, \bibinfo {author} {\bibfnamefont {J.~F.}\ \bibnamefont
  {Steiner}}, \bibinfo {author} {\bibfnamefont {F.}~\bibnamefont {von Oppen}},
  \ and\ \bibinfo {author} {\bibfnamefont {M.~M.}\ \bibnamefont {Deshmukh}},\
  }\bibfield  {title} {\emph {\enquote {\bibinfo {title} {High-temperature
  Josephson diode},}\ }}\href {\doibase 10.1038/s41563-024-01804-4} {\bibfield
  {journal} {\bibinfo  {journal} {Nature Materials}\ }\textbf {\bibinfo
  {volume} {23}},\ \bibinfo {pages} {612} (\bibinfo {year} {2024})}\BibitemShut
  {NoStop}%
\bibitem [{\citenamefont {\ifmmode~\check{S}\else \v{S}\fi{}mejkal}\ \emph
  {et~al.}(2022{\natexlab{a}})\citenamefont {\ifmmode~\check{S}\else
  \v{S}\fi{}mejkal}, \citenamefont {Sinova},\ and\ \citenamefont
  {Jungwirth}}]{Smejkal_PRX_1}%
  \BibitemOpen
  \bibfield  {author} {\bibinfo {author} {\bibfnamefont {L.}~\bibnamefont
  {\ifmmode~\check{S}\else \v{S}\fi{}mejkal}}, \bibinfo {author} {\bibfnamefont
  {J.}~\bibnamefont {Sinova}}, \ and\ \bibinfo {author} {\bibfnamefont
  {T.}~\bibnamefont {Jungwirth}},\ }\bibfield  {title} {\emph {\enquote
  {\bibinfo {title} {Beyond Conventional Ferromagnetism and Antiferromagnetism:
  A Phase with Nonrelativistic Spin and Crystal Rotation Symmetry},}\ }}\href
  {\doibase 10.1103/PhysRevX.12.031042} {\bibfield  {journal} {\bibinfo
  {journal} {Phys. Rev. X}\ }\textbf {\bibinfo {volume} {12}},\ \bibinfo
  {pages} {031042} (\bibinfo {year} {2022}{\natexlab{a}})}\BibitemShut
  {NoStop}%
\bibitem [{\citenamefont {\ifmmode~\check{S}\else \v{S}\fi{}mejkal}\ \emph
  {et~al.}(2022{\natexlab{b}})\citenamefont {\ifmmode~\check{S}\else
  \v{S}\fi{}mejkal}, \citenamefont {Sinova},\ and\ \citenamefont
  {Jungwirth}}]{Smejkal_PRX_2}%
  \BibitemOpen
  \bibfield  {author} {\bibinfo {author} {\bibfnamefont {L.}~\bibnamefont
  {\ifmmode~\check{S}\else \v{S}\fi{}mejkal}}, \bibinfo {author} {\bibfnamefont
  {J.}~\bibnamefont {Sinova}}, \ and\ \bibinfo {author} {\bibfnamefont
  {T.}~\bibnamefont {Jungwirth}},\ }\bibfield  {title} {\emph {\enquote
  {\bibinfo {title} {Emerging Research Landscape of Altermagnetism},}\ }}\href
  {\doibase 10.1103/PhysRevX.12.040501} {\bibfield  {journal} {\bibinfo
  {journal} {Phys. Rev. X}\ }\textbf {\bibinfo {volume} {12}},\ \bibinfo
  {pages} {040501} (\bibinfo {year} {2022}{\natexlab{b}})}\BibitemShut
  {NoStop}%
\bibitem [{\citenamefont {Bhowal}\ and\ \citenamefont
  {Spaldin}(2024)}]{BhowalPRX2024}%
  \BibitemOpen
  \bibfield  {author} {\bibinfo {author} {\bibfnamefont {S.}~\bibnamefont
  {Bhowal}}\ and\ \bibinfo {author} {\bibfnamefont {N.~A.}\ \bibnamefont
  {Spaldin}},\ }\bibfield  {title} {\emph {\enquote {\bibinfo {title}
  {Ferroically Ordered Magnetic Octupoles in $d$-Wave Altermagnets},}\ }}\href
  {\doibase 10.1103/PhysRevX.14.011019} {\bibfield  {journal} {\bibinfo
  {journal} {Phys. Rev. X}\ }\textbf {\bibinfo {volume} {14}},\ \bibinfo
  {pages} {011019} (\bibinfo {year} {2024})}\BibitemShut {NoStop}%
\bibitem [{\citenamefont {Bai}\ \emph {et~al.}(2023)\citenamefont {Bai},
  \citenamefont {Zhang}, \citenamefont {Zhou}, \citenamefont {Chen},
  \citenamefont {Wan}, \citenamefont {Han}, \citenamefont {Zhu}, \citenamefont
  {Liang}, \citenamefont {Su}, \citenamefont {Han}, \citenamefont {Pan},\ and\
  \citenamefont {Song}}]{Bai_PRL_2023}%
  \BibitemOpen
  \bibfield  {author} {\bibinfo {author} {\bibfnamefont {H.}~\bibnamefont
  {Bai}}, \bibinfo {author} {\bibfnamefont {Y.~C.}\ \bibnamefont {Zhang}},
  \bibinfo {author} {\bibfnamefont {Y.~J.}\ \bibnamefont {Zhou}}, \bibinfo
  {author} {\bibfnamefont {P.}~\bibnamefont {Chen}}, \bibinfo {author}
  {\bibfnamefont {C.~H.}\ \bibnamefont {Wan}}, \bibinfo {author} {\bibfnamefont
  {L.}~\bibnamefont {Han}}, \bibinfo {author} {\bibfnamefont {W.~X.}\
  \bibnamefont {Zhu}}, \bibinfo {author} {\bibfnamefont {S.~X.}\ \bibnamefont
  {Liang}}, \bibinfo {author} {\bibfnamefont {Y.~C.}\ \bibnamefont {Su}},
  \bibinfo {author} {\bibfnamefont {X.~F.}\ \bibnamefont {Han}}, \bibinfo
  {author} {\bibfnamefont {F.}~\bibnamefont {Pan}}, \ and\ \bibinfo {author}
  {\bibfnamefont {C.}~\bibnamefont {Song}},\ }\bibfield  {title} {\emph
  {\enquote {\bibinfo {title} {Efficient Spin-to-Charge Conversion via
  Altermagnetic Spin Splitting Effect in Antiferromagnet
  ${\mathrm{RuO}}_{2}$},}\ }}\href {\doibase 10.1103/PhysRevLett.130.216701}
  {\bibfield  {journal} {\bibinfo  {journal} {Phys. Rev. Lett.}\ }\textbf
  {\bibinfo {volume} {130}},\ \bibinfo {pages} {216701} (\bibinfo {year}
  {2023})}\BibitemShut {NoStop}%
\bibitem [{\citenamefont {Lee}\ \emph {et~al.}(2024)\citenamefont {Lee},
  \citenamefont {Lee}, \citenamefont {Jung}, \citenamefont {Jung},
  \citenamefont {Kim}, \citenamefont {Lee}, \citenamefont {Seok}, \citenamefont
  {Kim}, \citenamefont {Park}, \citenamefont {\ifmmode~\check{S}\else
  \v{S}\fi{}mejkal}, \citenamefont {Kang},\ and\ \citenamefont
  {Kim}}]{Lee2024MnTe}%
  \BibitemOpen
  \bibfield  {author} {\bibinfo {author} {\bibfnamefont {S.}~\bibnamefont
  {Lee}}, \bibinfo {author} {\bibfnamefont {S.}~\bibnamefont {Lee}}, \bibinfo
  {author} {\bibfnamefont {S.}~\bibnamefont {Jung}}, \bibinfo {author}
  {\bibfnamefont {J.}~\bibnamefont {Jung}}, \bibinfo {author} {\bibfnamefont
  {D.}~\bibnamefont {Kim}}, \bibinfo {author} {\bibfnamefont {Y.}~\bibnamefont
  {Lee}}, \bibinfo {author} {\bibfnamefont {B.}~\bibnamefont {Seok}}, \bibinfo
  {author} {\bibfnamefont {J.}~\bibnamefont {Kim}}, \bibinfo {author}
  {\bibfnamefont {B.~G.}\ \bibnamefont {Park}}, \bibinfo {author}
  {\bibfnamefont {L.}~\bibnamefont {\ifmmode~\check{S}\else \v{S}\fi{}mejkal}},
  \bibinfo {author} {\bibfnamefont {C.-J.}\ \bibnamefont {Kang}}, \ and\
  \bibinfo {author} {\bibfnamefont {C.}~\bibnamefont {Kim}},\ }\bibfield
  {title} {\emph {\enquote {\bibinfo {title} {Broken Kramers Degeneracy in
  Altermagnetic MnTe},}\ }}\href {\doibase 10.1103/PhysRevLett.132.036702}
  {\bibfield  {journal} {\bibinfo  {journal} {Phys. Rev. Lett.}\ }\textbf
  {\bibinfo {volume} {132}},\ \bibinfo {pages} {036702} (\bibinfo {year}
  {2024})}\BibitemShut {NoStop}%
\bibitem [{\citenamefont {Lin}\ \emph {et~al.}(2025)\citenamefont {Lin},
  \citenamefont {Zhang}, \citenamefont {Lu},\ and\ \citenamefont
  {Xie}}]{Lin2025}%
  \BibitemOpen
  \bibfield  {author} {\bibinfo {author} {\bibfnamefont {H.-J.}\ \bibnamefont
  {Lin}}, \bibinfo {author} {\bibfnamefont {S.-B.}\ \bibnamefont {Zhang}},
  \bibinfo {author} {\bibfnamefont {H.-Z.}\ \bibnamefont {Lu}}, \ and\ \bibinfo
  {author} {\bibfnamefont {X.~C.}\ \bibnamefont {Xie}},\ }\bibfield  {title}
  {\emph {\enquote {\bibinfo {title} {Coulomb Drag in Altermagnets},}\ }}\href
  {\doibase 10.1103/PhysRevLett.134.136301} {\bibfield  {journal} {\bibinfo
  {journal} {Phys. Rev. Lett.}\ }\textbf {\bibinfo {volume} {134}},\ \bibinfo
  {pages} {136301} (\bibinfo {year} {2025})}\BibitemShut {NoStop}%
\bibitem [{\citenamefont {Ghorashi}\ \emph {et~al.}(2024)\citenamefont
  {Ghorashi}, \citenamefont {Hughes},\ and\ \citenamefont
  {Cano}}]{Ghorashi2024PRL}%
  \BibitemOpen
  \bibfield  {author} {\bibinfo {author} {\bibfnamefont {S.~A.~A.}\
  \bibnamefont {Ghorashi}}, \bibinfo {author} {\bibfnamefont {T.~L.}\
  \bibnamefont {Hughes}}, \ and\ \bibinfo {author} {\bibfnamefont
  {J.}~\bibnamefont {Cano}},\ }\bibfield  {title} {\emph {\enquote {\bibinfo
  {title} {Altermagnetic Routes to Majorana Modes in Zero Net Magnetization},}\
  }}\href {\doibase 10.1103/PhysRevLett.133.106601} {\bibfield  {journal}
  {\bibinfo  {journal} {Phys. Rev. Lett.}\ }\textbf {\bibinfo {volume} {133}},\
  \bibinfo {pages} {106601} (\bibinfo {year} {2024})}\BibitemShut {NoStop}%
\bibitem [{\citenamefont {Mondal}\ \emph
  {et~al.}(2025{\natexlab{c}})\citenamefont {Mondal}, \citenamefont {Pal},
  \citenamefont {Saha},\ and\ \citenamefont {Nag}}]{Mondal2025PRBL}%
  \BibitemOpen
  \bibfield  {author} {\bibinfo {author} {\bibfnamefont {D.}~\bibnamefont
  {Mondal}}, \bibinfo {author} {\bibfnamefont {A.}~\bibnamefont {Pal}},
  \bibinfo {author} {\bibfnamefont {A.}~\bibnamefont {Saha}}, \ and\ \bibinfo
  {author} {\bibfnamefont {T.}~\bibnamefont {Nag}},\ }\bibfield  {title} {\emph
  {\enquote {\bibinfo {title} {Distinguishing between topological Majorana and
  trivial zero modes via transport and shot noise study in an altermagnet
  heterostructure},}\ }}\href {\doibase 10.1103/PhysRevB.111.L121401}
  {\bibfield  {journal} {\bibinfo  {journal} {Phys. Rev. B}\ }\textbf {\bibinfo
  {volume} {111}},\ \bibinfo {pages} {L121401} (\bibinfo {year}
  {2025}{\natexlab{c}})}\BibitemShut {NoStop}%
\bibitem [{\citenamefont {Pal}\ \emph {et~al.}(2025)\citenamefont {Pal},
  \citenamefont {Mondal}, \citenamefont {Nag},\ and\ \citenamefont
  {Saha}}]{Pal2025Flq_Josephson}%
  \BibitemOpen
  \bibfield  {author} {\bibinfo {author} {\bibfnamefont {A.}~\bibnamefont
  {Pal}}, \bibinfo {author} {\bibfnamefont {D.}~\bibnamefont {Mondal}},
  \bibinfo {author} {\bibfnamefont {T.}~\bibnamefont {Nag}}, \ and\ \bibinfo
  {author} {\bibfnamefont {A.}~\bibnamefont {Saha}},\ }\bibfield  {title}
  {\emph {\enquote {\bibinfo {title} {Josephson current signature of Floquet
  Majorana and topological accidental zero modes in altermagnet
  heterostructures},}\ }}\href {\doibase 10.1103/prnx-47mk} {\bibfield
  {journal} {\bibinfo  {journal} {Phys. Rev. B}\ }\textbf {\bibinfo {volume}
  {112}},\ \bibinfo {pages} {L201408} (\bibinfo {year} {2025})}\BibitemShut
  {NoStop}%
\bibitem [{\citenamefont {Li}(2024)}]{Li2024}%
  \BibitemOpen
  \bibfield  {author} {\bibinfo {author} {\bibfnamefont {Y.-X.}\ \bibnamefont
  {Li}},\ }\bibfield  {title} {\emph {\enquote {\bibinfo {title} {Realizing
  tunable higher-order topological superconductors with altermagnets},}\
  }}\href {\doibase 10.1103/PhysRevB.109.224502} {\bibfield  {journal}
  {\bibinfo  {journal} {Phys. Rev. B}\ }\textbf {\bibinfo {volume} {109}},\
  \bibinfo {pages} {224502} (\bibinfo {year} {2024})}\BibitemShut {NoStop}%
\bibitem [{\citenamefont {Li}\ \emph {et~al.}(2024)\citenamefont {Li},
  \citenamefont {Liu},\ and\ \citenamefont {Liu}}]{Li_PRBL_2024}%
  \BibitemOpen
  \bibfield  {author} {\bibinfo {author} {\bibfnamefont {Y.-X.}\ \bibnamefont
  {Li}}, \bibinfo {author} {\bibfnamefont {Y.}~\bibnamefont {Liu}}, \ and\
  \bibinfo {author} {\bibfnamefont {C.-C.}\ \bibnamefont {Liu}},\ }\bibfield
  {title} {\emph {\enquote {\bibinfo {title} {Creation and manipulation of
  higher-order topological states by altermagnets},}\ }}\href {\doibase
  10.1103/PhysRevB.109.L201109} {\bibfield  {journal} {\bibinfo  {journal}
  {Phys. Rev. B}\ }\textbf {\bibinfo {volume} {109}},\ \bibinfo {pages}
  {L201109} (\bibinfo {year} {2024})}\BibitemShut {NoStop}%
\bibitem [{\citenamefont {Zhu}\ \emph {et~al.}(2023)\citenamefont {Zhu},
  \citenamefont {Zhuang}, \citenamefont {Wu},\ and\ \citenamefont
  {Yan}}]{Zhu2023}%
  \BibitemOpen
  \bibfield  {author} {\bibinfo {author} {\bibfnamefont {D.}~\bibnamefont
  {Zhu}}, \bibinfo {author} {\bibfnamefont {Z.-Y.}\ \bibnamefont {Zhuang}},
  \bibinfo {author} {\bibfnamefont {Z.}~\bibnamefont {Wu}}, \ and\ \bibinfo
  {author} {\bibfnamefont {Z.}~\bibnamefont {Yan}},\ }\bibfield  {title} {\emph
  {\enquote {\bibinfo {title} {Topological superconductivity in two-dimensional
  altermagnetic metals},}\ }}\href {\doibase 10.1103/PhysRevB.108.184505}
  {\bibfield  {journal} {\bibinfo  {journal} {Phys. Rev. B}\ }\textbf {\bibinfo
  {volume} {108}},\ \bibinfo {pages} {184505} (\bibinfo {year}
  {2023})}\BibitemShut {NoStop}%
\bibitem [{\citenamefont {Yin}\ \emph {et~al.}(2025)\citenamefont {Yin},
  \citenamefont {Li}, \citenamefont {Yan},\ and\ \citenamefont
  {Wan}}]{Yin2025PRB}%
  \BibitemOpen
  \bibfield  {author} {\bibinfo {author} {\bibfnamefont {Z.}~\bibnamefont
  {Yin}}, \bibinfo {author} {\bibfnamefont {H.}~\bibnamefont {Li}}, \bibinfo
  {author} {\bibfnamefont {Z.}~\bibnamefont {Yan}}, \ and\ \bibinfo {author}
  {\bibfnamefont {S.}~\bibnamefont {Wan}},\ }\bibfield  {title} {\emph
  {\enquote {\bibinfo {title} {Multifold Majorana corner modes arising from
  multiple pairs of helical edge states},}\ }}\href {\doibase
  10.1103/PhysRevB.111.085421} {\bibfield  {journal} {\bibinfo  {journal}
  {Phys. Rev. B}\ }\textbf {\bibinfo {volume} {111}},\ \bibinfo {pages}
  {085421} (\bibinfo {year} {2025})}\BibitemShut {NoStop}%
\bibitem [{\citenamefont {Alam}\ \emph {et~al.}()\citenamefont {Alam},
  \citenamefont {Pal}, \citenamefont {Dutta},\ and\ \citenamefont
  {Saha}}]{Alam2025}%
  \BibitemOpen
  \bibfield  {author} {\bibinfo {author} {\bibfnamefont {O.}~\bibnamefont
  {Alam}}, \bibinfo {author} {\bibfnamefont {A.}~\bibnamefont {Pal}}, \bibinfo
  {author} {\bibfnamefont {P.}~\bibnamefont {Dutta}}, \ and\ \bibinfo {author}
  {\bibfnamefont {A.}~\bibnamefont {Saha}},\ }\bibfield  {title} {\emph
  {\enquote {\bibinfo {title} {Proximity-induced superconductivity and emerging
  topological phases in altermagnet-based heterostructures},}\ }}\href
  {https://arxiv.org/abs/2510.26894} {\ }\Eprint
  {http://arxiv.org/abs/2510.26894}{arXiv:2510.26894
  [cond-mat.supr-con]}\BibitemShut {NoStop}%
\bibitem [{\citenamefont {Maeda}\ \emph {et~al.}(2025)\citenamefont {Maeda},
  \citenamefont {Fukaya}, \citenamefont {Yada}, \citenamefont {Lu},
  \citenamefont {Tanaka},\ and\ \citenamefont {Cayao}}]{Maeda2025}%
  \BibitemOpen
  \bibfield  {author} {\bibinfo {author} {\bibfnamefont {K.}~\bibnamefont
  {Maeda}}, \bibinfo {author} {\bibfnamefont {Y.}~\bibnamefont {Fukaya}},
  \bibinfo {author} {\bibfnamefont {K.}~\bibnamefont {Yada}}, \bibinfo {author}
  {\bibfnamefont {B.}~\bibnamefont {Lu}}, \bibinfo {author} {\bibfnamefont
  {Y.}~\bibnamefont {Tanaka}}, \ and\ \bibinfo {author} {\bibfnamefont
  {J.}~\bibnamefont {Cayao}},\ }\bibfield  {title} {\emph {\enquote {\bibinfo
  {title} {Classification of pair symmetries in superconductors with
  unconventional magnetism},}\ }}\href {\doibase 10.1103/PhysRevB.111.144508}
  {\bibfield  {journal} {\bibinfo  {journal} {Phys. Rev. B}\ }\textbf {\bibinfo
  {volume} {111}},\ \bibinfo {pages} {144508} (\bibinfo {year}
  {2025})}\BibitemShut {NoStop}%
\bibitem [{\citenamefont {Fukaya}\ \emph
  {et~al.}(2025{\natexlab{a}})\citenamefont {Fukaya}, \citenamefont {Lu},
  \citenamefont {Yada}, \citenamefont {Tanaka},\ and\ \citenamefont
  {Cayao}}]{Fukaya_2025}%
  \BibitemOpen
  \bibfield  {author} {\bibinfo {author} {\bibfnamefont {Y.}~\bibnamefont
  {Fukaya}}, \bibinfo {author} {\bibfnamefont {B.}~\bibnamefont {Lu}}, \bibinfo
  {author} {\bibfnamefont {K.}~\bibnamefont {Yada}}, \bibinfo {author}
  {\bibfnamefont {Y.}~\bibnamefont {Tanaka}}, \ and\ \bibinfo {author}
  {\bibfnamefont {J.}~\bibnamefont {Cayao}},\ }\bibfield  {title} {\emph
  {\enquote {\bibinfo {title} {Superconducting phenomena in systems with
  unconventional magnets},}\ }}\href {\doibase 10.1088/1361-648X/adf1cf}
  {\bibfield  {journal} {\bibinfo  {journal} {Journal of Physics: Condensed
  Matter}\ }\textbf {\bibinfo {volume} {37}},\ \bibinfo {pages} {313003}
  (\bibinfo {year} {2025}{\natexlab{a}})}\BibitemShut {NoStop}%
\bibitem [{\citenamefont {Zhang}\ \emph {et~al.}(2024)\citenamefont {Zhang},
  \citenamefont {Hu},\ and\ \citenamefont {Neupert}}]{Zhang2024}%
  \BibitemOpen
  \bibfield  {author} {\bibinfo {author} {\bibfnamefont {S.-B.}\ \bibnamefont
  {Zhang}}, \bibinfo {author} {\bibfnamefont {L.-H.}\ \bibnamefont {Hu}}, \
  and\ \bibinfo {author} {\bibfnamefont {T.}~\bibnamefont {Neupert}},\
  }\bibfield  {title} {\emph {\enquote {\bibinfo {title} {Finite-momentum
  Cooper pairing in proximitized altermagnets},}\ }}\href {\doibase
  10.1038/s41467-024-45951-3} {\bibfield  {journal} {\bibinfo  {journal}
  {Nature Communications}\ }\textbf {\bibinfo {volume} {15}},\ \bibinfo {pages}
  {1801} (\bibinfo {year} {2024})}\BibitemShut {NoStop}%
\bibitem [{\citenamefont {Lu}\ \emph {et~al.}(2024)\citenamefont {Lu},
  \citenamefont {Maeda}, \citenamefont {Ito}, \citenamefont {Yada},\ and\
  \citenamefont {Tanaka}}]{Lu2025PRL}%
  \BibitemOpen
  \bibfield  {author} {\bibinfo {author} {\bibfnamefont {B.}~\bibnamefont
  {Lu}}, \bibinfo {author} {\bibfnamefont {K.}~\bibnamefont {Maeda}}, \bibinfo
  {author} {\bibfnamefont {H.}~\bibnamefont {Ito}}, \bibinfo {author}
  {\bibfnamefont {K.}~\bibnamefont {Yada}}, \ and\ \bibinfo {author}
  {\bibfnamefont {Y.}~\bibnamefont {Tanaka}},\ }\bibfield  {title} {\emph
  {\enquote {\bibinfo {title} {$\ensuremath{\varphi}$ Josephson Junction
  Induced by Altermagnetism},}\ }}\href {\doibase
  10.1103/PhysRevLett.133.226002} {\bibfield  {journal} {\bibinfo  {journal}
  {Phys. Rev. Lett.}\ }\textbf {\bibinfo {volume} {133}},\ \bibinfo {pages}
  {226002} (\bibinfo {year} {2024})}\BibitemShut {NoStop}%
\bibitem [{\citenamefont {Fukaya}\ \emph
  {et~al.}(2025{\natexlab{b}})\citenamefont {Fukaya}, \citenamefont {Maeda},
  \citenamefont {Yada}, \citenamefont {Cayao}, \citenamefont {Tanaka},\ and\
  \citenamefont {Lu}}]{Fukaya2025PRB_JJ}%
  \BibitemOpen
  \bibfield  {author} {\bibinfo {author} {\bibfnamefont {Y.}~\bibnamefont
  {Fukaya}}, \bibinfo {author} {\bibfnamefont {K.}~\bibnamefont {Maeda}},
  \bibinfo {author} {\bibfnamefont {K.}~\bibnamefont {Yada}}, \bibinfo {author}
  {\bibfnamefont {J.}~\bibnamefont {Cayao}}, \bibinfo {author} {\bibfnamefont
  {Y.}~\bibnamefont {Tanaka}}, \ and\ \bibinfo {author} {\bibfnamefont
  {B.}~\bibnamefont {Lu}},\ }\bibfield  {title} {\emph {\enquote {\bibinfo
  {title} {Josephson effect and odd-frequency pairing in superconducting
  junctions with unconventional magnets},}\ }}\href {\doibase
  10.1103/PhysRevB.111.064502} {\bibfield  {journal} {\bibinfo  {journal}
  {Phys. Rev. B}\ }\textbf {\bibinfo {volume} {111}},\ \bibinfo {pages}
  {064502} (\bibinfo {year} {2025}{\natexlab{b}})}\BibitemShut {NoStop}%
\bibitem [{\citenamefont {Sun}\ \emph {et~al.}(2025)\citenamefont {Sun},
  \citenamefont {Zhang}, \citenamefont {Li},\ and\ \citenamefont
  {Trauzettel}}]{Sun2025_PRB}%
  \BibitemOpen
  \bibfield  {author} {\bibinfo {author} {\bibfnamefont {H.-P.}\ \bibnamefont
  {Sun}}, \bibinfo {author} {\bibfnamefont {S.-B.}\ \bibnamefont {Zhang}},
  \bibinfo {author} {\bibfnamefont {C.-A.}\ \bibnamefont {Li}}, \ and\ \bibinfo
  {author} {\bibfnamefont {B.}~\bibnamefont {Trauzettel}},\ }\bibfield  {title}
  {\emph {\enquote {\bibinfo {title} {Tunable second harmonic in altermagnetic
  Josephson junctions},}\ }}\href {\doibase 10.1103/PhysRevB.111.165406}
  {\bibfield  {journal} {\bibinfo  {journal} {Phys. Rev. B}\ }\textbf {\bibinfo
  {volume} {111}},\ \bibinfo {pages} {165406} (\bibinfo {year}
  {2025})}\BibitemShut {NoStop}%
\bibitem [{\citenamefont {Ouassou}\ \emph {et~al.}(2023)\citenamefont
  {Ouassou}, \citenamefont {Brataas},\ and\ \citenamefont
  {Linder}}]{Ouassou2023PRL}%
  \BibitemOpen
  \bibfield  {author} {\bibinfo {author} {\bibfnamefont {J.~A.}\ \bibnamefont
  {Ouassou}}, \bibinfo {author} {\bibfnamefont {A.}~\bibnamefont {Brataas}}, \
  and\ \bibinfo {author} {\bibfnamefont {J.}~\bibnamefont {Linder}},\
  }\bibfield  {title} {\emph {\enquote {\bibinfo {title} {dc Josephson Effect
  in Altermagnets},}\ }}\href {\doibase 10.1103/PhysRevLett.131.076003}
  {\bibfield  {journal} {\bibinfo  {journal} {Phys. Rev. Lett.}\ }\textbf
  {\bibinfo {volume} {131}},\ \bibinfo {pages} {076003} (\bibinfo {year}
  {2023})}\BibitemShut {NoStop}%
\bibitem [{\citenamefont {Banerjee}\ and\ \citenamefont
  {Scheurer}(2024{\natexlab{b}})}]{Banerjee2024_PRB}%
  \BibitemOpen
  \bibfield  {author} {\bibinfo {author} {\bibfnamefont {S.}~\bibnamefont
  {Banerjee}}\ and\ \bibinfo {author} {\bibfnamefont {M.~S.}\ \bibnamefont
  {Scheurer}},\ }\bibfield  {title} {\emph {\enquote {\bibinfo {title}
  {Altermagnetic superconducting diode effect},}\ }}\href {\doibase
  10.1103/PhysRevB.110.024503} {\bibfield  {journal} {\bibinfo  {journal}
  {Phys. Rev. B}\ }\textbf {\bibinfo {volume} {110}},\ \bibinfo {pages}
  {024503} (\bibinfo {year} {2024}{\natexlab{b}})}\BibitemShut {NoStop}%
\bibitem [{\citenamefont {Chakraborty}\ and\ \citenamefont
  {Black-Schaffer}(2025)}]{Chakraborty2025_PRL}%
  \BibitemOpen
  \bibfield  {author} {\bibinfo {author} {\bibfnamefont {D.}~\bibnamefont
  {Chakraborty}}\ and\ \bibinfo {author} {\bibfnamefont {A.~M.}\ \bibnamefont
  {Black-Schaffer}},\ }\bibfield  {title} {\emph {\enquote {\bibinfo {title}
  {Perfect Superconducting Diode Effect in Altermagnets},}\ }}\href {\doibase
  10.1103/cv8s-tk4c} {\bibfield  {journal} {\bibinfo  {journal} {Phys. Rev.
  Lett.}\ }\textbf {\bibinfo {volume} {135}},\ \bibinfo {pages} {026001}
  (\bibinfo {year} {2025})}\BibitemShut {NoStop}%
\bibitem [{\citenamefont {Samanta}\ and\ \citenamefont
  {Ghosh}()}]{Samanta2025}%
  \BibitemOpen
  \bibfield  {author} {\bibinfo {author} {\bibfnamefont {D.}~\bibnamefont
  {Samanta}}\ and\ \bibinfo {author} {\bibfnamefont {S.~K.}\ \bibnamefont
  {Ghosh}},\ }\bibfield  {title} {\emph {\enquote {\bibinfo {title} {Field-free
  Superconducting Diode Effect and Topological Fulde-Ferrell-Larkin-Ovchinnikov
  Superconductivity in Altermagnetic Shiba Chains},}\ }}\href
  {https://arxiv.org/abs/2507.21446} {\ }\Eprint
  {http://arxiv.org/abs/2507.21446}{arXiv:2507.21446
  [cond-mat.supr-con]}\BibitemShut {NoStop}%
\bibitem [{\citenamefont {Debnath}\ and\ \citenamefont
  {Dutta}(2025)}]{Debnath_2025}%
  \BibitemOpen
  \bibfield  {author} {\bibinfo {author} {\bibfnamefont {D.}~\bibnamefont
  {Debnath}}\ and\ \bibinfo {author} {\bibfnamefont {P.}~\bibnamefont
  {Dutta}},\ }\bibfield  {title} {\emph {\enquote {\bibinfo {title} {Field-free
  Josephson diode effect in interacting chiral quantum dot junctions},}\
  }}\href {\doibase 10.1088/1361-648X/adbeaf} {\bibfield  {journal} {\bibinfo
  {journal} {Journal of Physics: Condensed Matter}\ }\textbf {\bibinfo {volume}
  {37}},\ \bibinfo {pages} {175301} (\bibinfo {year} {2025})}\BibitemShut
  {NoStop}%
\bibitem [{\citenamefont {Fu}\ \emph {et~al.}(2025)\citenamefont {Fu},
  \citenamefont {Lv}, \citenamefont {Xu}, \citenamefont {Cayao}, \citenamefont
  {Liu},\ and\ \citenamefont {Yu}}]{FuPeiHo2025}%
  \BibitemOpen
  \bibfield  {author} {\bibinfo {author} {\bibfnamefont {P.-H.}\ \bibnamefont
  {Fu}}, \bibinfo {author} {\bibfnamefont {Q.}~\bibnamefont {Lv}}, \bibinfo
  {author} {\bibfnamefont {Y.}~\bibnamefont {Xu}}, \bibinfo {author}
  {\bibfnamefont {J.}~\bibnamefont {Cayao}}, \bibinfo {author} {\bibfnamefont
  {J.-F.}\ \bibnamefont {Liu}}, \ and\ \bibinfo {author} {\bibfnamefont
  {X.-L.}\ \bibnamefont {Yu}},\ }\bibfield  {title} {\emph {\enquote {\bibinfo
  {title} {All-electrically controlled spintronics in altermagnetic
  heterostructures},}\ }}\href {\doibase 10.1038/s41535-025-00827-7} {\bibfield
   {journal} {\bibinfo  {journal} {npj Quantum Materials}\ } (\bibinfo {year}
  {2025}),\ 10.1038/s41535-025-00827-7}\BibitemShut {NoStop}%
\bibitem [{\citenamefont {Wickramaratne}\ and\ \citenamefont
  {Mazin}(2023)}]{Wickramaratne2023}%
  \BibitemOpen
  \bibfield  {author} {\bibinfo {author} {\bibfnamefont {D.}~\bibnamefont
  {Wickramaratne}}\ and\ \bibinfo {author} {\bibfnamefont {I.~I.}\ \bibnamefont
  {Mazin}},\ }\bibfield  {title} {\emph {\enquote {\bibinfo {title} {Ising
  superconductivity: A first-principles perspective},}\ }}\href {\doibase
  10.1063/5.0153345} {\bibfield  {journal} {\bibinfo  {journal} {Applied
  Physics Letters}\ }\textbf {\bibinfo {volume} {122}},\ \bibinfo {pages}
  {240503} (\bibinfo {year} {2023})}\BibitemShut {NoStop}%
\bibitem [{\citenamefont {Zhang}\ \emph {et~al.}(2025)\citenamefont {Zhang},
  \citenamefont {Shavit}, \citenamefont {Ma}, \citenamefont {Han},
  \citenamefont {Siu}, \citenamefont {Mukherjee}, \citenamefont {Watanabe},
  \citenamefont {Taniguchi}, \citenamefont {Hsieh}, \citenamefont
  {Lewandowski}, \citenamefont {von Oppen}, \citenamefont {Oreg},\ and\
  \citenamefont {Nadj-Perge}}]{Zhang2025_Ising}%
  \BibitemOpen
  \bibfield  {author} {\bibinfo {author} {\bibfnamefont {Y.}~\bibnamefont
  {Zhang}}, \bibinfo {author} {\bibfnamefont {G.}~\bibnamefont {Shavit}},
  \bibinfo {author} {\bibfnamefont {H.}~\bibnamefont {Ma}}, \bibinfo {author}
  {\bibfnamefont {Y.}~\bibnamefont {Han}}, \bibinfo {author} {\bibfnamefont
  {C.~W.}\ \bibnamefont {Siu}}, \bibinfo {author} {\bibfnamefont
  {A.}~\bibnamefont {Mukherjee}}, \bibinfo {author} {\bibfnamefont
  {K.}~\bibnamefont {Watanabe}}, \bibinfo {author} {\bibfnamefont
  {T.}~\bibnamefont {Taniguchi}}, \bibinfo {author} {\bibfnamefont
  {D.}~\bibnamefont {Hsieh}}, \bibinfo {author} {\bibfnamefont
  {C.}~\bibnamefont {Lewandowski}}, \bibinfo {author} {\bibfnamefont
  {F.}~\bibnamefont {von Oppen}}, \bibinfo {author} {\bibfnamefont
  {Y.}~\bibnamefont {Oreg}}, \ and\ \bibinfo {author} {\bibfnamefont
  {S.}~\bibnamefont {Nadj-Perge}},\ }\bibfield  {title} {\emph {\enquote
  {\bibinfo {title} {Twist-programmable superconductivity in
  spin--orbit-coupled bilayer graphene},}\ }}\href {\doibase
  10.1038/s41586-025-08959-3} {\bibfield  {journal} {\bibinfo  {journal}
  {Nature}\ }\textbf {\bibinfo {volume} {641}},\ \bibinfo {pages} {625}
  (\bibinfo {year} {2025})}\BibitemShut {NoStop}%
\bibitem [{\citenamefont {Lu}\ \emph {et~al.}(2015)\citenamefont {Lu},
  \citenamefont {Zheliuk}, \citenamefont {Leermakers}, \citenamefont {Yuan},
  \citenamefont {Zeitler}, \citenamefont {Law},\ and\ \citenamefont
  {Ye}}]{Lu2015}%
  \BibitemOpen
  \bibfield  {author} {\bibinfo {author} {\bibfnamefont {J.~M.}\ \bibnamefont
  {Lu}}, \bibinfo {author} {\bibfnamefont {O.}~\bibnamefont {Zheliuk}},
  \bibinfo {author} {\bibfnamefont {I.}~\bibnamefont {Leermakers}}, \bibinfo
  {author} {\bibfnamefont {N.~F.~Q.}\ \bibnamefont {Yuan}}, \bibinfo {author}
  {\bibfnamefont {U.}~\bibnamefont {Zeitler}}, \bibinfo {author} {\bibfnamefont
  {K.~T.}\ \bibnamefont {Law}}, \ and\ \bibinfo {author} {\bibfnamefont
  {J.~T.}\ \bibnamefont {Ye}},\ }\bibfield  {title} {\emph {\enquote {\bibinfo
  {title} {Evidence for two-dimensional Ising superconductivity in gated
  MoS<sub>2</sub>},}\ }}\href {\doibase 10.1126/science.aab2277} {\bibfield
  {journal} {\bibinfo  {journal} {Science}\ }\textbf {\bibinfo {volume}
  {350}},\ \bibinfo {pages} {1353} (\bibinfo {year} {2015})}\BibitemShut
  {NoStop}%
\bibitem [{\citenamefont {Skliannyi}\ \emph {et~al.}(2025)\citenamefont
  {Skliannyi}, \citenamefont {Oreg},\ and\ \citenamefont
  {Stern}}]{Skliannyi2025}%
  \BibitemOpen
  \bibfield  {author} {\bibinfo {author} {\bibfnamefont {D.}~\bibnamefont
  {Skliannyi}}, \bibinfo {author} {\bibfnamefont {Y.}~\bibnamefont {Oreg}}, \
  and\ \bibinfo {author} {\bibfnamefont {A.}~\bibnamefont {Stern}},\ }\bibfield
   {title} {\emph {\enquote {\bibinfo {title} {Gapfull and gapless
  one-dimensional topological superconductivity in spin-orbit-coupled bilayer
  graphene},}\ }}\href {\doibase 10.1103/217r-1fyc} {\bibfield  {journal}
  {\bibinfo  {journal} {Phys. Rev. B}\ }\textbf {\bibinfo {volume} {112}},\
  \bibinfo {pages} {134504} (\bibinfo {year} {2025})}\BibitemShut {NoStop}%
\bibitem [{\citenamefont {Xie}\ \emph {et~al.}(2023)\citenamefont {Xie},
  \citenamefont {Lantagne-Hurtubise}, \citenamefont {Young}, \citenamefont
  {Nadj-Perge},\ and\ \citenamefont {Alicea}}]{Xie2023}%
  \BibitemOpen
  \bibfield  {author} {\bibinfo {author} {\bibfnamefont {Y.-M.}\ \bibnamefont
  {Xie}}, \bibinfo {author} {\bibfnamefont {E.}~\bibnamefont
  {Lantagne-Hurtubise}}, \bibinfo {author} {\bibfnamefont {A.~F.}\ \bibnamefont
  {Young}}, \bibinfo {author} {\bibfnamefont {S.}~\bibnamefont {Nadj-Perge}}, \
  and\ \bibinfo {author} {\bibfnamefont {J.}~\bibnamefont {Alicea}},\
  }\bibfield  {title} {\emph {\enquote {\bibinfo {title} {Gate-Defined
  Topological Josephson Junctions in Bernal Bilayer Graphene},}\ }}\href
  {\doibase 10.1103/PhysRevLett.131.146601} {\bibfield  {journal} {\bibinfo
  {journal} {Phys. Rev. Lett.}\ }\textbf {\bibinfo {volume} {131}},\ \bibinfo
  {pages} {146601} (\bibinfo {year} {2023})}\BibitemShut {NoStop}%
\bibitem [{\citenamefont {Zhao}\ \emph {et~al.}(2023)\citenamefont {Zhao},
  \citenamefont {Sun}, \citenamefont {Tang},\ and\ \citenamefont
  {Zeng}}]{Zhao2023_Ising}%
  \BibitemOpen
  \bibfield  {author} {\bibinfo {author} {\bibfnamefont {D.}~\bibnamefont
  {Zhao}}, \bibinfo {author} {\bibfnamefont {J.}~\bibnamefont {Sun}}, \bibinfo
  {author} {\bibfnamefont {W.}~\bibnamefont {Tang}}, \ and\ \bibinfo {author}
  {\bibfnamefont {Y.-J.}\ \bibnamefont {Zeng}},\ }\bibfield  {title} {\emph
  {\enquote {\bibinfo {title} {Ising magnetoresistance induced by Ising
  spin-orbit coupling},}\ }}\href {\doibase 10.1103/PhysRevB.108.094420}
  {\bibfield  {journal} {\bibinfo  {journal} {Phys. Rev. B}\ }\textbf {\bibinfo
  {volume} {108}},\ \bibinfo {pages} {094420} (\bibinfo {year}
  {2023})}\BibitemShut {NoStop}%
\bibitem [{sup()}]{supp}%
  \BibitemOpen
  \href@noop {} {}\bibinfo {note} {See the Supplemental Material (SM) at
  XXXXXXXXXXX for detailed discussions for vanishing net magnetization of the
  system, Topological phase boundary in the normal state Hamiltonian,
  Validation of weak-coupling BCS approximation, Origin of four MZMs with
  $\mathcal{N}_x=2$, advantage of utilizing self-consistent solutions,
  Superconducting order parameter in real space with OBC, Effect of chemical
  potential on the TSC phase and diode efficiency, Superconductivity in SSH and
  1D BHZ model}\BibitemShut {NoStop}%
\bibitem [{\citenamefont {Chiu}\ \emph {et~al.}(2016)\citenamefont {Chiu},
  \citenamefont {Teo}, \citenamefont {Schnyder},\ and\ \citenamefont
  {Ryu}}]{ChiuRMP2016}%
  \BibitemOpen
  \bibfield  {author} {\bibinfo {author} {\bibfnamefont {C.-K.}\ \bibnamefont
  {Chiu}}, \bibinfo {author} {\bibfnamefont {J.~C.~Y.}\ \bibnamefont {Teo}},
  \bibinfo {author} {\bibfnamefont {A.~P.}\ \bibnamefont {Schnyder}}, \ and\
  \bibinfo {author} {\bibfnamefont {S.}~\bibnamefont {Ryu}},\ }\bibfield
  {title} {\emph {\enquote {\bibinfo {title} {Classification of topological
  quantum matter with symmetries},}\ }}\href {\doibase
  10.1103/RevModPhys.88.035005} {\bibfield  {journal} {\bibinfo  {journal}
  {Rev. Mod. Phys.}\ }\textbf {\bibinfo {volume} {88}},\ \bibinfo {pages}
  {035005} (\bibinfo {year} {2016})}\BibitemShut {NoStop}%
\bibitem [{\citenamefont {Ryu}\ \emph {et~al.}(2010)\citenamefont {Ryu},
  \citenamefont {Schnyder}, \citenamefont {Furusaki},\ and\ \citenamefont
  {Ludwig}}]{RyuNJP2010}%
  \BibitemOpen
  \bibfield  {author} {\bibinfo {author} {\bibfnamefont {S.}~\bibnamefont
  {Ryu}}, \bibinfo {author} {\bibfnamefont {A.~P.}\ \bibnamefont {Schnyder}},
  \bibinfo {author} {\bibfnamefont {A.}~\bibnamefont {Furusaki}}, \ and\
  \bibinfo {author} {\bibfnamefont {A.~W.~W.}\ \bibnamefont {Ludwig}},\
  }\bibfield  {title} {\emph {\enquote {\bibinfo {title} {Topological
  insulators and superconductors: tenfold way and dimensional hierarchy},}\
  }}\href {\doibase 10.1088/1367-2630/12/6/065010} {\bibfield  {journal}
  {\bibinfo  {journal} {New Journal of Physics}\ }\textbf {\bibinfo {volume}
  {12}},\ \bibinfo {pages} {065010} (\bibinfo {year} {2010})}\BibitemShut
  {NoStop}%
\bibitem [{\citenamefont {Benalcazar}\ and\ \citenamefont
  {Cerjan}(2022)}]{Benalcazar2022_PRL}%
  \BibitemOpen
  \bibfield  {author} {\bibinfo {author} {\bibfnamefont {W.~A.}\ \bibnamefont
  {Benalcazar}}\ and\ \bibinfo {author} {\bibfnamefont {A.}~\bibnamefont
  {Cerjan}},\ }\bibfield  {title} {\emph {\enquote {\bibinfo {title}
  {Chiral-Symmetric Higher-Order Topological Phases of Matter},}\ }}\href
  {\doibase 10.1103/PhysRevLett.128.127601} {\bibfield  {journal} {\bibinfo
  {journal} {Phys. Rev. Lett.}\ }\textbf {\bibinfo {volume} {128}},\ \bibinfo
  {pages} {127601} (\bibinfo {year} {2022})}\BibitemShut {NoStop}%
\bibitem [{\citenamefont {Pal}\ and\ \citenamefont
  {Ghosh}(2025)}]{Pal2025_WSM}%
  \BibitemOpen
  \bibfield  {author} {\bibinfo {author} {\bibfnamefont {A.}~\bibnamefont
  {Pal}}\ and\ \bibinfo {author} {\bibfnamefont {A.~K.}\ \bibnamefont
  {Ghosh}},\ }\bibfield  {title} {\emph {\enquote {\bibinfo {title}
  {Multi-higher-order Dirac and nodal line semimetals},}\ }}\href {\doibase
  10.1103/PhysRevB.111.195429} {\bibfield  {journal} {\bibinfo  {journal}
  {Phys. Rev. B}\ }\textbf {\bibinfo {volume} {111}},\ \bibinfo {pages}
  {195429} (\bibinfo {year} {2025})}\BibitemShut {NoStop}%
\bibitem [{\citenamefont {Beenakker}(1997)}]{Beenakker1997}%
  \BibitemOpen
  \bibfield  {author} {\bibinfo {author} {\bibfnamefont {C.~W.~J.}\
  \bibnamefont {Beenakker}},\ }\bibfield  {title} {\emph {\enquote {\bibinfo
  {title} {Random-matrix theory of quantum transport},}\ }}\href {\doibase
  10.1103/RevModPhys.69.731} {\bibfield  {journal} {\bibinfo  {journal} {Rev.
  Mod. Phys.}\ }\textbf {\bibinfo {volume} {69}},\ \bibinfo {pages} {731}
  (\bibinfo {year} {1997})}\BibitemShut {NoStop}%
\bibitem [{\citenamefont {Coleman}(2015)}]{Coleman_2015}%
  \BibitemOpen
  \bibfield  {author} {\bibinfo {author} {\bibfnamefont {P.}~\bibnamefont
  {Coleman}},\ }\href@noop {} {\emph {\bibinfo {title} {Introduction to
  Many-Body Physics}}}\ (\bibinfo  {publisher} {Cambridge University Press},\
  \bibinfo {year} {2015})\BibitemShut {NoStop}%
\bibitem [{\citenamefont {Paul}\ \emph {et~al.}(2018)\citenamefont {Paul},
  \citenamefont {Saha},\ and\ \citenamefont {Das}}]{Paul2018}%
  \BibitemOpen
  \bibfield  {author} {\bibinfo {author} {\bibfnamefont {G.~C.}\ \bibnamefont
  {Paul}}, \bibinfo {author} {\bibfnamefont {A.}~\bibnamefont {Saha}}, \ and\
  \bibinfo {author} {\bibfnamefont {S.}~\bibnamefont {Das}},\ }\bibfield
  {title} {\emph {\enquote {\bibinfo {title} {Spin-selective coupling to
  Majorana zero modes in mixed singlet and triplet superconducting
  nanowires},}\ }}\href {\doibase 10.1103/PhysRevB.97.205446} {\bibfield
  {journal} {\bibinfo  {journal} {Phys. Rev. B}\ }\textbf {\bibinfo {volume}
  {97}},\ \bibinfo {pages} {205446} (\bibinfo {year} {2018})}\BibitemShut
  {NoStop}%
\bibitem [{\citenamefont {Su}\ \emph {et~al.}(1979)\citenamefont {Su},
  \citenamefont {Schrieffer},\ and\ \citenamefont {Heeger}}]{SSH_1979}%
  \BibitemOpen
  \bibfield  {author} {\bibinfo {author} {\bibfnamefont {W.~P.}\ \bibnamefont
  {Su}}, \bibinfo {author} {\bibfnamefont {J.~R.}\ \bibnamefont {Schrieffer}},
  \ and\ \bibinfo {author} {\bibfnamefont {A.~J.}\ \bibnamefont {Heeger}},\
  }\bibfield  {title} {\emph {\enquote {\bibinfo {title} {Solitons in
  Polyacetylene},}\ }}\href {\doibase 10.1103/PhysRevLett.42.1698} {\bibfield
  {journal} {\bibinfo  {journal} {Phys. Rev. Lett.}\ }\textbf {\bibinfo
  {volume} {42}},\ \bibinfo {pages} {1698} (\bibinfo {year}
  {1979})}\BibitemShut {NoStop}%
\bibitem [{\citenamefont {Bernevig}\ \emph {et~al.}(2006)\citenamefont
  {Bernevig}, \citenamefont {Hughes},\ and\ \citenamefont
  {Zhang}}]{Bernevig2006_BHZ}%
  \BibitemOpen
  \bibfield  {author} {\bibinfo {author} {\bibfnamefont {B.~A.}\ \bibnamefont
  {Bernevig}}, \bibinfo {author} {\bibfnamefont {T.~L.}\ \bibnamefont
  {Hughes}}, \ and\ \bibinfo {author} {\bibfnamefont {S.-C.}\ \bibnamefont
  {Zhang}},\ }\bibfield  {title} {\emph {\enquote {\bibinfo {title} {Quantum
  Spin Hall Effect and Topological Phase Transition in HgTe Quantum Wells},}\
  }}\href {\doibase 10.1126/science.1133734} {\bibfield  {journal} {\bibinfo
  {journal} {Science}\ }\textbf {\bibinfo {volume} {314}},\ \bibinfo {pages}
  {1757} (\bibinfo {year} {2006})}\BibitemShut {NoStop}%
\bibitem [{\citenamefont {Faúndez}\ \emph {et~al.}()\citenamefont {Faúndez},
  \citenamefont {Fontenele}, \citenamefont {dos Anjos Sousa-Júnior},
  \citenamefont {Assaad},\ and\ \citenamefont {Costa}}]{Faundez2025}%
  \BibitemOpen
  \bibfield  {author} {\bibinfo {author} {\bibfnamefont {J.}~\bibnamefont
  {Faúndez}}, \bibinfo {author} {\bibfnamefont {R.~A.}\ \bibnamefont
  {Fontenele}}, \bibinfo {author} {\bibfnamefont {S.}~\bibnamefont {dos Anjos
  Sousa-Júnior}}, \bibinfo {author} {\bibfnamefont {F.~F.}\ \bibnamefont
  {Assaad}}, \ and\ \bibinfo {author} {\bibfnamefont {N.~C.}\ \bibnamefont
  {Costa}},\ }\bibfield  {title} {\emph {\enquote {\bibinfo {title} {The
  Two-Dimensional Rashba-Holstein Model: A Quantum Monte Carlo Approach},}\
  }}\href {https://arxiv.org/abs/2411.07119} {\ }\Eprint
  {http://arxiv.org/abs/2411.07119}{arXiv:2411.07119
  [cond-mat.str-el]}\BibitemShut {NoStop}%
\bibitem [{\citenamefont {Ruthvik}\ and\ \citenamefont {Nag}()}]{Ruthvik2025}%
  \BibitemOpen
  \bibfield  {author} {\bibinfo {author} {\bibfnamefont {S.~S.}\ \bibnamefont
  {Ruthvik}}\ and\ \bibinfo {author} {\bibfnamefont {T.}~\bibnamefont {Nag}},\
  }\bibfield  {title} {\emph {\enquote {\bibinfo {title} {Field-free diode
  effects in one-dimensional superconductor: a complex interplay between
  Fulde-Ferrell pairing and altermagnetism},}\ }}\href
  {https://arxiv.org/abs/2512.01415} {\ }\Eprint
  {http://arxiv.org/abs/2512.01415}{arXiv:2512.01415
  [cond-mat.mes-hall]}\BibitemShut {NoStop}%
\end{thebibliography}%
	
\clearpage	
\normalsize\clearpage

\begin{onecolumngrid}
		\begin{center}
			{\fontsize{12}{12}\selectfont
				\textbf{Supplemental Material for ``Topological superconductivity and superconducting diode effect mediated via unconventional magnet and Ising spin-orbit coupling''\\[5mm]}}
			{\normalsize  Amartya Pal\orcidA,$^{1,2}$ Debashish Mondal\orcidD{},$^{1,2}$ Tanay Nag\orcidB{},$^{4}$ and  Arijit Saha\orcidC{},$^{1,2}$ \\[1mm]}
			{\small $^1$\textit{Institute of Physics, Sachivalaya Marg, Bhubaneswar-751005, India}\\[0.5mm]}
			{\small $^2$\textit{Homi Bhabha National Institute, Training School Complex, Anushakti Nagar, Mumbai 400094, India}\\[0.5mm]}
			{\small $^4$\textit{Department of Physics, BITS Pilani-Hydrabad Campus, Telangana 500078, India}\\[0.5mm]}
			{}
		\end{center}
		
		\newcounter{defcounter}
		\setcounter{defcounter}{0}
		\setcounter{equation}{0}
		\setcounter{page}{1}
		\pagenumbering{roman}
		
		\renewcommand{\thesection}{S\arabic{section}}
		
		\newcounter{specialsection}
		\renewcommand{\thespecialsection}{S\arabic{specialsection}}
		
        \newcommand{\specialsection}[1]{
			\refstepcounter{specialsection}
			\addcontentsline{stoc}{section}
			{\protect\numberline{\thespecialsection}#1}
			\bigskip
			\noindent
			{\normalsize\bfseries
				\thespecialsection #1\quad}
			\par\medskip
		}

\section*{Contents}

\makeatletter
\@starttoc{stoc}
\makeatother

		\newcommand{\DN}{\downarrow}
		\newcommand{\UP}{\uparrow}
\begin{center}
\specialsection{\, Vanishing net magnetization of the system} 
\end{center}		
\label{sec:Magneization}

\begin{figure}[h]
	\centering
	\includegraphics[scale=0.65]{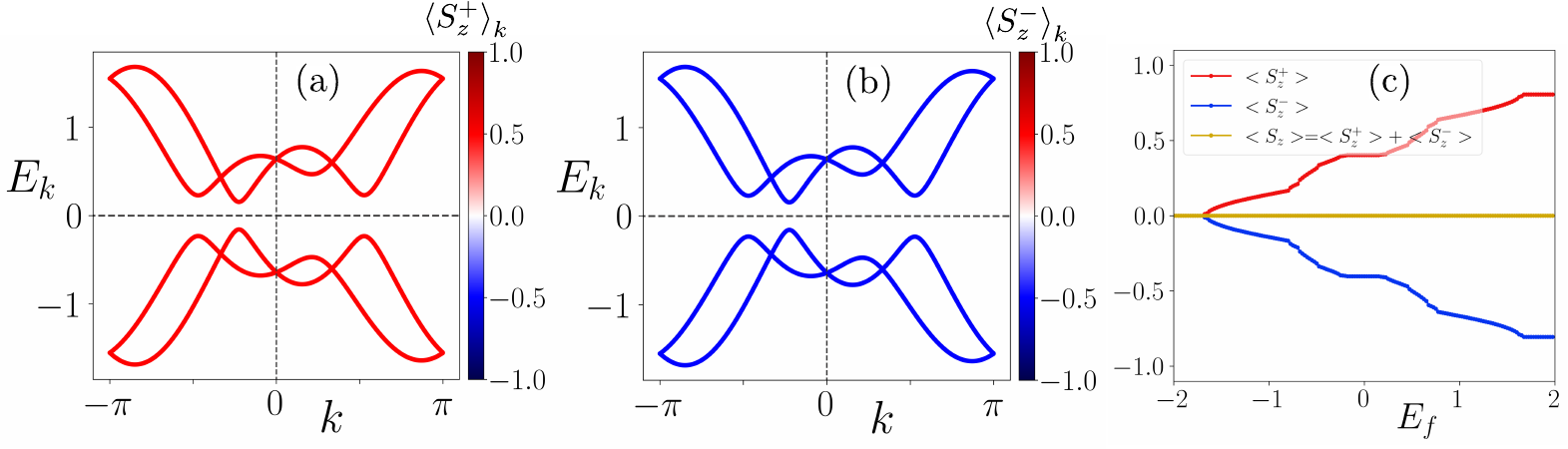}
	\caption{\textbf{Net magnetization of the normal state Hamiltonian:} Panels (a) and (b) highlight the energy-dispersion $E_k$ as a function of $k$ along with the projected spin-polarizations $\braket{S_z^+}_k $ and $\braket{S_z^-}_k $, respectively. In panel (c), we depcit the variation of net magnetization along with projected net-magnetization as a function of the Fermi energy $E_f$. The model parameters are chosen as $(m_0,\lambda_R,J_A,J_I)=(0.5t,0.5t,0.4t,0.25t)$.}
	\label{Fig_Ref2}
\end{figure}

In the main text, we have mentioned that our system is not associated with any net magnetization. To verify this explicitly, we compute the net magnetization of our normal state model Hamiltonian $\mc{H}(k)$ (see Eq.~(1) of the main text) as, 
\begin{equation}
	\braket{S_z}_k = \bra{\Psi_k}\sigma_3 \tau_0\ket{\Psi_k} = \bra{\Psi_k}\sigma^+_3 \tau_0\ket{\Psi_k} + \bra{\Psi_k}\sigma^-_3 \tau_0\ket{\Psi_k} = \braket{S_z^+}_k + \braket{S_z^-}_k\ ,
\end{equation}
where, $\sigma^+_3 = \begin{pmatrix} 1 & 0 \\ 0 & 0 \end{pmatrix}$ and $\sigma^-_3 = \begin{pmatrix} 0 & 0 \\ 0 & -1 \end{pmatrix}$. Here, $\braket{S_z^+}_k$ 
and $\braket{S_z^-}_k$ correspond to the spin-polarization, $\braket{S_z}_k$, projected onto the spin-up and spin-down subspaces of the eigenvector $\Psi_k$ 
with energy eigenvalue $E_k$. In addition, we also compute the net magnetization $\braket{S_z}$ as a function of the Fermi energy $E_f$ of the system as,
\begin{equation}
	\braket{S_z} (E_f) = \sum_{k} \braket{S_z}_k / N_k =  \sum_{k ; E_k\le Ef} \braket{S_z^+}_k / N_k +  \sum_{k ; E_k\le Ef} \braket{S_z^-}_k / N_k =  \braket{S_z^+} (E_f)  + \braket{S_z^-} (E_f) \ ,
\end{equation}
where, $N_k$ is the number of energy-eigenstates at $E=E_f$. The Fermi energy $E_f$ controls the filling of the electronic bands. We first depict the bulk band spectum $E_k$, obtained by diagonalizing $\mc{H}(k)$, and also highlight the variation of $\braket{S_z^+}_k$ in Fig.~\ref{Fig_Ref2}(a) and $\braket{S_z^-}_k$ 
in Fig.~\ref{Fig_Ref2}(b) respectively. We observe that for any state with enegy $E_k$, the value of $\braket{S_z^+}_k$ and $\braket{S_z^-}_k$ remain equal and opposite, thus indicating the net manetization to be zero. Furthermore, we depict $\braket{S_z} (E_f)$ as a function of $E_f$ in Fig.~\ref{Fig_Ref2}(c). We clearly observe that for every value of $E_f$, the magnitude of $\braket{S_z^+}$ and $\braket{S_z^-}$ are exactly equal and opposite to each other, thus the 
net magnetization, $\braket{S_z}$ becomes zero. These results clearly demonstrate that our proposed system does not exhibit any net magnetization.

\begin{center}
	\specialsection{\, Topological phase boundary of the normal state Hamiltonian} 
\end{center}
\label{sec:Topology_normal_state}
In the main text, we have discussed the band topology of the normal state system and present the variation of winding number $\mc{N}_x$ in the $m_0-J_A$ and $J_I-J_A$ plane (see Fig.\,(1) of the main text). We have also mentioned the phase boundary lines in the main text. Here, we analytically derive the topological phase boundary, and the `$q(k)$' matrix to compute $\mc{N}_x$ as mentioned in Eq.\,(2) of the main text. The normal state Hamiltonian, $\mc{H}(k)$ is chiral symmetric with chiral symmetry, $\mc{S}=\sigma_x\tau_x$. Utilizing the chiral symmetry, we rewrite $\mc{H}(k)$ into the diagonal basis of $\mc{S}$ which converts $\mc{H}(k)$ into an anti-diagonal block form as, 		
\begin{eqnarray}
	\mc{H}_D(k)= U_{s}^{\dagger}\mc{H}(k)U_{s}=\begin{pmatrix}
		0&q(k)\\
		q^{\dagger}(k) &0\\
	\end{pmatrix} \label{eq:anti_block},
\end{eqnarray}
where, $U_s$ is the matrix that diagonalizes $\mc{S}$ \ie $U_s^\dagger \mc{S} U_s = {\rm diag} (-1,-1,1,1)$. The $2\times2$ matrix $q(k)$, used in the Eq.\,(2) of the main text, is given by		
\begin{eqnarray}
	q(k)=\!\!\begin{pmatrix}
		m_0	-t\cos k & -J_A \cos k - (J_I-i\lambda_{R})\sin k\\
		-J_A \cos k - (J_I-i\lambda_{R})\sin k& -m_0 + t\cos k 
	\end{pmatrix} \label{eq:qk_supp}.
\end{eqnarray}

In order to find the topological phase boundary, we obtain the expression of gapless lines (the lines on which the bulk becomes gapless) in the model parameter space by computing possible solutions of the equation, 
\begin{eqnarray*}
	&&{\rm det}\left[  \mc{H}(k)\right]=0 \\ &\implies& |{\rm det}\left[  q(k)\right]|=0 
	\\ &\implies& \left[ (m_0 - \cos k + \lambda_{R} \sin k)^2 + (J_A \cos k + J_I\sin k)^2 \right]\left[ (m_0 - \cos k - \lambda_{R} \sin k)^2 + (J_A \cos k + J_I\sin k)^2 \right]=0
\end{eqnarray*}
for any value of $k\in [-\pi,\pi]$. We consider the following three situations which have been discussed in the main text and write the expression of lines over which the bulk is gapless.

\begin{enumerate}
	\item[(i)] When $J_I=J_A=0$ the bulk is gapless only if $m_0\le \sqrt{\lambda_{R}^2 + t^2}$.
	\item[(ii)] When $J_I=0, J_A\ne0$ the bulk is gapless at $k=\pi/2$ only if $m_0=\lambda_{R}$ for any nonzero value of $t$ and $J_A$.
	\item[(iii)] When $J_I\ne0, J_A\ne0$ and $m_0=\lambda_{R}\ne 0$, the bulk is gappless at $k=\tan^{-1}(-J_A/J_I)$ only if $J_A = \frac{t^2-m_0^2}{2m_0t}J_I$ for any nonzero value of $t$.
\end{enumerate}

These expressions match exactly with the topological phase boundary as illustrated using dashed lines in the Fig.\,1(b) and (d) of the main text.

In addition, the normal state Hamiltonian hosts four zero energy localized modes when $J_I=0$ corresponding to $\mc{N}_x=2$ phase (see Fig.\,1(b) of the main text). Here, we analytically show that our model can be block diagonalized into two $2\times 2$ Hamiltonians, which carries the same matrix structure as of the Su–Schrieffer–Heeger (SSH) model. We obtain a fully block-diagonalised matrix form of $\mc{H}(k)$ by performing another unitary tranformation with $\tilde{U}_s$ \ie $\mc{H}_{\rm diag}(k)= \tilde{U}_s \mc{H}(k) \tilde{U}_s^\dagger $.  The unitary matrix $\tilde{U}_s$ and $\mc{H}_{\rm diag}(k)$ is given as, 
\begin{equation}
	\tilde{U}_{s}=\frac{1}{\sqrt{2}}\begin{pmatrix}
		-1 & i & -i& 1\\
		1 & i & i& 1\\
		-1 & -i & i& 1\\
		1 & -i & -i& 1\\
	\end{pmatrix} \label{Eq.Us_tilde}\ ,
\end{equation}, 
\begin{equation}
	\mc{H}_{\rm diag}(k)=\begin{pmatrix}
		0 & \alpha(k)+ \beta(k) + i \gamma(k) & 0& 0\\
		\alpha(k) + \beta(k) - i\gamma(k)  & 0 & 0& 0\\
		0 & 0 &0&\alpha(k)- \beta(k) - i \gamma(k) \\
		0 & 0 & \alpha(k)-\beta(k) +i \gamma(k) &0 \\
	\end{pmatrix} \label{Eq.H_diag},
\end{equation} 
with $\alpha(k)=(m_0-t\cos k)$, $ \beta (k)= \lambda_R \sin k$, and  $\gamma(k)=J_A\cos k$. We can recast the block-diagonal matrix $\mc{H}_{\rm diag}(k)$ in terms of its two block-diagonal matrices as,
\begin{equation}
	\mc{H}_{\rm diag}(k)=\begin{pmatrix}
		\mc{H}^+(k) & 0 \\
		0 & \mc{H}^-(k) 
	\end{pmatrix}\ ,
\end{equation}
where,
\begin{equation}
	\mc{H^\pm}(k)=\left[\alpha(k) \pm \beta(k)\right]\sigma_x  \mp \gamma (k)\sigma_y \ .
\end{equation}

Thus, one can infer that the normal state Hamiltonian, $\mc{H}(k)$ (Eq.\,(1) of the main text) can be decomposed into two Hamiltonians which possess the similar matrix structure as of the SSH model. Under appropiate parameter choice, this system hosts four zero energy modes (characterized by $\mc{N}_x=2$).

\begin{center}
	\specialsection{\, Validation of weak-coupling BCS approximation} 
\end{center}
\label{sec:BCS_mean_field}
%
\begin{figure}[h]
	\centering
	\includegraphics[scale=0.61]{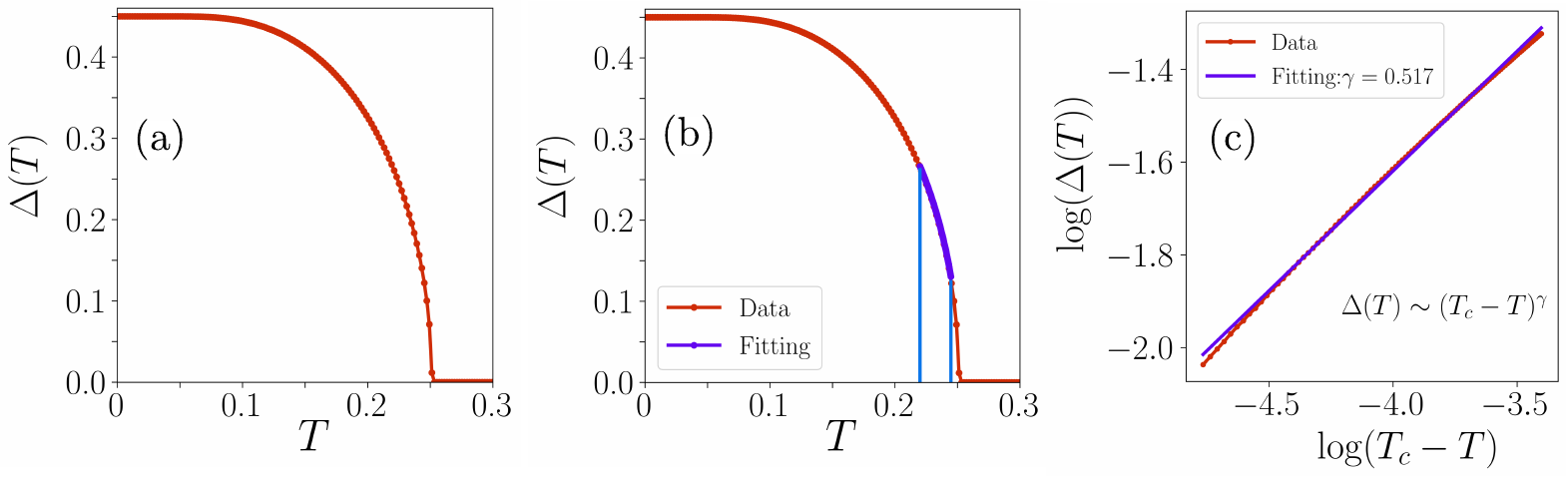}
	\caption {\textbf{Validation of weak-interaction approximation:} Panel (a) depicts the behavior of superconducting order paramter $\Delta(T)$ as a function temperature $T$. In panel (b), we highlight the region bounded by the vertical blue lines, close the critical temperature $T_c$, which is fitted to extract the mean-field exponent $\gamma$. In panel (c), we showcase ${\rm log}(\Delta(T))$ as a function of ${\rm log}(T_c-T)$ and also show the curve fitting having slope $\gamma=0.517$. In all the panels, we choose the model parameters as, $(m_0,\lambda_R,J_A, J_I,\mu)=(0.25t,0.25t,0,0,0)$.}
	\label{Fig_Ref3_Q2}
\end{figure}

In the main text, we consider the strength of attractive interaction, $U=1.5t$, considering into account the weak-coupling BCS approximation. To verify this, 
we investigate two key properties expected for weak-coupling BCS superconductivity.

(i) The value of the ratio $\Delta(T=0)/T_c\sim 1.76$ [P. Coleman, Introduction to Many-Body Physics (Cambridge University Press, 2015)] where $\Delta(T=0)$ is zero temperature superconducting gap 
and $T_c$ denotes the critical temperature above which superconductivity is destroyed. To verify this point, we calculate $\Delta(T)$ as a function of temperature $T$ (see Fig.~\ref{Fig_Ref3_Q2}(a)). We find $\Delta(T=0)= 0.451$ and $T_c=0.254$ leading to the value $\Delta(T=0)/T_c \simeq 1.775$, which is in very good agreement with the weak-coupling approximation.

(ii) The mean field scaling behaviour of $\Delta(T)$ near the critical temperature $T_c$ \ie $\Delta(T\lesssim T_c) \sim (T_c-T)^\gamma$ where the exponent $\gamma=0.5$ for the conventional BCS superconductor. To address this point, we examine the scaling behavior of $\Delta(T)$ near $T\lesssim T_c$ as shown in Fig.~\ref{Fig_Ref3_Q2}(b) (the region bounded by the vertical blue lines). To determine the exponent $\gamma$, we plot ${\rm log}(\Delta(T))$ as a function of ${\rm log} (T_c-T)$ in Fig.~\ref{Fig_Ref3_Q2}(c) and fit the data with a straight line $y = \gamma x + c$. From the slope of this line, we obtain $\gamma \approx 0.517$, which is again in excellent agreement with the expected weak-coupling behavior. Therefore, the choice of our interaction strength is physically justified.

In our mean-field treatment, the superconducting order parameter $\Delta$ depends on the interaction strength $U$ as $ \Delta_{q}^\alpha = -\frac{U}{N} \sum_k \braket{c_{-k+\frac{q}{2} \alpha \DN} c_{k+\frac{q}{2}\alpha\UP}}$ (as mentioned in the main text). As a result, increasing $U$ generally leads to an enhancement in the magnitude of $\Delta$. However, the interaction strength must remain within the weak-coupling regime for the mean-field approximation to remain valid.

\begin{center}
	\specialsection{\, Origin of four MZMs with $\mc{N}_x=2$} \label{sec:MZMs_Nx2}
\end{center}
\begin{figure}[h]
	\centering
	\includegraphics[scale=0.64]{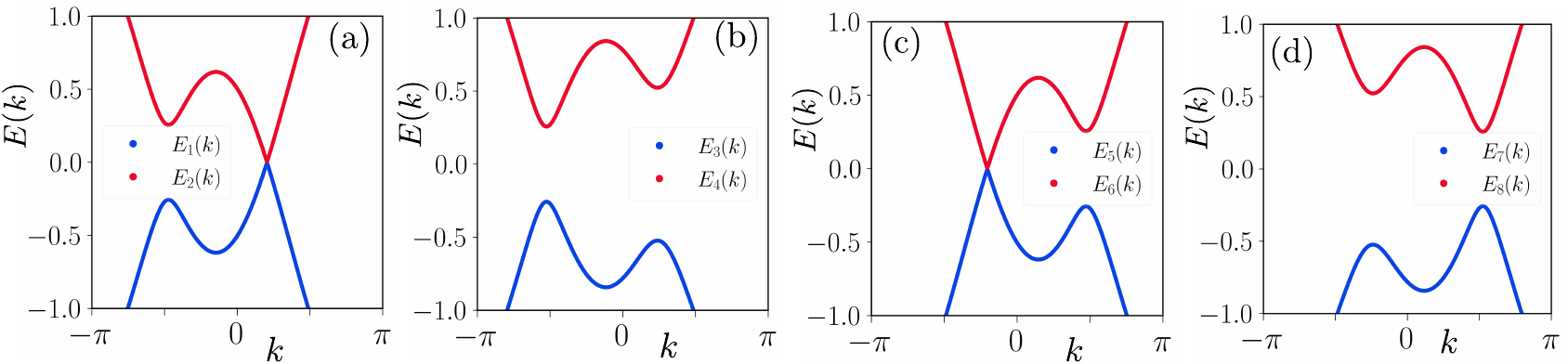}
	\caption{\textbf{Bulk spectrum in the BCS channel close to the topological phase transition points:} In panels (a),(b),(c), and (d), we depict the variation of bulk eigen-spectra ($E_1(k),E_2(k)$), ($E_3(k),E_4(k)$), ($E_5(k),E_6(k)$), and ($E_7(k),E_8(k)$) as a function of momentum $k$, respectively. 
		The model paramters are chosen as $(J_A,\Delta_0,m_0,\lambda_R,J_I,\mu)=(0.33t,0.27t,0.5t,0.5t,0,0)$.}
	\label{Fig.Ref1}
\end{figure}

In the main text, we have shown the existence of four MZMs, corresponding to the $\mc{N}_x=2$ value. Altough, two of them, remain localized at the same end of the 1D system. However, they do not couple to each other due to the multiband nature of our proposed model. Particularly, we consider an eight-band model Hamiltonian, $\mc{H}_{\rm BdG}(k,q)$ (see Eq.\,(2) of the main text). The four MZMs appear from the bulk band gap closing-and-reopening of four different bands with distinct momentum values in the energy spectrum. We observe the appearance of four MZMs in the Bardeen–Cooper–Schrieffer (BCS) channel with $\mc{N}_x=2$ (see Fig.\,2 of the main text). To validate our claim, we obtain the energy eigenvalues of the system ater diagonalizing the $\mc{H}_{\rm BdG}(k,q)$ in the BCS channel ($q_0=0$) as,
\begin{eqnarray}
	E_{1[2]}(k) = -[+]\sqrt{(J_A \cos k  - \Delta_0)^2 + (\lambda_R \sin k - t\cos k +m_0)^2 }\ , \\ 
	E_{3[4]}(k) = -[+]\sqrt{(J_A \cos k  + \Delta_0)^2 + (\lambda_R \sin k - t\cos k +m_0)^2 }\ , \\ 
	E_{5[6]}(k) = -[+]\sqrt{(J_A \cos k  - \Delta_0)^2 + (\lambda_R \sin k + t\cos k -m_0)^2 }\ , \\ 
	E_{7[8]}(k) = -[+]\sqrt{(J_A \cos k  + \Delta_0)^2 + (\lambda_R \sin k + t\cos k -m_0)^2 }\ .  
\end{eqnarray}
Here, all the model parameters $(J_A,\Delta_0,\lambda_R,m_0,t)$ are defined in the main text. In Fig.\,\ref{Fig.Ref1}, we depict these 8 bulk bands corresponding to the topological phase transition point $(J_A,\Delta_0)=(0.33t,0.27t)$ with $(m_0,\lambda_R,J_I,\mu)=(0.5t,0.5t,0,0)$ (as also chosen in Fig.\,2 of the main text). Importantly, we have used the self-consistently obtained value of $\Delta_0$. We find that $E_1(k)$, $E_2(k)$, $E_5(k)$, and $E_6(k)$ feature gapless spectrum while the other bands remain gapped. From these plots, we infer that one pair of MZMs localized at the opposite ends of the chain, appears from the bulk bandgap closing-and-reopening of the $E_1(k)$ and $E_2(k)$ bands, whereas the other pair originates from the gap closing of $E_5(k)$ and $E_6(k)$ bands. Interestingly, the bandgap closing-and-reopening takes place at two different momentum values as clearly observed from Figs.\,\ref{Fig.Ref1}(a) and (c). Therefore the two MZMs, although localized at the same end of the chain, originated from different bands at two different momentum values. As a result, they do not couple to each other and do not split away from zero energy~\cite{Paul2018}.

\vskip +5.0cm
\begin{center}
	\specialsection{\, Advantage of utilizing self-consistent solutions} \label{sec:TSC_w_wo_selfconsistency}
\end{center}

In our main text, we have mentioned that the topological regime in the BCS channel is significantly enhanced when using the self-consistently obtained superconducting order parameter, $\Delta_0$, compared to the non-self-consistent case. Here, we elaborate on this point by systematically investigating topological superconductivity in the BCS channel and comparing the results obtained from both self-consistent and non-self-consistent analyses. 
In the BCS channel, the true ground state $\Delta_0$ is determined self-consistently by minimizing the condensation energy density, $\Omega(q=0,\Delta)$, with respect to $\Delta$. In contrast, in the non-self-consistent case, $\Delta_0$ is fixed to an arbitrary chosen value. To compare these two situations, we compute the winding number $\mc{N}_x$ in the $\Delta$--$J_A$ plane, allowing $\Delta$ to vary freely regardless of the strength of $J_A$. The variation of $\mc{N}_x$ is shown in Fig.\,\ref{Fig.S1}(a), where we highlight two cases: (i) $\Delta_0$ obtained self-consistently (yellow line), and (ii) $\Delta_0$ fixed without self-consistency (green dashed line). For both cases, we set $\Delta_0 = 0.415t$ at $J_A = 0$.

From Fig.\,\ref{Fig.S1}(a), it is evident that when $\Delta_0$ follows the self-consistent solution, the onset of topological superconductivity occurs at a smaller value of $J_A$ compared to the non-self-consistent case. This clearly demonstrates that self-consistency significantly enhances the topological regime. Furthermore, as discussed in the main text, when the system  enters into the topological phase, $\Delta_0$ exhibits a discontinuous jump. This discontinuity can be attributed to the bulk gap closing and reopening of the BdG Hamiltonian [Eq.\,(6) in the main text], which signifies a topological phase transition.

To substantiate this, we explicitly compute the bulk gap $G$ in the $\Delta$--$J_A$ plane, as shown in Fig.\,\ref{Fig.S1}(b). The discontinuous change in $\Delta_0$ indeed coincides with the point where the bulk gap closes and reopens, confirming the topological nature of the transition.

\begin{figure}[h]
	\centering
	\includegraphics[scale=0.7]{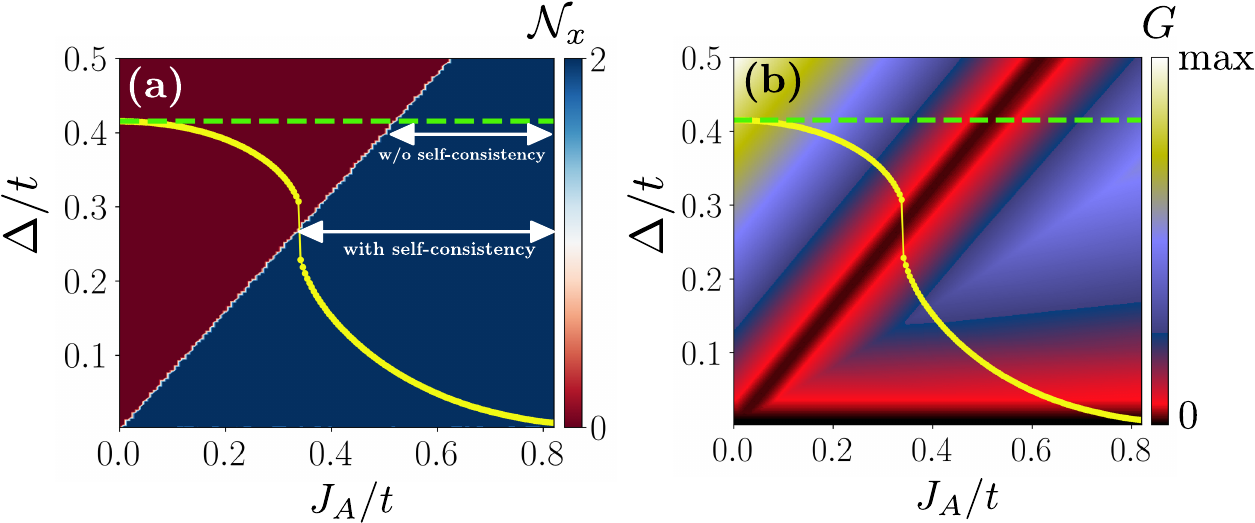}
	\caption{Panel (a): Variation of winding number, $\mc{N}_x$, is depicted in the $\Delta/t-J_A/t$ plane. 
		The variation of true superconducting order $\Delta_0$, obtained self-consistently, is hightlighted by yellow line, while the green dashed line represents the same without (w/o) self-consistent solution. Panel (b): Behavior of bulk gap, $G$ of the BdG Hamiltonian is illustrated, and the dependence of $\Delta_0$ for both self-consistent and w/o self-consistent cases are also highlighted. We choose the other model parameters as $m_0=\lambda_{R}=0.5t, \mu=J_I=0,U=1.5t,t=1$.}
	\label{Fig.S1}
\end{figure}

In short, our intention is to highlight the difference between two commonly used approaches in the literature:

(i) A fully self-consistent mean-field treatment, where the superconducting order parameter $\Delta$ is obtained by minimizing the condensation energy 
(as implemented in our main analysis).

(ii) A phenomenological approach in which $\Delta$ is treated as a fixed parameter and the superconducting properties are analyzed without considering the self-consistent mean field approach and without taking into account the modification of $\Delta$ as one changes various model parameters. Such non self-consistent analysis are present in the literature~\cite{Ghorashi2024PRL,Li_PRBL_2024}. In our work, we only intend to illustrate the differences between these two approaches. Importantly, this comparison does not imply that the interaction strength $U$ is varied rapidly in order to keep $\Delta$ fixed.

\vskip +5.0cm

\begin{center}
	\specialsection{\, Superconducting order parameter in real space with OBC}~\label{sec:Real_space}
\end{center}

%

\begin{figure}[h]
	\centering
	\includegraphics[scale=0.62]{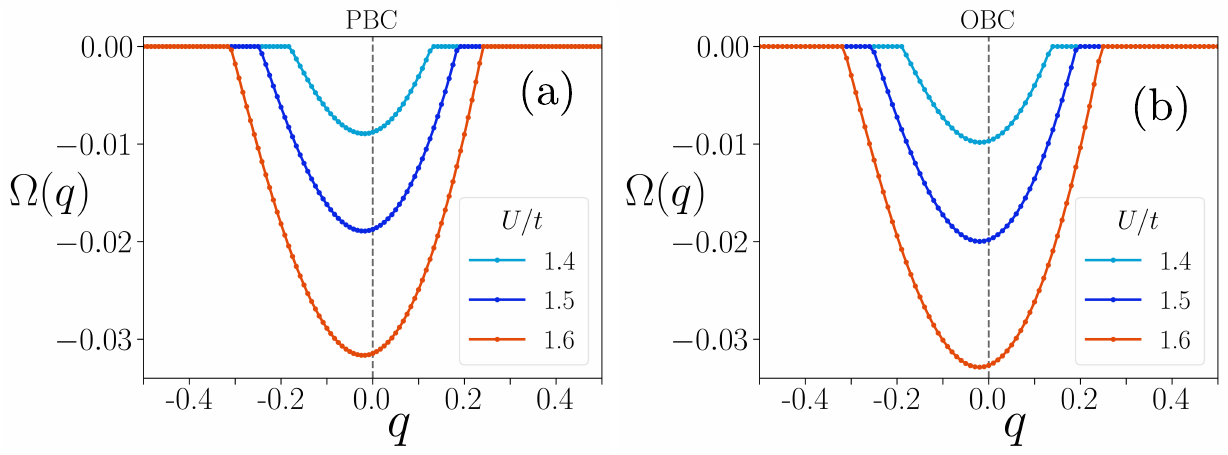}
	\caption{\textbf{Variation of condensation energy density in the FF ground state:} Panels (a) and (b) depict the variation of the condensation energy densiy $\Omega(q)$ with respect to $q$ considering PBC and OBC, respectively for three different choices of $U/t$. The model parameters are chosen as, $(m_0,\lambda_R,J_A, J_I,\mu)=(0.5t,0.5t,0.2t,0.2t,0)$. For OBC results, we consider a finite system with $L_x=250$ lattice sites.}
	\label{FigRef3_U_effect}
\end{figure}
In the main text, the eigenvalues of $H_{\rm BdG}(k,q)$ in real space is obtained after performing a lattice regularization of the momentum space Hamiltonian.  We can write the corresponding Hamilonian in real space on a one-dimensional lattice as (assuming the lattice spacing, $a=1$),  
\begin{eqnarray}
	\mc{H}_{\rm RS} &=& \sum_{x}^{L_x} \sum_{\alpha,\alpha',s,s'} c^{\dagger}_{x\alpha,s} [\frac{t}{2} \sigma_0 \tau_z + \frac{J_A-iJ_I}{2}\sigma_z\tau_x + \frac{-i\lambda_R}{2}\sigma_y \tau_0]_{\alpha s;\alpha' s'} c_{x+1,\alpha',s'} + {\rm H.c.} +  c^{\dagger}_{x\alpha,s} [m_0\sigma_0\tau_z]_{\alpha s;\alpha' s'} c_{x\alpha',s'} \nonumber \\
	&& + \sum_{x}^{L_x} \sum_{\alpha} c^{\dagger}_{x\alpha,\UP}[\Delta_0 e^{-iq_0x} ] c^{\dagger}_{x\alpha,\DN} + {\rm H.c.}\ , \label{Eq._Ham_RS}
\end{eqnarray} 
where, $c_{x,\alpha,s}$ represents the annhilation operator of the electron at site `$x$', in the orbital `$\alpha$' and with spin `$s$'. All the model parameters are already defined in Eq.\,(1) and Eq.\,(6) of the main text. Note that, in Eq.\,(\ref{Eq._Ham_RS}), $\Delta_0$ and $q_0$ correspond to the self-consistently obtained order parameters of the superconducting ground state in the FFLO channel. We have used the above Hamiltonian to compute the energy eigenvalues in real space to establish the presence of Majorana zero energy modes in the system as presented in Fig.\,2(b) of the main text. 

In our work, we first compute the superconducting order parameter $\Delta(q)$ 
self-consistently in momentum space (referred to as periodic boundary condition (PBC)) and using the same order parameter, we diagonalize the real-space BdG Hamiltonian $H_{\rm BdG}$ employing open boundary conditions (OBC), assuming that boundary effects on $\Delta$ are negligible as it is a bulk property.

To verify this assumption, we also compute the condensation energy density $\Omega(q)$ considering both PBC and OBC. The procedure to obtain $\Omega(q)$ 
in the PBC is mentioned in the main text. We use the following relation to find $\Omega(q)$ in real-space employing OBC, $\Omega (q,\Delta) = F(q,\Delta) - F(q,0)$ with $F(q,\Delta)$ is the free energy density of the superconductor at zero temperature obtained using the relation,  $\displaystyle{F(q,\Delta) = \frac{1}{L_x} \sum_{m, E_{m}<0}\!\!\!\! E_{m} + \frac{2\Delta^2}{U}}$ with $\{ E_{m}\}$ being the energy eigenvalues of the Hamiltonian in real space $\mc{H}_{RS}$ defined in Eq.~\eqref{Eq._Ham_RS}. We compare the results in Fig.~\ref{FigRef3_U_effect}(a) and (b) corresponding to the PBC and OBC cases. 
We find that in both situations, $\Omega(q)$ exhibits a single minimum at a finite value of $q$, indicating that the superconducting ground state corresponds 
to a FF phase.


\begin{figure}
	\centering
	\includegraphics[scale=0.62]{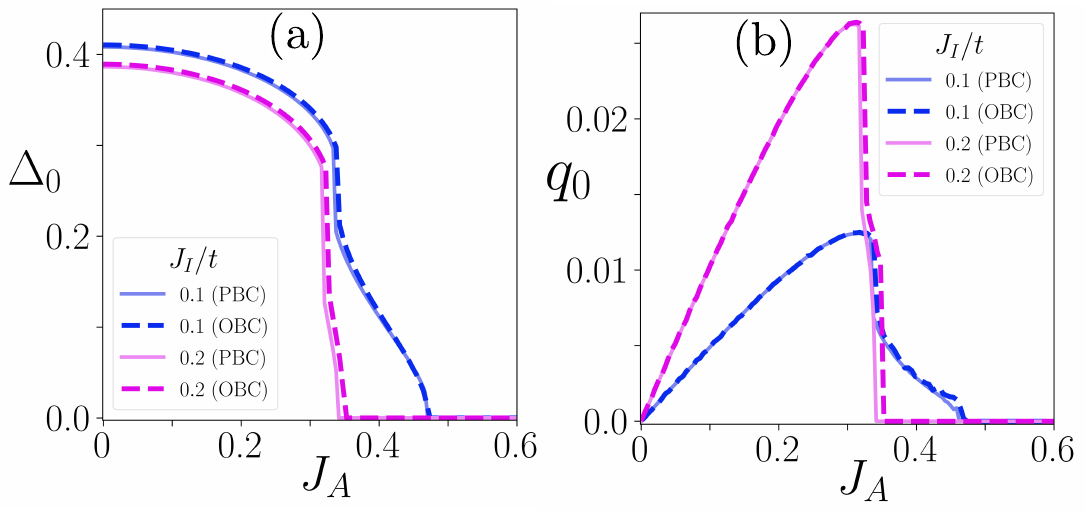}
	\caption{\textbf{Comparision between results obtained via PBC and OBC:} Panels (a) and (b) depict the variation of $\Delta_0$ and $q_0$ as a function of $J_A$, respectively. The model parameters are chosen as, $(m_0,\lambda_R,\mu)=(0.5t,0.5t,0)$. For OBC results, we consider a finite system with $L_x=250$ lattice sites.}
	\label{FigRef3_OBC_PBC}
\end{figure}


We also recalculate the order paramters $\Delta_0$ and $q_0$ in real space employing OBC and compare to the results obtained with PBC in Fig.\,\ref{FigRef3_OBC_PBC}. The variation of $\Delta_0$ and $q_0$ as a function of $J_{A}$ is presented in Fig.\,\ref{FigRef3_OBC_PBC}(a) and \ref{FigRef3_OBC_PBC}(b), respectively. We find that the superconducting order parameters obtained with OBC exhibit excellent agreement with those obtained 
under PBC, thereby justifying our approach. Therefore, our results presented in the manuscript remain valid for the real-space system with OBC.  

These results suggest that, within our parameter regime, the FF state remains the energetically favored solution even when OBC are imposed. This is consistent with the fact that the superconducting ground state is primarily determined by bulk energetics, while boundary conditions introduce only minor quantitative modifications in finite size systems.

We also would like to emphasize that superconducting order parameters, computed with the site-dependent mean-field order paramater in $\Delta_i$ should be consistent with our PBC results apart from some minor quantitative changes. This is beacuse our normal state model Hamiltonian do not have any site-dependent terms e.g., disorder or magnetic impurities, and thus preserves translational symmetry.  In addition, the attractive Hubbard interaction `$U$', from which the superconducting order parameter is generated, is also assumed to be uniform throughout the system. With these considerations, we can approximate the superconducting order paramter $\Delta_i$ to be site-independent and uniformly distributed along our one-dimensional system of interest.

%
\begin{center}
	\specialsection{\, Effect of chemical potential on the TSC phase}~\label{sec:mu_TSC_phase}
\end{center}

\begin{figure}
	\centering
	\includegraphics[scale=0.6]{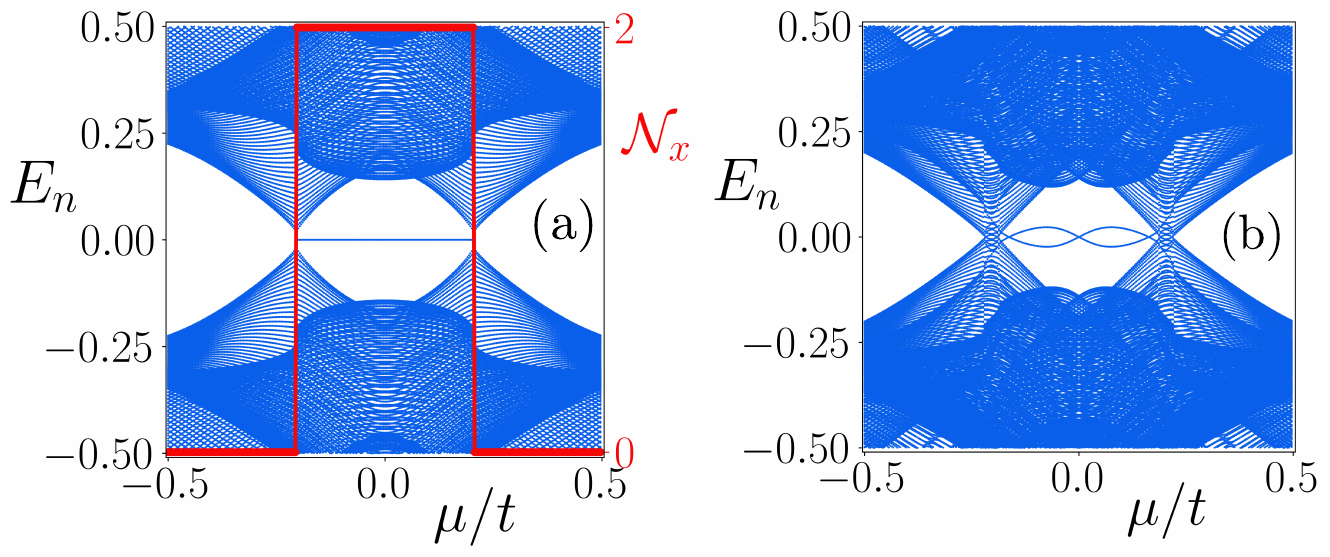}
	\caption{\textbf{Effect of chemical potential on the TSC phases:}  We depict the variation of eigenvalue spectrum $E_n$ as function of chemical potential $\mu/t$ corresponding to BCS and FFLO channels in panels (a) (left axis) and (b), respectively. In panel (a) (right axis), we display the winding number $\mc{N}_x$ as a function of $\mu/t$ to characterize the zero modes in BCS channel. We choose $(J_A,J_I)=(0.4t,0)$ in panel (a), and $(J_A,J_I)=(0.4t,0.1t)$ in panel (b) considering a system size of $L_x=250$ lattice sites for both panels. The other model paramters are chosen as $(m_0,\lambda_R,U)=(0.5t,0.5t,1.5t)$ and $\Delta_0$ and $q_0$ are obtained self-consistently by minimizing the condensation energy density.}
	\label{Fig.Ref1_mu}
\end{figure}

In our work, the altermagnetic order opens a gap near the crossing points of the Rashba bands, which effectively mimics the effect of a perpendicular magnetic field. Therefore, similar to the Rashba nanowire model~\cite{Alicea_2012}, the topological phases are sensitive to the chemical potential. 

To further illustrate this point, we investigate the effect of the chemical potential $\mu$ on the topological superconducting phase in both the BCS and FFLO channels. In particular, we first compute the energy eigenspectra in real space under OBC to examine the presence of zero-energy modes, and subsequently characterize these modes topologically by the appropriate topological index, if present. In Fig.\,\ref{Fig.Ref1_mu}(a) (left axis), we depict the energy eigenvalues $E_n$ as a function of $\mu/t$ in the BCS channel. We find the presence of zero-energy modes even at $\mu=0$, which are topologically characterized by a nonzero value of winding number $\mathcal{N}_x=2$, as shown in Fig.\,\ref{Fig.Ref1_mu}(a) (right axis).

Then, for the FFLO channel, we display the variation of $E_n$ with respect to $\mu/t$ in Fig.\,\ref{Fig.Ref1_mu}(b). Interestingly, we find that the MZMs are 
now gapped out if $\mu\ne 0$, in sharp contrast to the BCS channel. This behavior can be understood as follows. In our model, the MZMs are topologically protected by the chiral symmetry, which is broken for $\mu \neq 0$ in the full BdG Hamiltonian, particularly in the FFLO channel. However, such topological protection is preserved in the BCS channel. As a result, the MZMs in the FFLO channel are not protected by chiral symmetry and are therefore gapped away 
from zero energy at finite $\mu$.


\begin{center}
	\specialsection{\, Effect of chemical potential on the diode efficiency}~\label{sec:mu_eta_dependence}
\end{center}		

\begin{figure}
	\centering
	\includegraphics[scale=0.6]{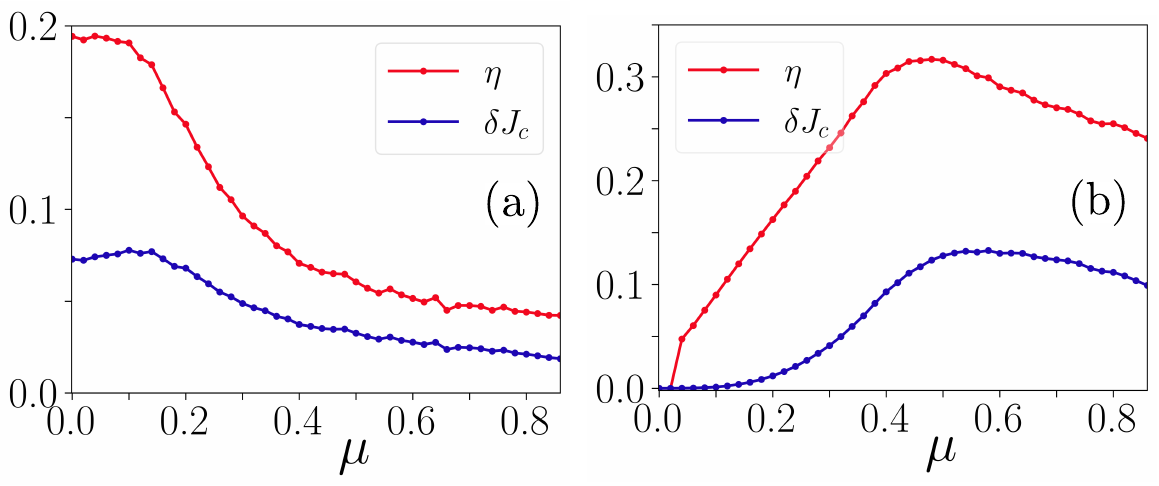}
	\caption{\textbf{Variation of diode efficiency with chemical potential:} In panels (a) and (b), we depict the variation of $\eta$ and $\delta J_c$ as function of chemical potential $\mu$ for two paramaeter sets. We fix $(J_A,J_I)=(0.4t,0.1t)$ in panel (a), and $(J_A,J_I)=(0.4t,0.4t)$ in panel (b). The other model paramters are chosen as $(m_0,\lambda_R,U)=(0.15t,0.15t,1.5t)$.}
	\label{Fig.Ref1_Q2}
\end{figure}

We note that tuning of chemical potential can change the diode efficiency. However, it is not guaranteed that the efficiency follows a fixed trend or pattern with chemical potential irrespective of the choice of other parameters. The reason behind this non-univeral behavior is the following: the imbalance between the number of left and right moving Cooper pairs would determine the diode current and subsequently the efficiency. These numbers change for each and every non-degenrate bands while their dispersions are controlled by the tight-binding parameters $J_A, J_I, m_0, \lambda_R$. The chemical potential determines how many of such bands to be considered for the transport. Therefore, by changing chemical potential, one can tune the total number of left and right movers originating from all the filled bands till that chemical potential. As a result, the diode current changes with $\mu$. Below we demonstrate two set of parameter regime where diode current/ efficiency changes in an opposite manner with respect to $\mu$.

In the main text, we have optimized the diode efficiency $\eta$ by suitably varying the model paramters such as, $J_A$, $J_I$, $m_0$, $\lambda_R$, and $\mu$. In this procedure, we have found the a large diode efficiency of $\eta \sim 65\%$, for $\mu=0.7t$ (as shown in Fig.\,4(d) of the main text) compared to $\mu=0$ where the maximum diode efficency is $\sim 29\%$ (as shown in Fig.\,4(c) of the main text). This may suggest that tuning the chemical potential enhances the diode efficiency. However, this is not true in general as the diode efficiency strongly depends on other model paramaters as well. To further illustrate this point, we consider two parameter sets, (i) $(J_A,J_I)=(0.4t,0.1t)$ and (ii) $(J_A,J_I)=(0.4t,0.4t)$ and choose other fixed parameters as $(m_0,\lambda_R,U)=(0.15t,0.15t,1.5t)$. For these two sets, we compute the diode efficiency $\eta$ and difference between critical current densities $\delta J_c=|J_c^+-J_c^-|$ (defined in Eq.(8) of the main text) for various values of $\mu$, and depict the corresponding behavior in Fig.\,\ref{Fig.Ref1_Q2}. 
In Fig.\,\ref{Fig.Ref1_Q2}(a), we observe both $\eta$ and $\delta J_c$ decrease with increasing the chemical potential $\mu$, whereas in Fig.\,\ref{Fig.Ref1_Q2}(b) we observe the complete opposite behavior. From these observations, we infer that tuning only the chemical potential does not always enhance the diode efficiency, rather it strongly depends on the other model parameters as well.

\newpage		
\begin{center}
	\specialsection{\, Superconductivity in SSH and 1D BHZ model} \label{sec:TSC_SSH_BHZ}
\end{center}

\begin{figure}
	\centering
	\includegraphics[scale=0.95]{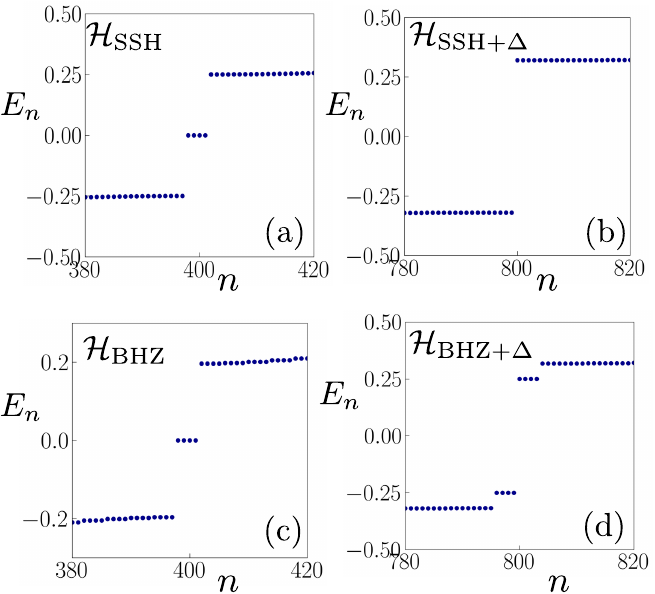}
	\caption{Variation of real-space energy eigenvalue spectra $E_n$ is shown as a function of state index $n$ employing open boundary condition. Panels (a) and (b) refer to the spinful SSH model in the normal state with model paramters $(t_1,t_2=0.25,0.5)$ and in the superconducting state with model parameters as $(t_1,t_2,\Delta=0.25,0.5,0.2)$. Panels (c) and (d) corresponds to 1D BHZ model in the normal state and in the superconducting state, respectively with model paramaters chosen as, $(\lambda,\tilde{m},t=0.2,0.2,1)$ and $(\lambda,\tilde{m},t,\Delta=0.2,0.2,1,0.25)$ and finite system size $L_x=200$. }
	\label{Fig.S2}
\end{figure}

Our model Hamiltonian, as introduced in the main text, retains its topological order in the normal state as well as in the superconducting state. Specifically, in the normal state our system can be represented as a topological insulator while in the superconducting state it behaves as a TSC. However, this intricate feature is absent in the well-known models of topological insulator in one dimension. We consider two model Hamiltonians, (i) Su–Schrieffer–Heeger (SSH) model~\cite{SSH_1979} and (ii) 1D Bernevig–Hughes–Zhang (BHZ) model~\cite{Bernevig2006_BHZ}, both hosting topological insulating phase in the normal state and it has been shown that in presence of superconducting order, these models loose their topological character.
\vskip 0.2cm
$\bullet$ \underline{SSH model:}
The model Hamiltonian in the normal state for the spinfull SSH model is given by,
\begin{equation}
	\mc{H}_{\rm SSH} (k_x) = (t_1 + t_2\cos k_x)\sigma_xs_0 + t_2\sin k_x\, \sigma_ys_0\ ,
\end{equation}
where, $t_1$ and $t_2$ denote the strength of the intra-cell and inter-cell hopping amplitudes. Here, `$\sigma$' and `$s$' denote the Pauli matrices in orbital and spin space. Theis system hosts topological insulating phase when $t_1<t_2$ with two zero energy modes. Introducing a regular $s$-wave superconducting pairing in the system, the SSH model can be written in the BdG basis as, 
\begin{equation}
	\mc{H}_{\rm SSH + \Delta} (k) = \begin{pmatrix}
		\mc{H}_{\rm SSH} (k) & -is_y \Delta \\
		is_y\Delta & -\mc{H}_{\rm SSH}^T (-k)
	\end{pmatrix}\ ,
\end{equation} where, $\Delta$ is the superconducting pairing amplitude. 

\vskip 0.2cm
$\bullet$ \underline{1D BHZ model:}
BHZ model represents a topological insulating phase in 2D~\cite{Bernevig2006_BHZ} and 
can be writen as, 
\begin{equation}
	\mc{H}_{\rm BHZ}^{\rm 2D}(k_x,k_y) = \lambda (\sin k_x \sigma_x s_z + \sin k_y \sigma_y s_0) + (m-t\cos k_x -t\cos k_y) \sigma_z s_0 \ ,	
\end{equation}
where, $\lambda, m,t$ represent the spin-orbit coupling, crystal-field splitting, and nearest-neighbour hopping amplitudes. As before, `$\sigma$' and `$s$' denote the Pauli matrices in orbital and spin space. To derive the 1D model, we set $k_y=0$ and obtain the following model Hamiltonian in the normal state as, 
\begin{equation}
	\mc{H}_{\rm BHZ}(k_x,k_y=0) = \lambda \sin k_x \sigma_x s_z + (\tilde{m}-t\cos k_x) \sigma_z s_0\ ,	
\end{equation}
with $\tilde{m} = m-t$. Furthermore, to investigate the topological order in presence of superconductivity, we introduce a superconducting pairing term in the 1D BHZ model and write as,
\begin{equation}
	\mc{H}_{\rm BHZ + \Delta} (k) = \begin{pmatrix}
		\mc{H}_{\rm BHZ} (k) & -is_y \Delta \\
		is_y\Delta & -\mc{H}_{\rm BHZ}^T (-k)
	\end{pmatrix}\ ,
\end{equation} 
where $\Delta$ is the superconducting pairing amplitude. 
\vspace{0.3cm}

We compute the real space energy eigenvalues under the open-boundary condition and explore the presence of zero energy modes in both normal and superconducting states. We display the real space energy eigenvalues $E_n$ as a function of state index $n$ in Fig.\,\ref{Fig.S2}(a)-(d). We choose the model paramaters in such a way that the normal system hosts zero energy modes which has been illustrated in Fig.\,\ref{Fig.S2}(a) (SSH model) and in Fig.\,\ref{Fig.S2}(c) (1D BHZ model). In case of spinful SSH model we clearly see the presence of four zero energy modes due to two copies. However, in the supeconducting state these zero modes are completely gapped out leaving behind a tivial superconductor with no topological order (see Fig.\,\ref{Fig.S2}(b)). Similarly, in case of 1D BHZ model, while the normal state has four zero modes, these modes are completely gapped out in presence of superconducting order as shown in Fig.\,\ref{Fig.S2}(d). Thus, 
we establish the special feature of our model Hamiltonian (compared to 1D SSH and BHZ model) where both normal and superconducting state posses topological character in the system.

\end{onecolumngrid}	
	
\end{document}